\documentclass[a4paper11pt]{article}
\parskip \bigskipamount
\usepackage{latexsym,epsf,bm}
\usepackage{amsmath}
\usepackage{amsfonts}
\usepackage{amssymb}
\usepackage{caption}
\usepackage{subcaption}
\usepackage{graphicx}
\usepackage{tikz}
\usepackage{float}
\usepackage{appendix} 

\newcommand{\dyad}[1]{\overline{\overline{\bf {#1}}}}

\begin{document}

\title{Energy Absorption Interferometry }

\author{Stafford Withington$^1$ and Willem Jellema$^2$ \\ \\
$^1$Department of Physics, University of Oxford, UK \\
stafford.withington@physics.ox.ac.uk \\
$^2$ Space Research Organisation Netherlands, Groningen, The Netherlands \\  
W.Jellema@sron.nl}

\date{\today}

\maketitle


\vspace{5mm}

\hspace{6mm} \begin{minipage}{155mm}
 
{\it {\bf Abstract:} Energy Absorption Interferometry (EAI) is a technique for measuring the responsivities and complex-valued spatial polarimetric forms of the individual degrees of freedom through which a many-body system can absorb energy. It was originally formulated using the language of quantum correlation functions, making it applicable to different kinds of excitation (electromagnetic, elastic and acoustic fields).  EAI has been applied in a variety of theoretical and experimental ways. It is particularly effective at characterising the multimode behaviour of ultra-low-noise far-infrared and optical devices, imaging arrays, and complete instruments, where it can be used to ensure that a system is maximally responsive to those partially coherent fields that carry signal whilst avoiding those that only carry noise. Despite its utility there is no comprehensive overview of electromagnetic EAI. In this paper we describe the theoretical foundations of the method, and present a range of new techniques in areas relating to sampling, phase referencing, mode reconstruction and noise. We present, for the first time, an analysis of how noise propagates through an experiment resulting in errors and artefacts on spectral and modal plots. A noise model is essential, because it determines the signal to noise ratio needed to ensure a given level of experimental fidelity.
}

\end{minipage}


\section{Introduction}

Energy Absorption Interferometry (EAI) \cite{With1,With2} is a way of determining the spatial and polarimetric forms of the individual degrees of freedom through which a many-body system can absorb energy. The method probes directly the dynamical forms of the excitations present at any chosen frequency, rather than inferring dynamical behaviour from spectral features. The basic idea is to illuminate the system under test (SUT) with two coherent phase-locked sources, and then to record the complex visibility of the fringe in the total power absorbed as the relative phase between the sources is varied. If the complex visibility is measured for a range of source locations, and where appropriate polarisations, the resulting correlation matrix can be diagonalised to yield the complex-valued forms of the natural modes to which the structure is sensitive, and their individual absorption efficiencies. The recovered modes are intimately related to the collective dynamical relaxation processes present in the structure, such the electrical currents in a film, the flexural  vibrations of a nano-mechanical resonator immersed in a fluid, or spin waves in a magnetic material.  Conceptually, EAI can be viewed in a variety of ways: (i) It measures the spatial state of coherence of the field to which a system is maximally responsive. (ii) It is a generalisation of holography, in the sense that the source and reference are both scanned spatially as an experiment proceeds. (iii) It is the dual of astronomical aperture synthesis interferometry, but rather than measuring the spatial correlations of radiated fields, it measures the fringe in the power absorbed when a structure is illuminated.

Since the original formulation \cite{With1,With2}, EAI has been applied in a variety of ways. It was first demonstrated experimentally, over the frequency range 195–270 GHz, by Thomas \cite{Thom1}, where the basic principles were illustrated. Later, a fibre-based experiment was assembled, using solid-state laser sources, and studied extensively at 1500 nm by Moinard \cite{Moin1,Moin2}, showing that the technique can be used at optical wavelengths. In 2023, Veenendaal and colleagues \cite{Veen1} built a system using optical-laser photomixers, operating over the frequency range 0.835 to 2.72 THz (360 µm to 110 µm), demonstrating that the technique can be implemented at long infrared wavelengths and for polarimetry \cite{Yates1}. This far-infrared system has now been refined, and is being used to characterise the ultra-low-noise imaging arrays and spectrometers being developed for the next generation of far-infrared space telescopes \cite{Roos1,Roos2,Shahab1}. Indeed, EAI can be used to measure the behaviour of complete instruments, in addition to individual devices. 

EAI can also be used as a powerful numerical tool, in combination with rigorous electromagnetic modelling software, to study the properties of complex structures. For example, in the case of a multilayer arrays of far-infrared detectors, EAI is able to use the outputs of the detectors to yield a great deal of information about the multimoded behaviour of the individual pixels, and the spatial forms of the currents that couple detectors and thereby introduce crosstalk \cite{With3}. An elegant application is decribed in the PhD. thesis of Tihon \cite{Tihon1}, and in a series of publications \cite{Tihon2,Tihon4,Tihon5}, where EAI was used with comprehensive electromagnetic modelling to study the absorptive properties of 3D plasmonic materials, including periodic arrays of rods and discs. This work was extended to consider the relationship between the partially coherent radiation and absorption  patterns of thermal and active emitters, such as LEDs \cite{Tihon6,Tihon7,Greff2,Greff3}. A free space EAI experiment was built and used to measure the modal {\em absorption} patterns of a biased light emitting diode (LED) \cite{Tihon8}, opening up a range of intruiging applications, including testing generalisations of Kirchoff's Law for luminescent structures \cite{Tihon9,Greff1,Rau1}. Additionally, the technique is related to other experimental methods such as Phase Shifting Interferometry \cite{Brad1,Rey1} and pump probe sensing \cite{Mar1,Ver1}. 
 
EAI was first described in terms of quantum correlation functions, making it applicable to different kinds of excitation, such as electromagnetic, elastic and acoustic fields. Simulations show that it is effective at probing the forms of spin-coupled systems and waves. It can also be applied, in principle at least, using different kinds of sources simultaneously to yield the spatial forms of dynamical processes that couple to different kinds of field simultanesouly \cite{With1}. The technique has the advantage that because the SUT is illuminated by coherent sources, it is able to reveal correlations that are undetectable in weak radiated thermal fields. An interesting possibility, but one that has not yet been addressed, is to measure the dynamical behaviour of systems driven into nonlinear states, which again is possible because high-power external sources can be used. Finally, there is the possibility of using more than two sources to measure higher-order spatial correlations. 
 
The purpose of this paper is threefold: (i) The original publication was described in terms of quantum correlation functions and generalised forces, and so is not convenient for those wanting to use the method for millimetre-wave, far-infrared and optical characterisation. This update serves to provide a general overview of the method as applied to electromagnetic experiments, together with physical insights that can help guide interested users. (ii) There is now a good understanding of the strengths and weaknesses of different experimental configurations and data analysis procedures, but these have not been presented in unified way. This paper discusses practical considerations that should be taken into account when planning and using EAI. (iii) To date, no one has carried out a noise anaysis describing how thermal background and readout noise propagate through an experiment, and so this paper presents a noise model.  A noise model is central to designing an EAI experiment, because it determines the signal to noise ratio needed to ensure a given level of fidelity, and allows error bars to be put on spectral and modal plots. 
 
 
\section{Power Absorption} 
\label{sec_met}

If an electromagnetic field, ${\bf E}({\bf r},t), {\bf H}({\bf r},t)$, is incident on a structure,  the instantaneous rate at which work is done is given by
\begin{align}
\label{eqn_1a}
P(t)  & =  \int_{\cal T} \int_{\cal V} \int_{\cal V} {\bf E}({\bf r},t) \cdot \dyad{\bm \chi}^{EE} ({\bf r},t;{\bf r}',t') \cdot {\bf E}({\bf r}',t') \, d^3{\bf r} d^3{\bf r}' dt'  \\ \nonumber
& +   \int_{\cal T} \int_{\cal V} \int_{\cal V} {\bf H}({\bf r},t) \cdot \dyad{\bm \chi}^{HH}({\bf r},t;{\bf r}',t') \cdot {\bf E}({\bf r}',t') \, d^3{\bf r} d^3{\bf r}' dt', 
\end{align}
where ${\cal V}$ is a reference volume that contains the SUT, and ${\cal T}$ extends over the period for which the source is turned on. $\dyad{\bm \chi}^{EE} ({\bf r},t;{\bf r}',t')$ and $\dyad{\bm \chi}^{HH} ({\bf r},t;{\bf r}',t')$ are dyadic, possibly nonlocal, response tensors that take into account polarisation and dissipative currents. For example,  $\dyad{\bm \chi}^{EE} ({\bf r},t;{\bf r}',t')$ gives the current at $({\bf r},t)$ as a consequence of having an electric field at $({\bf r}',t')$. As formulated, the fields in (\ref{eqn_1a}) are the total fields after scattering has been taken into consideration, and the response tensors can be calculated using Greens functions, which is discussed in Section \ref{sec_sct}. Strictly, (\ref{eqn_1a}) should be supplemented with cross terms that allow an electric field to generate a magnetic response and magnetic field to generate electric response: as in magnetoelectric materials. In some cases, the response tensor is nonlinear, and perturbation theory can be used to yield expressions that involve higher-order moments of the illuminating field. 

Suppose that we have some measure of the rate of work done, which is derived through some filtering process characterised by the impulse response function $h(t_0 - t)$. Ordinarily, the impulse response function will be determined by time-constants associated with relaxation processes in the structure, and any filtering imposed by post-detection readout electronics. Concentrating on the electric term in (\ref{eqn_1a}),
\begin{align}
\label{eqn_1b}
P(t_0)  & =  \int_{-\infty}^{+\infty} \int_{-\infty}^{+\infty} \int_{\cal V} \int_{\cal V} {\bf E}({\bf r},t) \cdot \dyad{\bm \chi}^{EE} (t_0;{\bf r},t;{\bf r}',t') \cdot {\bf E}({\bf r}',t') \, d^3{\bf r} d^3{\bf r}' dt' dt  \\ \nonumber
\end{align}
where the integral over ${\cal T}$ has been extended to infinity, and 
\begin{align}
\label{eqn_1c}
\dyad{\bm \chi}^{EE} (t_0;{\bf r},t;{\bf r}',t') & =  h(t_0 - t) \, \dyad{\bm \chi}^{EE} ({\bf r},t;{\bf r}',t') \\ \nonumber
\end{align}
In this way, it can be can assumed that the signal at the ouput $P(t_0)$ is proportional the average power absorbed. For example, if the recorded power is simply time averaged for duration $T$, 
\begin{align}
\label{eqn_1}
P(t_0)  & =  \frac{1}{T} \int_{t_0-T/2}^{t_0+T/2}  \int_{-\infty}^{+\infty}  \int_{\cal V} \int_{\cal V} {\bf E}({\bf r},t) \cdot \dyad{\bm \chi}^{EE} ({\bf r},t;{\bf r}',t') \cdot {\bf E}({\bf r}',t') \, d^3{\bf r} d^3{\bf r}' dt' dt.
\end{align}

Adopting the $ F(\omega) e^{-i \omega t}$ convention for time dependence, the response tensor can be cast into the Fourier domain 
\begin{align}
\label{eqn_2}
\dyad{\bm \chi} (t_0;{\bf r},t;{\bf r}',t') & = \frac{1}{(2 \pi)^2} \int \int \dyad{\bm \chi} (t_0;{\bf r},\omega;{\bf r}',\omega')  e^{-i \omega t} e^{+i\omega' t'} \, d\omega d\omega',
\end{align}
to give a complex-valued power:
\begin{align}
\label{eqn_3}
P(t_0)  & = \frac{1}{(2 \pi)^2} \int \int \int_{\cal V} \int_{\cal V}  {\bf E}^\ast({\bf r},\omega) \cdot \dyad{\bm \chi}^{EE} (t_0;{\bf r},\omega;{\bf r}',\omega') \cdot {\bf E}({\bf r}',\omega') \, d^3{\bf r} d^3{\bf r}' d \omega d\omega',
\end{align}
and root mean square (rms) field amplitudes are assumed throughout. It is straightforward to show that if the response is constant parameter, and so depends only on time difference $\overline{\overline{\bm \chi}} ({\bf r},t;{\bf r}',t') \equiv \overline{\overline{\bm \chi}} ({\bf r},{\bf r}',t-t')$, the spectral representation is diagonal $ \overline{\overline{\bm \chi}} ({\bf r},\omega; {\bf r}',\omega') = \overline{\overline{\bm \chi}} ({\bf r},{\bf r}',\omega) 2 \pi \delta ( \omega - \omega')$, and the power spectral density measured on the input side becomes
\begin{align}
\label{eqn_4}
P(t_0,\omega)  & = \int_{\cal V} \int_{\cal V}  {\bf E}^\ast ({\bf r},\omega) \cdot \dyad{\bm \chi} (t_0;{\bf r},{\bf r}',\omega) \cdot {\bf E}({\bf r}',\omega) \, d^3{\bf r} d^3{\bf r}'. 
\end{align}
In what follows, we shall use (\ref{eqn_4}) as the basic form, but not refer to $t_0$ explicitly. A similar expression is found for the magnetic term in (\ref{eqn_1a}). 

The response tensor can be expressed in terms of its hermitian and antihermitian parts,
\begin{align}
\label{eqn_5}
\overline{\overline{\bm \chi}}^{H/A} ({\bf r},{\bf r}',\omega) & = \left[ \frac{   \overline{\overline{\bm \chi}} ({\bf r},{\bf r}',\omega) \pm   \overline{\overline{\bm \chi}}^{\dagger} ({\bf r}',{\bf r},\omega)}{2} \right], 
\end{align}
where $\overline{\overline{\bm \chi}}^{\dagger} ({\bf r}',{\bf r},\omega)$ is the adjoint of $\overline{\overline{\bm \chi}} ({\bf r},{\bf r}',\omega)$. Remembering that $ \overline{\overline{\bm \chi}}^{A'} ({\bf r},{\bf r}',\omega) = -i \overline{\overline{\bm \chi}}^{A} ({\bf r},{\bf r}',\omega)$ is hermitian, it is straightforward to show that $\overline{\overline{\bm \chi}}^{H} ({\bf r},{\bf r}',\omega)$ results in a real-valued power $P^r (\omega)$, whereas $\overline{\overline{\bm \chi}}^{A} ({\bf r},{\bf r}',\omega)$ results in an imaginary-valued power $P^i  (\omega)$. $P^r  (\omega)$ corresponds to the average rate of energy flow into the system, whereas $P^i$ corresponds to the peak value of the energy sloshing in and out of the system, which are the tensorial equivalents of the real and reactive powers used in electrical circuit theory:  Appendix \ref{appendix_A}. For example, according to (\ref{eqn_4}), if a coherent source having frequency $\omega_0$ is used to illuminate the SUT, 
\begin{align}
\label{eqn_6}
& P (\omega_0)  =  \int_{\cal V} \int_{\cal V}  {\bf E}^\ast ({\bf r},\omega_0) \cdot \dyad{\bm \chi} ({\bf r},{\bf r}',\omega_0) \cdot {\bf E}({\bf r}',\omega_0) \, d^3{\bf r} d^3{\bf r}' + {\rm c.c.} \\ \nonumber
& = \int_{\cal V} \int_{\cal V}  {\bf E}^\ast ({\bf r},\omega_0) \cdot \left[ \dyad{\bm \chi}^H ({\bf r},{\bf r}',\omega_0) + \dyad{\bm \chi}^A({\bf r},{\bf r}',\omega_0) \right]  \cdot {\bf E}({\bf r}',\omega_0) \, d^3{\bf r} d^3{\bf r}' + {\rm c.c.} \\ \nonumber
& =  \left\{ \int_{\cal V} \int_{\cal V}  {\bf E}^\ast ({\bf r},\omega_0) \cdot \dyad{\bm \chi}^H ({\bf r},{\bf r}',\omega_0) \cdot {\bf E}({\bf r}',\omega_0) \, d^3{\bf r} d^3{\bf r}'  + i \int_{\cal V} \int_{\cal V}  {\bf E}^\ast ({\bf r},\omega_0) \cdot \dyad{\bm \chi}^{A'} ({\bf r},{\bf r}',\omega_0) \cdot {\bf E}({\bf r}',\omega_0) \, d^3{\bf r} d^3{\bf r}' \right\}  + {\rm c.c.}  \\ \nonumber
& =  \left( P + i  Q \right)  + {\rm c.c.}.
\end{align}
where $\omega_0$ is restricted to being positive. The last line follows because each of the integrals contracts to a real-valued scalar, which depends on how the field propagates through the response tensor. In measurement systems where the response of the readout is sufficiently slow, $T$ is large, the measured value is proportional to the real part of the complex power $S =  P + i  Q $, which is the time-averaged dissipated power. In modelling work when analytic signals are used,  $Q$ can be calculated, which gives the magnitude of the reactive power exchanged between the SUT and the source.

With these considerations in mind, it is convenient  to assume analytic signals such that the measured dissipated power spectral density is given by
\begin{align}
\label{eqn_6b}
P(\omega)  & = \int_{\cal V} \int_{\cal V}  {\bf E}^\ast ({\bf r},\omega) \cdot \dyad{\bm \chi}^H ({\bf r},{\bf r}',\omega) \cdot {\bf E}({\bf r}',\omega) \, d^3{\bf r} d^3{\bf r}'. 
\end{align}
If the incident field is randomly varying, the expectation value of the dissipated power is
\begin{align}
\label{eqn_7}
\langle P(\omega)  \rangle   & =  \int \int_{\cal V} \int_{\cal V}  \dyad{\bm \chi}^H ({\bf r},{\bf r}',\omega)  \cdot \cdot \, \dyad{E} ({\bf r},{\bf r}',\omega)   \, d^3{\bf r} d^3{\bf r}' d \omega, 
\end{align}
which has been obtained from (\ref{eqn_6b}) by taking the trace of both sides, noting that $P(\omega)$ is a scalar, and  cyclically rotating the right prior to calculating the ensemble average. The field correlation tensor $\dyad{E} ({\bf r},{\bf r}',\omega)  = \langle {\bf E}({\bf r},\omega) {\bf E}^\ast ({\bf r}' ,\omega) \rangle$ has been defined, which describes the correlations between different polarisations of the field at different locations; its adjoint is given by  $\dyad{E}^\dagger ({\bf r}',{\bf r},\omega)  = \langle {\bf E}({\bf r}',\omega) {\bf E}^\ast ({\bf r} ,\omega) \rangle$.  By comparing the matrix elements of $\dyad{E} ({\bf r},{\bf r}',\omega)$ and  $\dyad{E}^\dagger ({\bf r}',{\bf r},\omega)$ in some basis, it can be shown that the field correlation tensor is hermitian: $\dyad{E}^\dagger ({\bf r}',{\bf r},\omega)  = \dyad{E} ({\bf r},{\bf r}',\omega) $. The double-dot notation and spatial integrals indicate the full contraction of two tensor fields to a scalar. 

Equation (\ref{eqn_7}) can be viewed in a variety of ways: (i) Over an ensemble, the dissipative and reactive powers are calculated through the full contraction of two tensor fields to a scalar, which is not is not suprising because full contraction is the most natural way of creating a scalar from two tensors. (ii) In the abstract vector space of square-integrable hermitian operators,  (\ref{eqn_7}) describes the projection of an abstract vector  $\dyad{E}({\bf r},{\bf r}',\omega)$, which describes the state of coherence of the field,  onto the abstract vector $\dyad{\bm \chi}^H ({\bf r},{\bf r}',\omega)$, which describes the spatial state of coherence to which the SUT is maximally responsive. Indeed, it is possible to write  (\ref{eqn_7}) in the form
$\langle P(\omega)  \rangle =  \langle P(\omega)  \rangle_{\rm max} \cos \xi $, where $\xi$ provides a normalised measure of the power detected. The power is maximised when the two tensor fields are coaligned $\xi =0$. (iii)  The hermitian operators $\dyad{E}({\bf r},{\bf r}',\omega)$ and $\dyad{\bm \chi}^H ({\bf r},{\bf r}',\omega)$ can both be diagonalised to give the coupled-mode model \cite{Sak1}, which describes the way in which the natural modes of the field, which are individually fully coherent and mutually incoherent, couple to the natural modes of the SUT. The response tensor $\dyad{\bm \chi}^H ({\bf r},{\bf r}',\omega)$ therefore characterises the way in which a partially coherent field dissipates power in the SUT through the degrees of freedom available. 

The formulation used in this section is based on volume integration, which is necessary if the behaviour of a specific device is to be simulated, or volumetric field distributions need to be measured.  Ordinarily, however, a generic characterisation is sought where the detailed form of the SUT is not known or required. A fundamental theorem of electromagnetism states that if the tangential fields over a surface are known, the fields through the enclosed volume are also known. Therefore, energy flow can be reformulated, using Poyntings theorem, in terms of the tangential fields over a reference surface. Expressions identical to those above are found, but now the integrals become two-dimensional surface integrals. Additionally, it is often the case, that the response tensor  $\dyad{\bm \chi}^H ({\bf r},{\bf r}',\omega)$ is zero over much of the reference surface, which simplifies the analysis further. In some cases, such as planar absorbers and detectors, it is sufficient to use a flat reference plane, and then the formulation is particularly straightforward. 


\section{Energy Absorption Interferometry}
\label{sec_eai}

In order to characterise the energy absorbing properties of a structure using EAI, it is necessary to have some measure of the total average power absorbed, such as the temperature, but here we will focus on the response patterns of electromagnetic sensors, such as far-infrared detectors, or energy harvesting devices, where the output is an intrinsic measure of the power absorbed. 

\begin{figure}[H]
     \centering
     \begin{subfigure}[b]{0.45\textwidth}
         \centering
             \includegraphics[trim = 2cm 17cm 8cm 3.5cm, clip,width=90mm ]{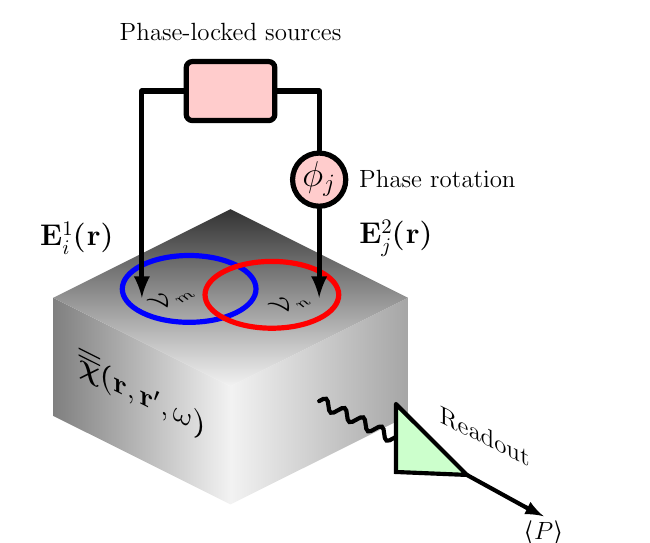}
         \caption{Experimental configuration. The grey box is the SUT; the coherent sources are shown in pale red, and the power readout in pale green.}
         \label{figure1_sub1} 
     \end{subfigure}
     \hfill
     \begin{subfigure}[b]{0.45\textwidth}
         \centering
                     \includegraphics[trim = 2cm 20cm 11cm 3.5cm, clip,width=90mm ]{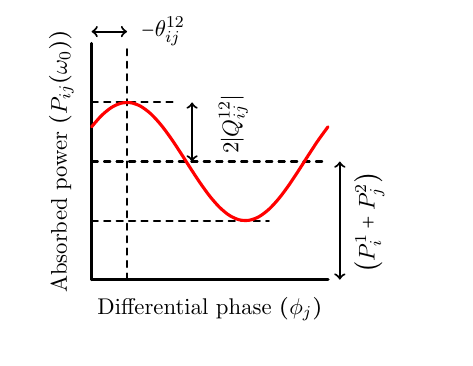}
           \caption{Characteristics of a typical fringe, showing the absorbed power $P_{ij}$, with the sources located at $i,j$, as the differential phase $\phi_j$ is varied.}    
            \label{figure1_sub2}
     \end{subfigure}
        \caption{Diagramatic representation of space-domain Energy Absorption Interferometry.  }
        \label{figure1}
\end{figure}

Consider the situation where a structure is illuminated by two phase-locked coherent sources having different locations: Fig \ref{figure1_sub1}. The power aborbed can be written in terms of a volume or surface integral depending on the application. In what follows, a reference volume is assumed without loss of generality. The total field at the device is given by
\begin{align}
\label{eqn_eai_1}
{\bf E} ({\bf r}, \omega_0) & = {\bf E}^1_i({\bf r}) +  {\bf E}^2_j({\bf r}) e^{i \phi_j},
\end{align}
where  ${\bf E}^1_i({\bf r})$ is the complex-valued field produced by source 1, which is at position (and in orientation) $i$, over the device, and ${\bf E}^2_j({\bf r})$ is the complex-valued field produced by source 2, which is at position $j$. Explicit reference to $\omega_0$ has been dropped in the fields for brevity. In this sense, the subscripts $i,j$ span the set of source positions and polarisations as an EAI experiment is performed. Superscripts $1,2$ indicate that the sources are not necessarily identical. An additional phase factor $\phi_j$ is introduced into one of the arms to allow the global phase of one of the source fields to be varied. This phase factor moves with the position of the second source, and is indexed by $j$ accordingly. In the case of localised source fields, the domains of integration may be different, $\nu_m, \nu_n$ (rather than $\nu$) as shown in Fig \ref{figure1_sub1}, which allows cross coupling to be measured. It what follows, we shall assume that the sources illuminate the same volume.
Then using (\ref{eqn_6})
\begin{align}
\label{eqn_eai_2}
P(\omega_0)  & = \int_{\cal V} \int_{\cal V}  \left(  {\bf E}^1_i({\bf r}) +  {\bf E}^2_j({\bf r}) e^{i \phi_j}  \right)^\ast
\cdot \dyad{\bm \chi}^H ({\bf r},{\bf r}',\omega_0) \cdot  \left(  {\bf E}^1_i({\bf r}') +  {\bf E}^2_j({\bf r}') e^{i \phi_j}  \right) \, d^3{\bf r} d^3{\bf r}' \\ \nonumber
 & = \int_{\cal V} \int_{\cal V}    {\bf E}^{1 \ast}_i({\bf r})\cdot \dyad{\bm \chi}^H ({\bf r},{\bf r}',\omega_0) \cdot   {\bf E}^1_i({\bf r}') + \int_{\cal V} \int_{\cal V}  {\bf E}^{2 \ast}_j({\bf r})\cdot \dyad{\bm \chi}^H ({\bf r},{\bf r}',\omega_0) \cdot  {\bf E}^2_j({\bf r}') \, d^3{\bf r} d^3{\bf r}' \\ \nonumber
 & +  \int_{\cal V} \int_{\cal V}  {\bf E}^{1 \ast} _i({\bf r}) \cdot \dyad{\bm \chi}^H ({\bf r},{\bf r}',\omega_0) \cdot  {\bf E}^2_j({\bf r}') e^{i \phi_j}
 + {\bf E}^{2 \ast}_j({\bf r}) e^{-i \phi_j} \cdot \dyad{\bm \chi}^H ({\bf r},{\bf r}',\omega_0) \cdot {\bf E}^1_i({\bf r}') \, d^3{\bf r} d^3{\bf r}'.
\end{align}
It is convenient to define
\begin{align}
\label{eqn_eai_3}
P^1_i & = \int_{\cal V} \int_{\cal V} {\bf E}^{1 \ast}_i({\bf r}) \cdot \dyad{\bm \chi}^H ({\bf r},{\bf r}',\omega_0) \cdot  {\bf E}^1_i({\bf r}') \, d^3{\bf r} d^3{\bf r}' \\ \nonumber
P^2_j & =   \int_{\cal V} \int_{\cal V}  {\bf E}^{2 \ast}_j({\bf r})\cdot \dyad{\bm \chi}^H ({\bf r},{\bf r}',\omega_0) \cdot  {\bf E}^2_j({\bf r}') \, d^3{\bf r} d^3{\bf r}'  \\ \nonumber
Q^{12}_{ij}  & =  \int_{\cal V} \int_{\cal V}  {\bf E}^{1 \ast} _i({\bf r}) \cdot \dyad{\bm \chi}^H ({\bf r},{\bf r}',\omega_0) \cdot  {\bf E}^2_j({\bf r}') \, d^3{\bf r} d^3{\bf r}' \\ \nonumber
R^{21}_{ji} & = \int_{\cal V} \int_{\cal V} {\bf E}^{2 \ast}_j({\bf r}) \cdot \dyad{\bm \chi}^H ({\bf r},{\bf r}',\omega_0) \cdot {\bf E}^1_i({\bf r}') \, d^3{\bf r} d^3{\bf r}',
\end{align}
where the third and fourth terms are complex conjugates of each other,  $R^{21}_{ji} = Q^{12 \ast}_{ij}$, which can be shown by taking the conjugate transpose of the second term, and remembering that the response is hermitian:
\begin{align}
\label{eqn_eai_4}
\left\{ {\bf E}^{2 \ast}_j({\bf r}) e^{-i \phi_j} \cdot \dyad{\bm \chi}^H ({\bf r},{\bf r}',\omega_0) \cdot {\bf E}^1_i({\bf r}') \right\}^\dagger & =
{\bf E}^{1 \ast}_i({\bf r}) \cdot \dyad{\bm \chi}^H ({\bf r},{\bf r}',\omega_0) \cdot {\bf E}^{2}_j({\bf r}') e^{+i \phi_j}.
\end{align}

The recorded average power for a given source configuration becomes 
\begin{align}
\label{eqn_eai_5}
P_{ij}(\omega_0)  & = P^1_i + P^2_j + Q^{12}_{ij} e^{i \phi_j} + Q^{12 \ast}_{ij}  e^{-i \phi_j} \\ \nonumber
& =  P^1_i + P^2_j + | Q^{12}_{ij} |  \left( e^{i (\theta^{12}_{ij} + \phi_j)}  +  e^{-i ( \theta^{12}_{ij} + \phi_j )} \right) \\ \nonumber
 & = P^1_{i}  + P^2_{j}  + 2 |Q^{12}_{ij}| \cos \left( \theta^{12}_{ij} + \phi_j \right).
\end{align}
where $Q = |Q| e^{i \theta}$. If the sources are identical ${\bf E}^{1}_i({\bf r}) =  {\bf E}^{2}_i({\bf r}) \, \forall i$,
\begin{align}
\label{eqn_eai_6}
P_{ij}(\omega_0)  & = P_{i}  + P_{j}  + 2 |Q_{ij}| \cos \left( \theta_{ij} + \phi_j \right).
\end{align}
The first term in (\ref{eqn_eai_5}), $P^1_i$  is the power that would be absorbed from source 1 at position $i$ if it were alone, and the second term  $P^2_j$ is the power that would be absorbed from source 2 at position $j$ if it were alone. The third term forms a fringe as $\phi_j$ is varied as a consequence of the fields combining over the reference volume or surface. A simple fringe is produced {\em regardless of whether the sources are identical or not}:  Fig. \ref{figure1_sub2}. Looking at (\ref{eqn_eai_3}), it can be seen that the terms are the discrete matrix elements of the continuous operator $\dyad{\bm \chi}^H ({\bf r},{\bf r}',\omega_0)$ in the basis of source beam patterns. This representation is valid as long as the reponse tensor is Hilbert-Schmidt, which is valid for physical systems.Therefore,
\begin{align}
\label{eqn_eai_7}
P_{ij}(\phi_j)  & = D_{ii}  + D_{jj}  + 2 |D_{ij} | \cos \left( \theta_{ij} + \phi_j \right) ,
\end{align}
where $D_{ij}$ is the $ij$'th element of the hermitian response matrix ${\bf \mathsf D}$, which leads to a complex visibility of $2 D_{ij}  / ( D_{ii}  + D_{jj})$. The off-diagonal matrix elements are given by the complex visibilities of the recorded fringes.

If enough positions and polarisations are used, the complex-valued response matrix can be determined from power measurements alone. There are many ways of populating  ${\mathsf D}$. Because ${\bf \mathsf D}$ is hermitian  $\theta_{ii} = \, 0 \, \forall i$, if the sources are placed, effectively, at the same location $i$ then $D_{ii} = P_{ii} (0) / 4$. Experimentally, however, it is generally not known whether $\phi_j=0$ corresponds to zero phase because of differential local phase errors and global offsets, and so it is better to use
\begin{align}
\label{eqn_eai_8}
 D_{ii} & = \frac{P_{ii} (0) +  P_{ii} (\pi)}{4}.
\end{align}
Alternatively, $\phi_i$ can be adjusted until the recorded power is at a maximum, and this gives a zero reference for the phase. In any case, phase reference measurements need to be made throughout an experiment to deal with phase drift: Section \ref{sec_ref}. The off-diagonal elements can be found by placing the sources at different locations, and in different polarisations:
\begin{align}
\label{eqn_eai_9}
\frac{P_{ij} (0) -  P_{ij} (\pi)}{4} & =  {\rm Re}   [ D_{ij}  ] = |D_{ij}| \cos \left( \theta_{ij} \right) \\ \nonumber
\frac{P_{ij} (3 \pi /2) -  P_{ij} (\pi/2)}{4} & =  {\rm Im}   [ D_{ij}  ]  = |D_{ij}| \sin \left( \theta_{ij} \right) \\ \nonumber
D_{ij} & = |D_{ij}| e^{i \theta_{ij}} = \frac{P (0) -  P (\pi)}{4} + i \frac{P (3 \pi /2) -  P (\pi/2)}{4}.
\end{align}
Other combinations are possible, for example fitting the data to a fringe, or using a lock-in amplifier to extract the time varying part of the fringe, $\Delta P_{ij} (0) = 2 |D_{ij}| \cos \left( \theta_{ij} \right)$, $\Delta P_{ij} (\pi/2) = - 2 |D_{ij}| \sin \left( \theta_{ij} \right)$, when the sources are run at slightly different frequencies \cite{Thom1}. It is also possible to calculate the Fourier transform of a sampled fringe \cite{Yates1}.

When this processes is carried out keeping one source fixed and moving the other, the detected ouput corresponds to sweeping out a correlation area, or indeed correlation volume, and so the second source only needs to be moved over the region where the visibility is nonzero. This behaviour is clearly demonstrated in reference \cite{Thom1}. It should be appreciated that, for far-field sources, when the phase of one arm is changed by $\phi_j = \pi /2$, the functional form form of the illuminating field `slides over' the SUT within the envelope of the overall beam pattern. As such, phase-shifted  measurements can be regarded as different field distributions, and analysis could proceed by including them as new entries in ${\bf \mathsf D}$, rather than using separate calculations as in (\ref{eqn_eai_9}).   

Once the sampled response ${\bf \mathsf D}$ has been measured, the actual continuous response $\dyad{\bm \chi}^H ({\bf r},{\bf r}',\omega_0)$ must be reconstructed.  According to (\ref{eqn_eai_3}), the measurements give the matrix elements in the basis of the set of shifted source fields, and therefore reconstruction needs to be carried out using those distributions, either throughout the reference volume
or over the reference surface. If the shifted source fields are orthogonal over the reference volume or surface, it is sufficient to use the source fields themselves, 
\begin{align}
\label{eqn_eai_12}
\dyad{\bm \chi}^H ({\bf r},{\bf r}',\omega_0) & = \sum_{ij}  {\bf E}_i ({\bf r})  D_{ij}  {\bf  E}^\ast_j ({\bf r}'),  
\end{align}
which can be appreciated by subsituting (\ref{eqn_eai_12}) in (\ref{eqn_eai_3}). Equivalently, the individual natural modes can be reconstructed through
\begin{align}
\label{eqn_eai_14}
{\bf u}^n({\bf r}) = \sum_{m}  u_{nm} {\bf E}_m ({\bf r}) .  
\end{align}
where $u_{nm}$ is the $m$'th element of the $n$'th eigenvector, ${\bf u}^n$, of  ${\bf \mathsf D}$. More generally, however, if the field patterns are not orthogonal, such as when the apertures do overlap, or the polarisation is rotated imperfectly on sampling, the basis is not orthogonal, can be overcomplete, and the dual set must be used \cite{With4}, 
\begin{align}
\label{eqn_eai_13}
\dyad{\bm \chi}^H ({\bf r},{\bf r}',\omega_0) & = \sum_{ij}  \tilde{\bf E}_i ({\bf r})  D_{ij}   \tilde{\bf E}^\ast_j ({\bf r}'),  
\end{align}
where $\tilde{E}_i ({\bf r})$ are the dual functions. The use of an overcomplete basis can, in principle at least, be used to achieve a level of super-resolution. The number of samples needed depends on the number of degrees of freedom in the response, which will be discussed in
Section \ref{sec_sam}, and the degree to which the source fields are matched to the natural modes of the response over the reconstruction surface. The ability of a shifted set of source fields to represent the forms of the natural modes is properly described using Frame Theory.

When considering reconstruction, it is important to be mindful of the surface being used. If the source aperture fields are used in the reconstruction, the reconstruction surface is over the surface mapped out by the apertures, which may be in the far field of the device; or if the far-field patterns of the sources are used, the reference surface may be over some surface or volume enclosing the detector. The measurements themselves do not determine the surface of the reconstruction; the reconstructed response can be anywhere between the detector and the sources, or beyond if time-reversed back propagation is used. There is a distinction between whether the measurement is in the near or far field, and whether the reconstruction is in the near or far field. Evanescent fields affect the former, but not the latter because of modal filtering on propagation. 

As an illustration, consider sources having top-hat aperture fields $u_n (r)$, such as the E-plane of a rectangular waveguide. In this case, the expansion coefficients are given by evaluating the function at the sample points, at the centre of each interval, but the reconstructed function is then a collection of top hats. The central points can be joined by a line on a plot, but this is only correct in the limit that the width of the top hat $\Delta r$ tends to zero, and the implied interpolation is acceptable. If the samples are tightly packed, which is not necessarily beneficial, Section \ref{sec_sam}, the reconstruction takes the form of a `staircase'. Likewise, a measurement based on cosine aperture fields, such as the H-plane of a rectangular waveguide, would lead to a reconstruction having the form of a set of displaced cosines. In the extreme, if it is assumed theoretically that true point sources are used, the reconstructed response would be a set of weighted displaced delta functions, which is valid because the response tensor is only defined under an integral as a way of calculating absorbed power. Ultimately, because the response tensor is Hilbert-Schmidt, the reconstructed field may be continuous even though a finite number of samples is taken.

This section has described how EAI measures the matrix elements of $\dyad{\bm \chi}^H ({\bf r},{\bf r}',\omega_0)$ in the basis of the illuminating fields ${\bf E}_i({\bf r},\omega_0)$. Once this has been done, the sampled response matrix ${\bf D}$ can be diagonalised to give the natural modes of the system, which can then be reconstructed in continuous form. In this sense, coherent illuminating fields are used to determine the form of the fully coherent response tensor of the SUT. In many cases, however, the response is itelf a stochastic quantity, such as when the structure comprises a randomly fluctuating medium. Then, EAI measures the set of modes that best represents the average behaviour of the response. In this case, the average response, in an optimal sense, of the SUT is being measured using a set of coherent sources. Equation (\ref{eqn_7}) then gives the power absorbed by a partially coherent system when illuminated by a partially coherent source. The mathematical details of the stochastic case will be described in an upcoming paper, including the recovery of the spatio-temporal behaviour of the response.


\section{K-Domain EAI}
\label{sec_refs}
 
\begin{figure}[H]
         \centering
            \includegraphics[trim = 2cm 16cm 8cm 3cm, clip,width=90mm ]{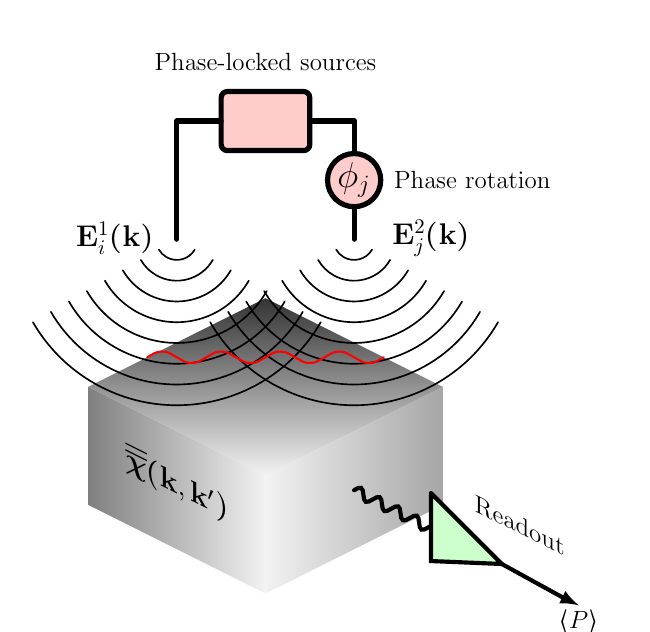}
        \caption{Diagramatic representation of k-domain Energy Absorption Interferometry. The far-field sources produce a fringe across the SUT that samples one Fourier component of the response. }
        \label{figure4}
\end{figure}

Section \ref{sec_eai} formulates EAI in the space domain, but the method can be described in other basis sets, which amounts, effectively, to changing the surface over which the sources are scanned, which is often the surface over which the continuous forms of the modes are recovered. The choice of best reference surface is related to whether near-field or far-field measurements are made, Fig. \ref{figure4}, although there are other reasons why a k-domain description might be best. One advantage is that for a translationally invariant system, which strictly speaking requires the system to have infinite extent, the response tensor can be written solely in terms of a single ${\bf k}$ variable. As discussed, it is possible, often desirable, to calculate the intrinsic properties of a system having infinite extent, and then to take into account the finite size of the actual sample through scattering. The important point is that response functions are often expressed in the k-domain and so it is desirable to describe EAI in the k-domain. A k-domain description can involve both propagating and evanescent modes \cite{Tihon1,Tihon2}, but in what follows, far-field illumination is assumed. 
Fourier transforming the space-domain response tensor, gives the k-domain response tensor:
\begin{align}
\label{eqn_refs_1}
\dyad{\bm \chi} ({\bf r}, {\bf r}')  = \frac{1}{(2 \pi)^{3}} \int_{-\infty}^{+\infty} \frac{1}{(2 \pi)^{3}}  \int_{-\infty}^{+\infty}   \dyad{\bm \chi} ({\bf k}, {\bf k}')  \exp \left[- i {\bf k}\cdot{\bf r} \right] \exp \left[ + i {\bf k}' \cdot {\bf r}' \right] \, d^{3}{\bf k}    d^{3} {\bf k}' \\ \nonumber,
\end{align}
and likewise for the field correlation tensor
\begin{align}
\label{eqn_refs_2}
\dyad{E} ({\bf r}, {\bf r}')  = \frac{1}{(2 \pi)^{3}} \int_{-\infty}^{+\infty} \frac{1}{(2 \pi)^{3}}  \int_{-\infty}^{+\infty}  \dyad{E} ({\bf k}, {\bf k}')  \exp \left[- i {\bf k}\cdot{\bf r} \right] \exp \left[ + i {\bf k}' \cdot {\bf r}' \right] \, d^{3}{\bf k} d^{3} {\bf k}' .
\end{align}

Substituting (\ref{eqn_refs_1}) and (\ref{eqn_refs_2}) in (\ref{eqn_4}) gives
\begin{align}
\label{eqn_refs_3}
P (\omega)  =    \frac{1}{(2 \pi)^{3}}  \int_{\infty}^{+\infty}  \frac{1}{(2 \pi)^{3}}  \int_{\infty}^{+\infty} \dyad{\bm \chi} ({\bf k},{\bf k}') \cdot \cdot \, \dyad{E}^{\dagger} ({\bf k}',{\bf k}) \, d^{3}{\bf k} d^{3}{\bf k}'.
\end{align}
which compares with (\ref{eqn_7}). Again, the total power absorbed takes the form of the contraction of two tensor fields to a scalar, but now the contraction is in the k-domain.

In the case of an interferometric measurement, Fig. \ref{figure4}, where the sources are fully coherent, (\ref{eqn_refs_3}) can be written
\begin{align}
\label{eqn_refs_4}
P (\omega_0)   = &  \frac{1}{(2 \pi)^{3}}  \int_{\infty}^{+\infty}  \frac{1}{(2 \pi)^{3}}  \int_{\infty}^{+\infty} {\bf E}^{\ast} ({\bf k},\omega_0) \cdot \dyad{\bm \chi} ({\bf k},{\bf k}',\omega_{0}) \cdot {\bf E} ({\bf k}',\omega_{0}) \, d^{3}{\bf k}  d^{3}{\bf k}' .
\end{align}
The matrix elements of the response tensor are now measured in the k domain. In those cases where the sources produce plane waves, an interferometric measurement records elements of the k-domain response tensor directly. 

More generally, for partially coherent fields,  (\ref{eqn_refs_3}) can be written in the convenient form 
\begin{align}
\label{eqn_refs_5}
P (\omega)  =    \int  \int  \dyad{\bf A}_e ({\bf \Omega},{\bf \Omega}', \omega) \cdot \cdot \, \dyad{B}^{\dagger} ({\bf \Omega}',{\bf \Omega},\omega) \, d{\bf \Omega} d{\bf \Omega}',
\end{align}
where the integrals are performed over the unit sphere. $\dyad{\bf A}_e ({\bf \Omega},{\bf \Omega}', \omega)$ is an {\em effective area tensor} and $\dyad{B} ({\bf \Omega},{\bf \Omega}', \omega)$ is a {\em brightness tensor}. This formulation, which includes polarisation, can be derived rigorously, and is beneficial because it provides a direct link with radiometry. Additionally, the magnetic self and cross terms can be included, giving a complete formulation of (\ref{eqn_1}) in the angular domain. 

In the case of plane-wave illumination, the smallest feature that can be resolved is determined by the wavelength of the incident field, which together with restrictions on the polarisation, effectively induces angular correlations in the response tensor of the SUT \cite{Chuss1,Thom2,Thom3}. In the context of detectors, the interferometric method measures the far-field angular response tensor, which can then be decomposed to give the amplitude, phase, polarisation patterns, and responsivities of modes through which the SUT absorbs power.


\section{Dual-Surface EAI}
\label{sec_dsurf}

Suppose that an SUT is illuminated by two coherent phase-locked sources, one of which moves over some surface ${\cal U}$, whilst the other moves over some surface ${\cal V}$. If ${\cal U} = {\cal V}$, the usual EAI is performed, but generally ${\cal U} \neq {\cal V}$. The field of the first source on ${\cal U}$, which is two dimensional, creates a three dimensional field on the device, and likewise for the second source. It is convenient to write
 \begin{align}
\label{eqn_dsurf_1}
{\bf E}^1 ({\bf r},\omega) & = \int_{{\cal S}^u} \dyad{T}^u ({\bf r}, {\bf r}^u) \cdot {\bf E}^u ({\bf r}^u) \, d^2 {\bf r}^u \\ \nonumber
{\bf E}^2 ({\bf r},\omega) & = \int_{{\cal S}^v} \dyad{T}^v ({\bf r}, {\bf r}^v) \cdot {\bf E}^v ({\bf r}^v) \, d^2 {\bf r}^v,
\end{align}
where  $\dyad{T}^u ({\bf r}, {\bf r}^u)$ and  $\dyad{T}^v ({\bf r}, {\bf r}^v)$ are operators that propagate the tangential fields
${\bf E}^u ({\bf r}^u)$ and ${\bf E}^v ({\bf r}^v)$, on ${\cal U}$ and ${\cal V}$, over the device. ${\bf E}^1 ({\bf r},\omega)$ is a three-dimensional vector field, whereas  ${\bf E}^u ({\bf r}^u)$ and ${\bf E}^v ({\bf r}^v)$ are two-dimensional fields. Likewise, ${\bf r}$ is a three-dimensional position vector, whereas ${\bf r}^u$ and ${\bf r}^v$  are two-dimensional. $\dyad{T}^u ({\bf r}, {\bf r}^u)$ and  $\dyad{T}^v ({\bf r}, {\bf r}^v)$ therefore map between two different vector spaces.

The power absorbed by the SUT is given by (\ref{eqn_eai_2}), which again results in 4 terms: The first term gives
\begin{align}
\label{eqn_dsurf_2}
P^1(\omega)  & = \int_{\cal V} \int_{\cal V} {\bf E}^{1 \ast} ({\bf r},\omega)  \cdot \dyad{\bm \chi} ({\bf r},{\bf r}',\omega) \cdot {\bf E}^1({\bf r}',\omega)  \, d^3{\bf r} d^3{\bf r}' \\ \nonumber 
& = \int_{{\cal S}^u} \int_{{\cal S}^u} {\bf E}^{u \ast}  ({\bf r}^u)  \cdot \dyad{\bm \chi}^u ({\bf r}^u,{\bf r}^{u'},\omega) \cdot {\bf E}^u ({\bf r}^{u'}) \, \, d^2 {\bf r}^ud^2 {\bf r}^{u'} \\ \nonumber  
\end{align}
where
\begin{align}
\label{eqn_dsurf_3}
\dyad{\bm \chi}^u ({\bf r}^u,{\bf r}^{u'},\omega) & = \int_{\cal V} \int_{\cal V}  \dyad{T}^{u \dagger} ({\bf r}^u, {\bf r})   \cdot \dyad{\bm \chi} ({\bf r},{\bf r}',\omega) \cdot \dyad{T}^u ({\bf r}', {\bf r}^{u'})\,  \, d^3{\bf r} d^3{\bf r}' 
\end{align}
is the newly referenced hermitian response tensor over ${\cal U}$. It is the quantity that would be measured by an EAI experiment if both sources were moved over ${\cal U}$. $P^1(\omega)$ is therefore the power absorbed from source 1 alone.  Likewise the second term becomes
\begin{align}
\label{eqn_dsurf_4}
P^2(\omega)  & = \int_{{\cal S}^v} \int_{{\cal S}^v} {\bf E}^{v \ast}  ({\bf r}^v)  \cdot \dyad{\bm \chi}^v ({\bf r}^v,{\bf r}^{v'},\omega) \cdot {\bf E}^v ({\bf r}^{v'}) \, \, d^2 {\bf r}^v d^2 {\bf r}^{v'} \\ \nonumber  
\end{align}
where
\begin{align}
\label{eqn_dsurf_5}
\dyad{\bm \chi}^v ({\bf r}^v,{\bf r}^{v'},\omega) & = \int_{\cal V} \int_{\cal V}  \dyad{T}^{v \dagger} ({\bf r}^v, {\bf r})   \cdot \dyad{\bm \chi} ({\bf r},{\bf r}',\omega) \cdot \dyad{T}^v ({\bf r}', {\bf r}^{v'})\,  \, d^3{\bf r} d^3{\bf r}', 
\end{align}
is the hermitian response tensor that would be measured if both sources were moved over surface ${\cal V}$. $P^2(\omega)$ is the power absorbed from source 2 alone. The remaining terms account for the interference 
\begin{align}
\label{eqn_dsurf_6}
P^{12}(\omega)  & = \int_{{\cal S}^u} \int_{{\cal S}^v} {\bf E}^{u \ast}  ({\bf r}^u)  \cdot \dyad{\bm \chi}^{uv} ({\bf r}^u,{\bf r}^{v},\omega) \cdot {\bf E}^v ({\bf r}^{v}) \, \, d^2 {\bf r}^u d^2 {\bf r}^{v} \\ \nonumber  
\end{align}
where
\begin{align}
\label{eqn_dsurf_7}
\dyad{\bm \chi}^{uv} ({\bf r}^u,{\bf r}^{v},\omega) & = \int_{\cal V} \int_{\cal V}  \dyad{T}^{u \dagger} ({\bf r}^u, {\bf r})   \cdot \dyad{\bm \chi} ({\bf r},{\bf r}',\omega) \cdot \dyad{T}^v ({\bf r}', {\bf r}^{v})\,  \, d^3{\bf r} d^3{\bf r}', 
\end{align}
and
\begin{align}
\label{eqn_dsurf_8}
P^{21}(\omega) & = \int_{{\cal S}^u} \int_{{\cal S}^v} {\bf E}^{v \ast}  ({\bf r}^v)  \cdot \dyad{\bm \chi}^{vu} ({\bf r}^v,{\bf r}^{u},\omega) \cdot {\bf E}^u ({\bf r}^{u}) \, \, d^2 {\bf r}^u d^2 {\bf r}^{v} \\ \nonumber  
\end{align}
where
\begin{align}
\label{eqn_dsurf_9}
\dyad{\bm \chi}^{vu} ({\bf r}^v,{\bf r}^{u},\omega) & = \int_{\cal V} \int_{\cal V}  \dyad{T}^{v \dagger} ({\bf r}^v, {\bf r})   \cdot \dyad{\bm \chi} ({\bf r},{\bf r}',\omega) \cdot \dyad{T}^u ({\bf r}', {\bf r}^{u})\,  \, d^3{\bf r} d^3{\bf r}'. 
\end{align}
It follows that
\begin{align}
\label{eqn_dsurf_10}
\dyad{\bm \chi}^{vu} ({\bf r}^v,{\bf r}^{u},\omega) & = \left[ \dyad{\bm \chi}^{uv} ({\bf r}^u,{\bf r}^{v},\omega) \right]^\dagger,
\end{align}
because the underlying response tensor of the device is hermitian.  The cross response tensors form a conjugate pair, but they are not individually hermitian. Because of (\ref{eqn_dsurf_10}), $P^{12}(\omega)$ and $P^{21}(\omega)$ are complex conjugates of each other. Once again, according to (\ref{eqn_eai_6}), a fringe is formed as the phase difference between the sources in varied. Now, however, the matrix elements are those of $\dyad{\bm \chi}^{uv} ({\bf r}^u,{\bf r}^{v},\omega)$ in the basis ${\bf E}^{u}_i  ({\bf r}^u)$ and  ${\bf E}^v_j ({\bf r}^{v})$. The matrix ${\bf D}_{ij}$ is not hermitian and not necessarily square, but its SVD can be used to calculate the modes on ${\cal U}$ and the modes on ${\cal V}$ that overlap and couple to the device, and so produce interference in the power absorbed.

Suppose that ${\bf D}^u$ and ${\bf D}^v$ are discretised response matrices associated with EAI measurements on ${\cal U}$ and ${\cal V}$ alone, and   ${\bf D}^{uv}$ and  ${\bf D}^{vu}$ come from dual-surface EAI. The question is whether the modes that come from diagonalising the single-surface measurements are the same as those that come from the SVD of the dual-surface measurement. If so, 
they can be assembled into the following hermitian block-matrix form:
\begin{align}
\label{eqn_dsurf_11}
\left[
\begin{matrix}
{\bf D}^u  &  {\bf D}^{uv} \\
{\bf D}^{vu}  & {\bf D}^v
\end{matrix}
\right] = 
\left[
\begin{matrix}
 {\bf U} {\boldsymbol \Sigma}^u {\bf U}^\dagger & {\bf U} {\boldsymbol \Sigma}^{uv} {\bf V}^\dagger \\
 {\bf V} {\boldsymbol \Sigma}^{vu} {\bf U}^\dagger & {\bf V} {\boldsymbol \Sigma}^v {\bf V}^\dagger
\end{matrix}
\right] = 
\left[
\begin{matrix}
 {\bf U} & {\bf 0} \\
{\bf 0} & {\bf V}
\end{matrix}
\right] 
\left[
\begin{matrix}
{\boldsymbol \Sigma}^u  & {\boldsymbol \Sigma}^{uv} \\
{\boldsymbol \Sigma}^{vu} & {\boldsymbol \Sigma}^v 
\end{matrix}
\right]
\left[
\begin{matrix}
 {\bf U} & {\bf 0} \\
{\bf 0} & {\bf V}
\end{matrix}
\right]^\dagger,
\end{align}
This seems reasonable because the intrinsic modes of the device are being measured on each surface, and so connected by a propagator. It seems that the information contained in each off-diagonal block is essentially the same as that contained in the on-diagonal blocks, apart from an additional propagation factor. In other words, the modes available for absorbtion on ${\cal U}$ are essentially the same as those available on ${\cal V}$.

Suppose that fields on ${\cal U}$ and ${\cal V}$ are connected by some propagator ${\bf T}$: ${\cal V} \rightarrow {\cal U}$. A dual-surface EAI experiment measures
\begin{align}
\label{eqn_dsurf_12}
{\bf D}^{uv} & = {\bf D}^u {\bf T}.
\end{align}
At this stage, it is not clear whether the SVD of the off-diagonal blocks can be written ${\bf D}^{uv} = {\bf U} {\boldsymbol \Sigma}^u {\bf V}^\dagger$; in other words, whether the singular vectors ${\bf U}$ and ${\bf V}$ are preserved on the left and right respectively. The left singular vectors of ${\bf D}^{uv}$ can be found by diagonalising 
\begin{align}
\label{eqn_int_3}
\left( {\bf D}^{uv} \right) \left( {\bf D}^{uv} \right)^\dagger  & = {\bf D}^u {\bf T}  {\bf T}^\dagger {\bf D}^{u \dagger} 
\end{align}
The left singular vectors are preserved, $ \left( {\bf D}^{uv} \right) \left( {\bf D}^{uv} \right)^\dagger = \left( {\bf D}^{u} \right) \left( {\bf D}^{u} \right)^\dagger$, as long as the propagator ${\bf T}$ is complete with respect to the eigenvalues  ${\bf U}$ having appreciable eigenvalues. Writing
\begin{align}
\label{eqn_dsurf_13}
{\bf T} {\bf T}^\dagger = {\bf I}_u,
\end{align}
where ${\bf I}_u$ is the identity operator spanning the range space of ${\bf D}^{u}$, and then
\begin{align}
\label{eqn_int_5}
\left( {\bf D}^{uv} \right) \left( {\bf D}^{uv} \right)^\dagger  & = {\bf D}^u {\bf D}^{u \dagger}, 
\end{align}
and so the left singular vectors are indeed given by the columns of ${\bf U}$. Equation (\ref{eqn_dsurf_13}) states that it must be possible to back propagate each of the the modes on ${\cal U}$ to ${\cal V}$, and then forward propagate them again to ${\cal U}$ without changing their forms, which seems a reasonable physical requirement. Crucially, the measured spatial spectrum is that of the device itself.

Likewise, the right singular vectors of ${\bf D}^{uv}$ can be found by diagonalising 
\begin{align}
\label{eqn_dsurf_14}
 \left( {\bf D}^{uv} \right)^\dagger \left( {\bf D}^{uv} \right)  & = {\bf T}^\dagger {\bf D}^{u \dagger}  {\bf D}^u {\bf T} \\ \nonumber
 & =  {\bf T}^\dagger {\bf U} (\Sigma^{u} )^2 {\bf U}^\dagger {\bf T} \\ \nonumber
  &= {\bf V} (\Sigma^{v} )^2 {\bf V}^\dagger, 
\end{align}
where
\begin{align}
\label{eqn_dsurf_15}
{\bf V}^\dagger & = {\bf U}^\dagger {\bf T} \\ \nonumber
{\bf V} & = {\bf T}^\dagger {\bf U} \\ \nonumber
{\bf U} & = {\bf T} {\bf V}
\end{align}
The second line describes the time reversed back propagation of the modes on ${\cal U}$  to the modes on ${\cal V}$, as expected. The last line again requires the propagator to be complete. It seems that (\ref{eqn_dsurf_11}) is valid as long as the propagator is unitary with respect to the subspace containing the appreciable modes of the SUT. 

Free-space propagation is unitary, and so this formulation is highly valuable. Given that the SVD of the dual-surface measurement gives ${\bf U}$ and ${\bf V}$, the propagator is given by 
\begin{align}
\label{eqn_dsurf_16}
{\bf T} & = {\bf U} {\bf V}^{\dagger},
\end{align}
because the eigenmodes are orthonormal. If point-source measurements are carried out on the two surfaces, the entries in ${\bf T}$ give the point to point propagation factors between the two surface, which may include spreading, and cross -polar coupling. This coupling is only defined to within the degrees of freedom that influence the measurement of the specific SUT, and so provides a smoothed coupling function. For example, if the two surfaces are the result of imperfect scanning geometries, where one source follows a slightly different surface to the other.

Consider some specific cases. If the surfaces are close, and the polarisation preserved, the point to point mapping is one to one, with a diagonal phase factor. Then
\begin{align}
\label{eqn_dsurf_17}
{\bf U} & =  {\bf T} {\bf V} \\ \nonumber
{\bf U} & = \boldmath{\Theta} {\bf V}, 
\end{align}
where $\boldmath{\Theta}$ is a diagonal matrix of phase factors. Each mode of ${\bf U}$ is the same as that of ${\bf V}$, but with a point to point multiplicative phase factor. It can also have a position independent loss.

If  ${\boldsymbol  \Sigma}^p = \eta \boldmath{\Theta}$, where $\eta$ is a common loss factor, then
\begin{align}
\label{eqn_dsurf_18}
{\bf D}^{uv} & =   {\bf U} {\boldsymbol \Sigma}^u {\bf U}^\dagger \,  \eta \boldmath{\Theta}  \\ \nonumber
& =  {\bf U} {\boldsymbol \Sigma}^{uv} {\bf V}^\dagger,
\end{align}
and a measurement determines $ {\bf U}$, ${\boldsymbol \Sigma}^{uv} = \eta {\boldsymbol \Sigma}^u $, and ${\bf V} =  \boldmath{\Theta}^\dagger {\bf U} $. Trivially, ${\boldsymbol \Sigma}^v   =  |\eta|^2 {\boldsymbol  \Sigma}^u$. Thus the aborptive modes of the SUT over ${\cal U}$ and ${\cal V}$ are given directly by a single EAI experiment, and crucially the singular values of the SVD preserve the spatial spectrum of the SUT's response to within a common loss factor

If the propagator is not unitary over the subspace of the modes, information can still be recovered. For example, suppose that the SUT has a broad spectrum of degenerate modes, then ${\boldsymbol  \Sigma}^u = \lambda {\bf I}_u$, and 
\begin{align}
\label{eqn_dsurf_19}
{\bf D}^{uv} & =  \lambda {\bf U} {\bf U}^\dagger \,  {\bf S} {\boldsymbol \Sigma}^p {\bf R}^\dagger \\ \nonumber
 & =  {\bf S}  \lambda {\boldsymbol  \Sigma}^p  {\bf R}^\dagger, 
\end{align}
then a measurement determines  the modes of the optical system: ${\bf S}$,  ${\boldsymbol  \Sigma}^p$, and  ${\bf R}^\dagger$. In general, neither the SUT nor the optical system dominates, and a dual-surface measurement characterises the composite behaviour. In this case, single-surafce and dual-surface measurements  could be carried out, and (\ref{eqn_dsurf_11}) diagonalised, but this is not seem worthwhile in the majority of cases.

Dual-surface EAI has a number of potential advantages: (i) The functional forms of the modes over the two surfaces are given directly through SVD. The continuous forms of the modes on the two surfaces, the response tensor,  and the propagator can be recovered using the basis functions of the scanned sources.  Ordinarily, it would be necessary to calculate the modes on one surface and then transform them using $ {\bf T}$ if both the far field and the near field forms are needed. Here, however, it is not necessary to carry out a numerical transform using some assumed form for the propagator; indeed the measured propagator is given by  (\ref{eqn_dsurf_12}). (ii) If the two surfaces are notionally, but not exactly, the same, the method allows the discretised forms of the modes to be found over the two, not equal, surfaces. This allows the propagator to be recovered, which includes the forms of any phase and polarisation errors accrued as a consequence of having imperfect scanning geometry. Even in the case of single-surface EAI, it seems prudent to use SVD to analyse a set of data. (iii) In certain cases, such as a  detector followed by a paraxial optical system, dual-surface EAI may provide a way of gaining access to the behaviour of the foreoptics. It may also be beneficial when measuring the behaviour of systems exhibiting exotic behaviour such as optical vortices. 


\section{Phase-Space EAI}
\label{sec_pseai}

\begin{figure}[H]
         \centering
          \includegraphics[trim = 2cm 16cm 8cm 3cm, clip,width=90mm ]{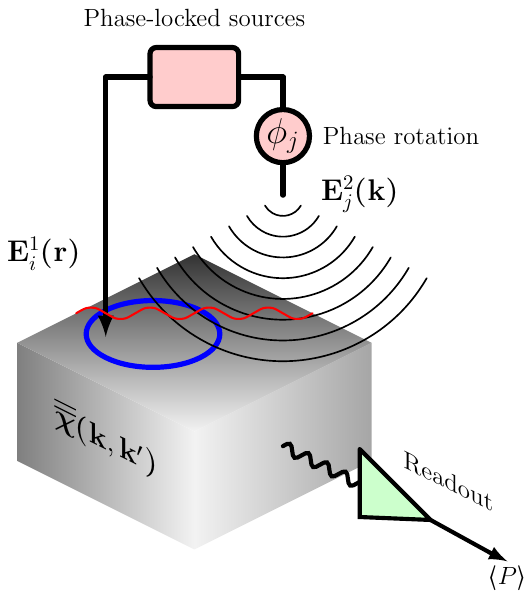}
        \caption{Diagramatic representation of phase-space Energy Absorption Interferometry, where far-field and near-field sources are used.}
        \label{figure4b}
\end{figure}

A particular example of dual-surface EAI is when one of the sources is in the near field and one in the far field of the device being studied: Fig. \ref{figure4b}. Consider a planar device,
\begin{align}
\label{eqn_cnt_1}
P(\omega)  & = \int_{\cal V} \int_{\cal V}  {\bf E}^\ast ({\bf r},\omega) \cdot \dyad{\bm \chi} ({\bf r},{\bf r}',\omega) \cdot {\bf E}({\bf r}',\omega) \, d^2{\bf r} d^2{\bf r}', 
\end{align}
where $\dyad{\bm \chi} ({\bf r},{\bf r}',\omega)$ is the spatial response over a reference plane just in front of the device. The mapping between the surface and the volume of the device has been omitted for brevity, but could be included easily. Suppose that we wish to characterise the device in terms of this near-field surface ${\cal U}$ and a far-field sphere ${\cal V}$. Then using
\begin{align}
\label{eqn_cnt_2}
{\bf E}({\bf r}',\omega)  & = \frac{1}{(2 \pi)^{2}}  \int  {\bf E}({\bf k},\omega) \exp \left[ - i {\bf k} \cdot {\bf r}' \right] \, d^{2} {\bf k},
\end{align}
and so 
\begin{align}
\label{eqn_cnt_3}
P(\omega)  & = \frac{1}{(2 \pi)^{2}}  \int_{\cal K} \int_{\cal V} \int_{\cal V}  {\bf E}^\ast ({\bf r},\omega) \cdot \dyad{\bm \chi} ({\bf r},{\bf r}',\omega) \cdot {\bf E}({\bf k},\omega)  \exp \left[ - i {\bf k} \cdot {\bf r}' \right] \, d^{2} {\bf k} d^2{\bf r} d^2{\bf r}' \\ \nonumber
  & = \frac{1}{(2 \pi)^{2}}  \int \int_{\cal V}  {\bf E}^\ast ({\bf r},\omega) \cdot \dyad{\bm \chi} ({\bf r},{\bf k},\omega) \cdot {\bf E}({\bf k},\omega)  \, d^{2} {\bf k} d^2{\bf r},
\end{align}
where
\begin{align}
\label{eqn_cnt_4}
\dyad{\bm \chi} ({\bf r}, {\bf k},\omega)  =  \int   \dyad{\bm \chi} ({\bf r}, {\bf r}', \omega)  \exp \left[ - i {\bf k} \cdot {\bf r}' \right] \, d^{2} {\bf r}'.
\end{align}
$\dyad{\chi} ({\bf r}, {\bf k},\omega)$ shall be referred to as the {\em phase-space response tensor}, because it characterises behaviour in a mixture of the ${\bf r}$ and ${\bf k}$ domains. Strictly the first integral in (\ref{eqn_cnt_3}) should extend to infinity, but for all practical purposes the response is spatially band limited. It then follows directly from Section \ref{sec_dsurf} that if a point source is moved over a near-field surface and a point source is moved over a far-field surface, the matrix elements become spatial samples of $\dyad{\bm \chi} ({\bf r}, {\bf k}',\omega)$. Remember that for each near-field position, the far-field probe only needs to be moved over some coherent solid angle. In fact, this phase-space representation is closely related to the Wigner-Weyl transform. The Wigner-Wyel representation of the beamed partially coherent reception pattern can be derived from the EAI phase-space measurement by a rotation of the spatial coordinates into sum and difference terms.

Finally, if $\langle {\bf E}({\bf k},\omega)  {\bf E}^\ast ({\bf r},\omega) \rangle$ is the phase-space field correlation tensor,  the power is given by a tensor contraction as before:
\begin{align}
\label{eqn_cnt_5}
P (\omega)  =      \frac{1}{(2 \pi)^{3}}  \int \int_{\cal V}\dyad{\bm \chi} ({\bf r},{\bf k}) \cdot \cdot \, \dyad{E}^{\dagger} ({\bf r},{\bf k}) \, d^{3}{\bf r} d^{3}{\bf k},
\end{align}
which is also true of the Wigner-Weyl form.


\section{Scattering and Crosstalk}
\label{sec_sct}

To appreciate EAI further, consider scattering. Here, scattering is described in terms of electromagnetic measurements, but the principles apply to other kinds of excitation also. Suppose that the fields radiated by the illuminating sources scatter off some other nearby, possibly absorbing, component, which may be a neighboring pixel in an imaging array.  The total field over the SUT,  ${\bf E} ({\bf r})$, is the sum of the incident field, ${\bf E}^i ({\bf r})$, and the scattered field, ${\bf E}^s ({\bf r})$, and so
\begin{align}
\label{eqn_sct_1}
{\bf E}^i ({\bf r}) & = {\bf E} ({\bf r}) - {\bf E}^s ({\bf r}) \\ \nonumber
& =  {\bf E} ({\bf r}) - \int_{\cal{\nu}'}  \dyad{G}({\bf r},{\bf r}') \cdot {\bf E} ({\bf r}') \, d^3{\bf r}' \\ \nonumber
& = \int_{\cal{V}'}  \, d^3{\bf r}'  \left[ \dyad{I} \delta( {\bf r} - {\bf r}') -   \dyad{G}({\bf r},{\bf r}')  \right]  \cdot {\bf E} ({\bf r}'),
\end{align}
where ${\cal V}'$ encloses all external regions where scattering currents may be present. $\dyad{G}({\bf r},{\bf r}')$ is a Green's function, which describes how the total field at ${\bf r}'$ leads to reradiation that modifies the field at ${\bf r}$.  In electromagnetics,  (\ref{eqn_sct_1}) is the Electric Field Integral Equation (EFIE), which can be inverted to give the total field, or equivalently the current at any point, in terms of the incident field. A complete scheme for calculating the power absorbed by patterned thin films over multiple planes is described by Withington \cite{With3}. 

For our purposes, it is sufficient to realise that  (\ref{eqn_sct_1}) can be inverted to give
\begin{align}
\label{eqn_sct_2}
{\bf E} ({\bf r}') & = \int \dyad{S}({\bf r}',{\bf r}) \cdot {\bf E}^i ({\bf r})  \, d^3{\bf r},
\end{align}
where  $\dyad{S}({\bf r}',{\bf r})$ is a scattering operator. The inversion needed to calculate  $\dyad{S}({\bf r},{\bf r}')$ reveals multi-path processes, which lead to nonuniform induced currents in films and screening in thick media. When there is no external scattering  $\dyad{G}({\bf r},{\bf r}') = \dyad{0}$, which is separate from the existence of internal scattering, which is already contained in $\dyad{\bm \chi}^H ({\bf s},{\bf s}',\omega_0)$, the total field incident on the device is simply the incident field. Alternatively, external and internal scattering may calculated simultaneously, giving multipath scattering between the device and external structures. 

Regardless, substituting  (\ref{eqn_sct_2}) in (\ref{eqn_4}) gives
\begin{align}
\label{eqn_sct_3}
P(\omega_0)  & = \int_{\cal V} \int_{\cal V}  {\bf E}^\ast ({\bf r},\omega_0) \cdot \dyad{\bm \chi}' ({\bf r},{\bf r}',\omega_0) \cdot {\bf E}({\bf r}',\omega_0) \, d^3{\bf r} d^3{\bf r}'. 
\end{align}
where the new reponse tensor, which includes scattering, is given by
\begin{align}
\label{eqn_sct_4}
 \dyad{\bm \chi}' ({\bf r},{\bf r}',\omega_0) & = \int \int  \dyad{S}^\dagger({\bf r},{\bf s}) \cdot \dyad{\bm \chi}^H ({\bf s},{\bf s}',\omega_0) \cdot  \dyad{S}({\bf s}',{\bf r}') \, d^3{\bf s} d^3{\bf s}'.
\end{align}
$\dyad{\bm \chi}' ({\bf r},{\bf r}',\omega_0)$ is the response tensor measured by an EAI experiment. Equation (\ref{eqn_sct_4}) shows that the scattering operator wraps around the undressed response tensor to create a dressed response tensor. In general terms, internal and external scattering can be included by adding repeated layers of wrapping.

Scattering may be produced by dissipative or reactive currents in the scatterer. Although the spatial modes associated with the undressed device are local to the volume or area of the device, the modes associated with the dressed device may extend beyond the volume of the device. For example, the effective area of a pixel in an array may be greater than the physical size of the pixel, and may involve other pixels or layers of the device. Any dissipative scattering dissipates power in the scatterer as well as the device. In the case of neighboring pixels in an array, the reception fields obtained by EAI may overlap, and so the pixels effectively overlap \cite{With3}. The experimental system itself may scatter fields, leading to multipath reflections, standing waves, and experimental strategies should be sought to minimise these effects.


\section{Scanning Errors}
\label{sec_sce}

The basic scheme described above does not do justice to the errors that can degrade an experiment. For example, ideally, the field produced over the device by source 2 alone when at position $i$ should be the same as the field produced by source 1 alone when at position $i$.  Ideally, when measuring the non-diagonal terms, it is required that $D_{ij} = D_{ji}^\ast$. Because the response matrix is hermitian, it may be possible to record only half the data, say the upper off-axis triangle of  ${\bf \mathsf D}$ , and then to infer the other, but it is usually be best to measure the whole of  ${\bf \mathsf D}$ in order to reveal any assymetries introduced by the experiment. If the sources are not identical, in the sense that swapping the sources over does not lead to a conjugate response, $|Q^{12}_{ji}| \neq |Q^{12}_{ij}|$ and $\theta^{12}_{ji} \neq -\theta^{12}_{ij}$, the antihermitian part, or the SVD, of the measured response can be calculated to reveal the form of any experimental errors. 

In any real experiment based on mechanical scanning, it is exceedingly difficult to ensure coaligned sources. Assume that sources 1 and 2 have position-dependent phase errors $\psi^1_i({\bf r})$ and $\psi^2_j({\bf r}')$ respectively, which vary with location $i,j$, then $P^1_i$ and  $P^2_j$ remain unchanged, but
\begin{align}
\label{eqn_eai_10}
Q^{12}_{ij}  & =  \int_{\cal V} \int_{\cal V}  {\bf E}^{1 \ast} _i({\bf r}) \cdot \dyad{\bm \chi}^H ({\bf r},{\bf r}',\omega_0) \cdot  {\bf E}^2_j({\bf r}')  e^{i \delta \psi_{ij} ( {\bf r}, {\bf r}')} \, d^3{\bf r} d^3{\bf r}' \\ \nonumber
R^{21}_{ji} & = \int_{\cal V} \int_{\cal V} {\bf E}^{2 \ast}_j({\bf r}) \cdot \dyad{\bm \chi}^H ({\bf r},{\bf r}',\omega_0) \cdot {\bf E}^1_i({\bf r}')  e^{-i \delta \psi_{ij} ( {\bf r}, {\bf r}')} \, d^3{\bf r} d^3{\bf r}'
\end{align}
where $\delta \psi_{ij} ( {\bf r}, {\bf r}') = -\psi_i^1({\bf r}) + \psi_j^2({\bf r}')$. Despite these phase errors, the measured data still shows a simple fringe, but the off-diagonal matrix elements accrue entry-dependent phase errors $e^{i \delta \psi_{ij} ( {\bf r}, {\bf r}')}$. An effective way of dealing with these errors is to calculate the antihermitian part of the response matrix, strip out those eigenvalues smaller than some threshold, reconstruct  the smoothed version of  $\dyad{\bm \chi}^A ({\bf r},{\bf r}',\omega_0)$, and then study the reconstruction for large-scale structure, such as a uniform offset, tilts and curvature. Altervatively, the SVD of the data can be calculated as described in Section \ref{sec_dsurf}. The needed corrections can then be applied to  $\dyad{\bm \chi} ({\bf r},{\bf r}',\omega_0)$ to form a response tensor that is as close to hermitian as possible. The hermitian part of the newly calibrated response tensor can then be calculated as the final result. 

A practical question is what is the effect of any small residual assymetry prior to the final hermitian part being calculated. For example although $\phi_j = 0$ establishes the phase reference, in the case where the path lengths of the two arms are mismatched by a constant phase $\phi_{0}$, the diagonal elements of the measured response are not real. Rather than using (\ref{eqn_eai_8}) to determine the diagonal elements, (\ref{eqn_eai_9}) can be used to measure the fringe even when $i=j$.  Each element becomes  $|D_{ij}| \cos \left( \theta_{ij} + \phi_0 \right) + i  |D_{ij}| \sin \left( \theta_{ij} + \phi_0 \right)$. Of course, $\theta_{ij} = 0$  when $i=j$, but even the diagonal elements are complex-valued when this phase offset is introduced. If the phase error is small, $ \phi_0 \approx 0$, then because $\cos(\phi_0) \approx 1$ and $\sin(\phi_0) \approx \phi_0$, imaginary-valued errors remain throughout. When the hermitian part is calculated, these additional contributions dissapear. If $\phi_0$ is appreciable, the whole measured response is multiplied by the phase factor $e^{i \phi_0}$:
and 
\begin{align}
\label{eqn_eai_11}
{\bf \mathsf D}' & = {\bf \mathsf D} \cos(\phi_0) + i {\bf \mathsf D} \sin(\phi_0). 
\end{align}
If only the hermitian part is retained, signal will be lost. A uniform phase correction can be applied to reduce the antihermitian part to zero, which is the same as ensuring that all of the signal is in one quadrature component. Large scale corrections of this kind are also beneficial when noise is present: Section \ref{sec_add}. Overall, many schemes are available for identifying and removing systematic errors, and these are closely related to calibration procedures.


\section{Reference Source}
\label{sec_ref}

A valuable experimental technique involves using a reference detector to overcome certain phase instabilities and drifts in the illuminating sources. A 
reference detector is particularly effective at improving the stabilities of optical measurements \cite{Moin1}. Schemes can be derived for increasing the stabilities of the sources themselves, but other effects still remain, such as phase fluctuations caused by the refractive indices of the free-space paths of the illuminating beams varying, and systematic phase errors caused by the flexing of cables or optical fibres as the sources are moved \cite{Moin1}. The method also allows systematic errors, caused by say the imperfect geometry of the mechanical scanning system, to be minimised.

The idea is to place an additional single-mode reference detector in the field of view of the sources; for example a detector that is offset from, or embedded in, the SUT. One realisation is to fabricate a small, single-mode reference detector at the centre of an array. This reference detector can then be used to accurately characteristise the individual and relative behaviours of the other pixels. As the sources are scanned, this additional detector records a fringe:  Fig.\ref{figure1_sub2}. The I-Q components of the reference fringe are then subtracted from those of the SUT. Real-time analogue differencing is possible, but it is best to record all outputs simultaneously, and then to carry out off-line processing. Large arrays can be characterised with just a single set of measurements.

The general benefits of using a reference detector can be appreciated as follows. Suppose that each source incurs a phase error of some kind, $e^{i \Delta_i}$ and $e^{i \Delta_j}$. These phase errors can be written in terms of their symmetric and antisymmetric parts:
\begin{align}
\label{eqn_ref_1a}
\Sigma_{ij} & = \Delta_i+\Delta_j \\ \nonumber
\Delta_{ij} & = \Delta_i-\Delta_j 
\end{align}
respectively.
According to (\ref{eqn_eai_3}), the elements of the measured response matrices become
\begin{align}
\label{eqn_ref_1b}
\tilde{D}_{ij} & = D_{ij} e^{-i \Delta_{ij}} \\ \nonumber
\tilde{R}_{ij} & = R_{ij} e^{-i \Delta_{ij}},
\end{align}
where $D_{ij}$ are those of the SUT's ideal response, and $R_{ij}$ are those of the reference. $\Delta_{ij}$ is the differential phase error between the two arms, which can vary from one spatial sample to the next. Common-mode errors do not appear in any of the elements of the measured response, and in particular this reduces the effects of phase noise in the signal source that drives the two arms. Differential errors, both static and time varying, only appear in the off-diagonal elements. The same differential phase errors also appear in the reference channel, because they are measured simultaneousy with the SUT, which is distinct from additive noise, which generally leads to uncorrelated phase errors: Section \ref{sec_add}. Lissajous figures can be formed by combining the output of the reference channel with that of the SUT to give a visual display of any time-dependent differential phase fluctuations present. In fact, this formalism is closely related to driving the two arms at slightly different frequencies, and recording the I-Q channels using a lock-in amplifier. 

Because the reference channel is single mode, its response matrix can be written ${\bf \mathsf R}  = \lambda_r {\bf \mathsf r} {\bf \mathsf r}^\dagger$, and so
\begin{align}
\label{eqn_ref_2}
\tilde{R}_{ij} & = \lambda_r r_i r_j e^{i(\theta_i - \theta_j)} e^{-i \Delta_{ij}},
\end{align}
where $r_i e^{i \theta_i}$ is the intrinsic complex-valued reception pattern of the reference, and $\lambda_r$ the responsivity. 
As long as the amplitude response of the reference is appreciable over the scanning range of the sources, it is straightforward to calculate the phase factor of the reference detector's measured response, (\ref{eqn_eai_9}),
\begin{align}
\label{eqn_ref_3}
\tilde{R}^\phi_{ij} & = e^{i(\theta_i - \theta_j)} e^{-i \Delta_{ij}}.
\end{align}
Here, the amplitude of the response has been explicitly removed. Equation (\ref{eqn_ref_3}) includes information about the phase error present on each sample, and it can be used to correct the data from the SUT:
\begin{align}
\label{eqn_ref_4}
\tilde{D}^c_{ij} & = D_{ij} e^{-i \Delta_{ij}} ( \tilde{R}^{\phi}_{ij} )^\ast \\ \nonumber
& =  e^{-i \theta_i} D_{ij}  e^{+i \theta_j} \\ \nonumber
{\bf \mathsf D}^c & = {\bf \mathsf \Phi}^\dagger {\bf \mathsf D} {\bf \mathsf \Phi},
\end{align}
where  ${\bf \mathsf \Phi}$ is a unitary diagonal matrix comprising the phase reponse of the reference detector, over the reference surface of the measurement. ${\bf \mathsf D}^c$ does not incude the factor $e^{i \Delta_{ij}}$. If the amplitude $\lambda_r r_i r_j$ had been left in, it would degrade ${\bf \mathsf D}^c$ in those parts of the field of view where the reference detector is insensitive.

${\bf \mathsf D}^c$ is hermitian and so can be diagonalised, but how do the eigenvalues and eigenvectors compare with those of an ideal set of measurements? Writing the actual response in terms of its eigenmodes
\begin{align}
\label{eqn_ref_5}
{\bf \mathsf D}^c & = {\bf \mathsf \Phi}^\dagger  {\bf \mathsf U}  {\bf \mathsf \Sigma}  {\bf \mathsf U}^\dagger  {\bf \mathsf \Phi}.
\end{align}
The individual modes take the form
\begin{align}
\label{eqn_ref_6}
d^{c}_{n,i} & = u_{n,i} e^{-i \theta_i},
\end{align}
and so every mode is recovered with differential phase errors and instabilities removed, but with the phase fronts of the modes now referenced to the phase front of the reference. This situation is analogous to making microwave measurements with a vector network analyser, where a calibration kit establishes the reference planes of multiport scattering parameters. Here, however, a {\em phase reference surface} has been established, which for a point-like reference detector can be near-spherical. 

This approach is valuable when using a flat scanning surface in the far field of an SUT at short wavelengths, because the phase of the response changes rapidly, and so the experimenter must use a large number of sample points to overcome the ambiguity of phase wrapping. A reference detector, however, removes this need. A reference detector is also valuable when characterising arrays, because the responses the individual pixels can be  referenced to a single phase front.  Although the referencing scheme has been described in terms of the discretised response, it can be written in terms of its reconstructed continuous response with the same conclusion.


\section{Sampling}
\label{sec_sam}

When implementing EAI, it is important to choose the sample points carefully to ensure that the degrees of freedom in the response are well covered by the measurements. However, too many measurements should be avoided as this merely introduces noise, which eventually degrades the data without adding information: Section \ref{sec_add}. To achieve the best compromise, the illumination patterns of the sources should be reasonably well matched to sampling the coherence areas of the SUT.

The sampling strategy needed depends on the geometry of the SUT.  The options available do not merely relate to moving the sources over a scanning surface, but could also include moving the sources away from the SUT. Usually, the sources are scanned over the field of view of the SUT,  or around some volume occupied by the SUT. For example, in the case of an infrared array, one may be interested in the number and efficiencies of any modes that can couple straylight into the sides or back of the device. When scanning the sources around a device, a spherical coordinate surface may be used, perhaps by having the sources on gimbles; or when scanning over a plane, a Cartesian or polar coordinate system could be used. Gaining access to the diagonal and near-diagonal elements is important, and so it can be beneficial to use a beam splitter so that the sources can, effectively, be placed at the same location. Attempts at interpolating data to give the diagonal elements of the response matrix have not been as effective as actually measuring the diagonal elements directly. The Cartesian and polar scanning strategies correspond to different kinds of measurement: one based on orthogonal linear scans, and one based on radial and angular scans. The scheme should be chosen, if possible, to match any known dominent device symmetries: for example picking up the principle radial and angular modes of a circular aperture, or the Cartesian modes of a square aperture. 

For low-throughput devices, the number of sample points needed is low, Section \ref{sec_sim}, and for high-throughput devices, the number of sample points increases according to the number of dominant eigenvalues. On this basis, strategies can be devised to minimise the number of measurements needed. For example, it can be shown that the electromagnetic modes of thin-film detectors are low-order multipole moments, due to the modal currents in the films \cite{Thom3}, and so it is effective to use sampling methods on a sphere \cite{Dris1}, such as the Gauss-Legendre sampling theorem \cite{Skuk1},  spherical harmonic sampling \cite{Ewen1}, or even those based on spin-weighted spherical harmonics \cite{Dur1}. Tihon \cite{Tihon2} used EAI, the Method of Moments, and carefully chosen sampling to calculate the behaviour of lossy dielectric spheres and complicated periodic plasmonic structures.

A closely related consideration is that, in some cases, it is sufficient to assume that the dynamical modes of the SUT are separable in some particular coordinate system. For example, if the modes have Cartesian symmetry, one might hope that two linear scans are sufficient to recover the spatial forms of each significant degree of freedom. This would happen in the case of a square pixel having a spatially incoherent absorber, or when the absorbing element is illuminated by a paraxial field. A paraxial free-space path introduces Cartesian separability by virtue of the form of the paraxial wave equation. Separability  can be assumed for any device if it is known that the device will only be used with paraxial illumination. Separability over the full sphere is also inherent in certain of the sampling schemes.

How can separability reduce the number of sample points needed? Suppose that the spatial form of each mode can be separated whilst retaining the requirement that the overall response is self adjoint:
\begin{align}
\label{eqn_sep_1}
\dyad{Z} (x,y;x',y')  & = \sum_{mm'}  X_m(x,x') Y_{m'}(y,y')  \hat{\bf e}_m \hat{\bf e}_{m'},
\end{align}
where $\hat{\bf e}_m$ are unit vectors in the Cartesian directions.

The matrix elements can be calculated with respect to a set of basis functions that are also separable:
\begin{align}
\label{eqn_sep_2}
{\bf E}_{ij,k} (x,y) & =  E^x_i (x)  E^y_{j} (y)  \hat{\bf f}_k,
\end{align}
where $\hat{\bf f}_k$ is the unit polarisation vector of the source. The functional form form of (\ref{eqn_sep_2}) is typical of many sources; for example, the field produced by a rectangular waveguide probe. The matrix elements then become
\begin{align}
\label{eqn_sep_3}
Z_{ii',jj',kk'} & =   \int_{x} \int_{y} \int_{x'} \int_{y'}  {\bf E}_{ij,k}^{\ast} (x,y)  \cdot  \dyad{Z} (x,y;x',y')  \cdot  {\bf E}_{i'j',k'} (x',y')  \, dx dy  \, dx' dy'  \\ \nonumber
& =  \sum_{mm'} \left[   \int_{x}  \int_{x'}  E^{x \ast} _i (x)  X_m(x,x')   E^x_{i'} (x') \, dx dx'  \right] \left[ \int_{y}  \int_{y'}
E^{\ast y}_{j} (y) Y_{m'} (y,y')   E_{j'}^y (y') \, dy dy'  \right]  ( \hat{\bf f}_k \cdot \hat{\bf e}_m ) ( \hat{\bf e}_{m'}  \hat{\bf f}_{k'} ) \\ \nonumber
& =  \sum_{mm'}   \alpha_{ii',m} \beta_{jj',m'} \gamma_{kk',mm'}.
\end{align}

Suppose that indices $i,i'$ correspond to the sources being moved along a line in the $x$ direction, in such a way that  the $x$ dependence $E^x_i (x)$ changes, whilst the $y$ dependence $E^y_{jr} (y)$ does not. Here $jr$ indexes the position of the $x$-directed scan line on the $y$ axis. Conversely, the roles can be swapped, where the indices $j,j'$ correspond to the sources being moved along a line in the $y$ direction, in such a way that  the $y$ dependence $E^y_j (y)$ changes, whilst the $x$ dependence $E^x_{ir} (x)$ does not.  $ir$ indexes the position of the $y$-directed scan line on the $x$ axis. It is convenient, but not essential, if in each case, the two scans share a common reference position: $ir \in i = 1, \cdots I$ and  $jr \in j = 1, \cdots J$, so that they cross at $(ir,jr)$. If the response tensor is separable in polar coordinates, the scheme would correspond to scanning the two sources around a circle having constant radius, and then scanning both sources along a radial line at some constant angle. In either case, this reduced sampling corresponds to only measuring a subspace of the whole set of data that would be needed to scan the full 2D surface; the number of sample points now increases as $I + J$ rather than $I \times J$.

If each source is linearly polarised, there are 4 unique combinations of polarisation-rotated sources. For the specific combination $k = m$ and  $k' = m'$, 
\begin{align}
\label{eqn_sep_4}
Z_{ii',jj',kk'} & =  \alpha_{ii',k} \beta_{jj',k'}
\end{align}
where 
\begin{align}
\label{eqn_sep_5}
\alpha_{ii',k} & =   \int_{x}  \int_{x'}  E^{x \ast} _i (x)  X_k(x,x')   E^x_{i'} (x') \, dx dx' \\ \nonumber
\beta_{jj',k'} & = \int_{y}  \int_{y'}
E^{\ast y}_{j} (y) Y_{k'} (y,y')   E_{j'}^y (y') \, dy dy'. 
\end{align}

From (\ref{eqn_sep_4}), the sets of data measured during the $x$-directed and $y$-directed scans are
\begin{align}
\label{eqn_sep_6}
Z_{ii',jr jr,kk'}^x & =  \alpha_{ii',k} \beta_{jr jr,k'} \\ \nonumber
Z_{ir ir,jj',kk'}^y & =  \alpha_{ir ir,k} \beta_{jj',k'} 
\end{align}
respectively, and when both sources are at the crossing point $(ir,jr)$,
\begin{align}
\label{eqn_sep_7}
Z_{ir ir,jr jr,kk'} & =  \alpha_{ir ir,k} \beta_{jr jr,k'} = P_{kk'},
\end{align}
which is a real-valued power.

From (\ref{eqn_sep_6})
\begin{align}
\label{eqn_sep_8}
\alpha_{ii',k} & = \frac{Z_{ii',jr jr,kk'}}{ \beta_{jr jr,k'} } \\ \nonumber
\beta_{jj',k'} & = \frac{Z_{ir ir,jj',kk'}}{ \alpha_{ir ir,k}},
\end{align}
and so using (\ref{eqn_sep_4})
\begin{align}
\label{eqn_sep_9}
Z_{ii',jj',kk'} & =   \frac{Z_{ii',jr jr,kk'}^x}{ \beta_{jr jr,k'} } \frac{Z_{ir ir,jj',kk'}^y}{ \alpha_{ir ir,k}} \\ \nonumber
& =  \frac{Z_{ii',jr jr,kk'}^x Z_{ir ir,jj',kk'}^y}{ P_{kk'} }. 
\end{align}
The overall discretised response is given by the tensor product of the the two linear scans, where each measurement subspace (the two line scans) has be normalised to $\sqrt{P_{kk'}}$. 

If the normalised matrix elements of the $x$-directed scan are assembled into a matrix ${\bf \mathsf Z}^x_{kk'}$, and normalised matrix elements of the $y$-directed scan are assembled into a matrix ${\bf \mathsf Z}^y_{kk'}$, the overall discretised response matrix is
\begin{align}
\label{eqn_sep_10}
{\bf \mathsf Z}_{kk'} & = {\bf \mathsf Z}^x_{kk'} \otimes {\bf \mathsf Z}^y_{kk'},
\end{align}
where $\otimes$ indicates the tensor product. Each of  ${\bf \mathsf Z}^x_{kk'}$ and ${\bf \mathsf Z}^y_{kk'}$ can be diagonalised individually to give a set of natural modes in each of the two directions.  Thus, (\ref{eqn_sep_10}) becomes
\begin{align}
\label{eqn_sep_11}
{\bf \mathsf Z}_{kk'} & = \left[ \sum_i \lambda_i {\bf \mathsf u}^x_i {\bf \mathsf u}^{x \dagger}_i \right] \otimes \left[ \sum_j \lambda_j {\bf \mathsf u}^y_j {\bf \mathsf u}^{y \dagger}_j \right] \\ \nonumber
& = \sum_{ij} \lambda_i \lambda_j \left( {\bf \mathsf u}^x_i  \otimes {\bf \mathsf u}^y_j \right) \left({\bf \mathsf u}^{x}_i \otimes {\bf \mathsf u}^{y}_j \right)^\dagger \\ \nonumber
& = \sum_n \lambda_n  {\bf \mathsf u}_n {\bf \mathsf u}_n^{\dagger}
\end{align}
for the co-polar and cross-polar measurements. The penultimate line shows that  the overall set of modes is given by the product of each $x$-directed mode and each $y$-directed mode, with the associated eigenvalue given by the product of the individual eigenvalues. Once the continuous forms of the modes of the indiviual scans have been reconstructed, using the beam patterns of the sources, the continuous modes of the 2D response can also be formed. 

Notice that no attempt has been made to measure the cross terms, when one source is on the $x$-directed line and the other is on the $y$-directed line. In principle, these are not needed, and so reduce the measurement time, but in the case when the measurements are only sufficient to first order, these additional measurements may add further information, and certainly help retain phase integrity. If the cross terms are measured, then one would diagonalise  
\begin{align}
\label{eqn_sep_12}
\begin{bmatrix}
{\bf \mathsf Z}_{xx} & {\bf \mathsf Z}_{xy} \\
{\bf \mathsf Z}_{yx} & {\bf \mathsf Z}_{yy},
\end{bmatrix}
\end{align}
which gives the overall vector-valued natural modes. In summary, if the response tensor separates according to (\ref{eqn_sep_1}), or some equivalent form in another coordinate system, the modal response can be calculated from two line scans. It can be shown that the resulting reconstruction is then the same as the Hilbert-Schmidt decomposition of the original response tensor.

In general, it is not known apriori whether a reduced set of measurements spans the modes having appreciable responsivities. It is often the case that although symmetries can be used to indicate what is needed, the actual requirement is not known. In this case, an elegant method is to use incremental, or non-uniform adaptive, sampling. Here, everytime a new sample point is added, the SVD is calculated in real time to determine the spectrum. Also recall that once a source has been placed, the other source only needs to be moved around within its coherence area, and this can help guide where the next sample point should be placed; effectively hopping between coherence areas. The process is continued until all of the degrees of freedom have been found down into the noise: Section \ref{sec_sim}. In this case, a good sampling scheme might be to build up the separated scans and then add in other points as needed to break symmetries. Incremental SVD  \cite{Brnd1,Nik1} is particularly helpful, because the complete response matrix does not have to be diagonalised at each step; it is sufficient to update the existing SVD using the new measurement. In this way,  the process stops as soon as all of the degrees of freedom have been found, and before the dynamic range on the eigenvalues starts to degrade. 

%
\begin{figure}[H]
     \centering
     \begin{subfigure}[b]{0.45\textwidth}
         \centering
                   \includegraphics[trim = 2cm 16cm 8cm 3cm, clip,width=80mm ]{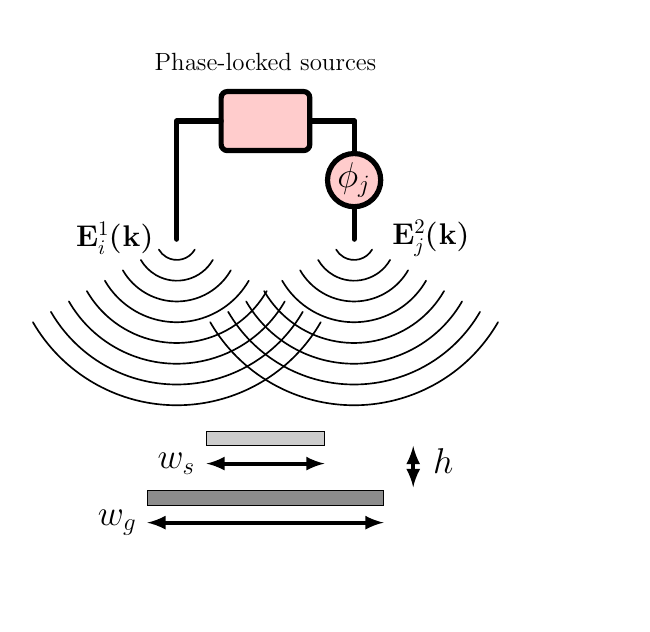}
         \caption{Resistive strip absorber (light grey) having width $w_s$ over a perfectly conducting ground plane (dark gray) having width $w_g$ separated by height $h$.}
         \label{fig_far_field_strip}
     \end{subfigure}
     \hfill
     \begin{subfigure}[b]{0.45\textwidth}
         \centering
\includegraphics[trim = 1cm 1cm 8cm 18cm, clip,width=80mm ]{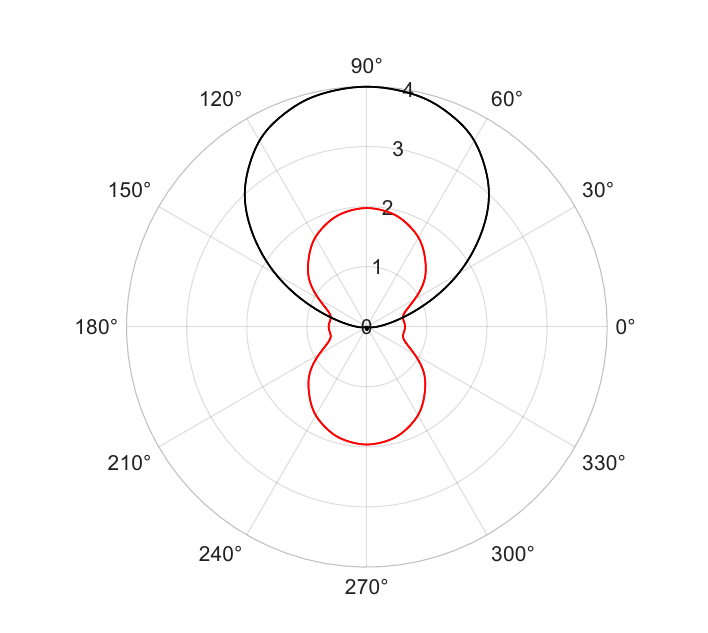}
\caption{Polar power pattern of a 188 $\Omega$  absorbing thin film having a physical width, $A_p$, of 4 wavelengths (red). A perfectly conducting ground plane having a width of 6 wavelength was placed behind the film whose sheet impedance was 377 $\Omega$ (black). The radial scale shows the responsivity as an effective area, $A_e$, where maximum apsorption efficiency corresponds to  $A_e = A_p$.}
\label{fig_powabs_film_prog1_plot1}
  \end{subfigure}
 \caption{Far-field EAI of thin-film strip absorber.}
\end{figure}
\section{Strip Absorber}
\label{sec_flm}

To illustrate the principles of EAI, a simple thin-film strip absorber, $w_s=$ 4$\lambda$ wide, was modelled numerically by solving the EFIE. An infinitely long strip was used, so that one-dimensional currents and power patterns could be displayed and interpreted easily. Additionally, a perfectly conducting ground plane, $w_g=$ 6$\lambda$ wide, was placed $h=$ 0.25$\lambda$ behind the absorbing strip to act as a matching backshort. This configuration is typical of pixels in ultra-low-noise far-infrared imaging arrays. To `measure' the response, linearly-polarised plane-wave sources were used, Fig. \ref{fig_far_field_strip}, corresponding to k-domain far-field measurement: Section \ref{sec_refs}.

Figure \ref{fig_powabs_film_prog1_plot1} shows the power pattern of a free-space strip having a sheet impedance of 188 $\Omega$ (red). The radial scale shows the responsivity as an effective area in wavelengths, $A_e$, where maximum apsorption efficiency corresponds to  $A_e = A_p$. The effect of including a perfectly conducting ground plane is also shown (black); in this case, the sheet impedance of the film was increased to 377 $\Omega$ to ensure an optimum match. It can be shown, using a simple equivalent circuit, that if edge effects are ignored, and for face-on illumination, the absorption efficiency is given by
\begin{equation}
\label{eqn_flm_1}
\frac{P_{abs}}{P_{avs}} = \frac{4 \widetilde{R}}{(2 \widetilde{R} + 1)^{2}}
\mbox{,}
\end{equation}
where $\widetilde{R}$ is the sheet resistance normalised to the impedance of free space. This function has a maximum value of 0.5 when $\widetilde{R} = 0.5$, and so for wide films having a surface resistance of 188 $\Omega$, the effective area is half of the physical area. In this case, 0.5 of the available power is absorbed, 0.25 is transmitted and 0.25 reflected. Figure \ref{fig_powabs_film_prog1_plot1} replicates this behaviour, showing an effective width of 2$\lambda$, which is half the physical area (red line). Additionally, the dependence  on film impedance was found to follow (\ref{eqn_flm_1}). 
\begin{figure}[H]
     \centering
     \begin{subfigure}[b]{0.45\textwidth}
         \centering
        \includegraphics[trim = 1cm 1cm 8cm 19cm, clip,width=70mm ]{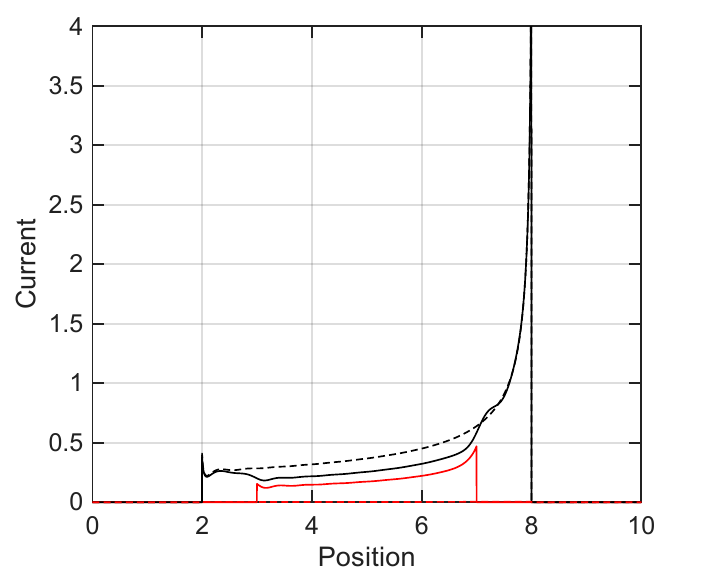}
         \caption{Angle 0 degrees}
         \label{fig_curr_grd_1}
     \end{subfigure}
     \hfill
     \begin{subfigure}[b]{0.45\textwidth}
         \centering
          \includegraphics[trim = 1cm 1cm 8cm 19cm, clip,width=70mm ]{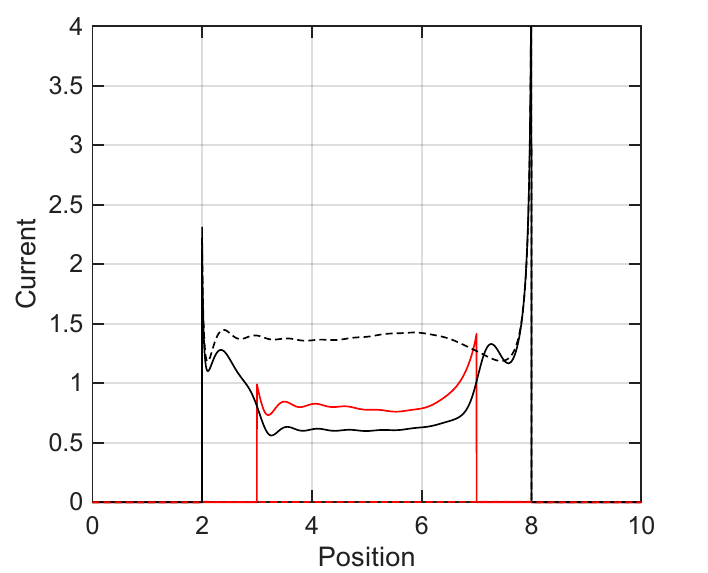}
           \caption{Angle 45 degrees}    
            \label{fig_curr_grd_2}
     \end{subfigure}
          \hfill
     \begin{subfigure}[b]{0.45\textwidth}
         \centering
             \includegraphics[trim = 1cm 1cm 8cm 19cm, clip,width=70mm ]{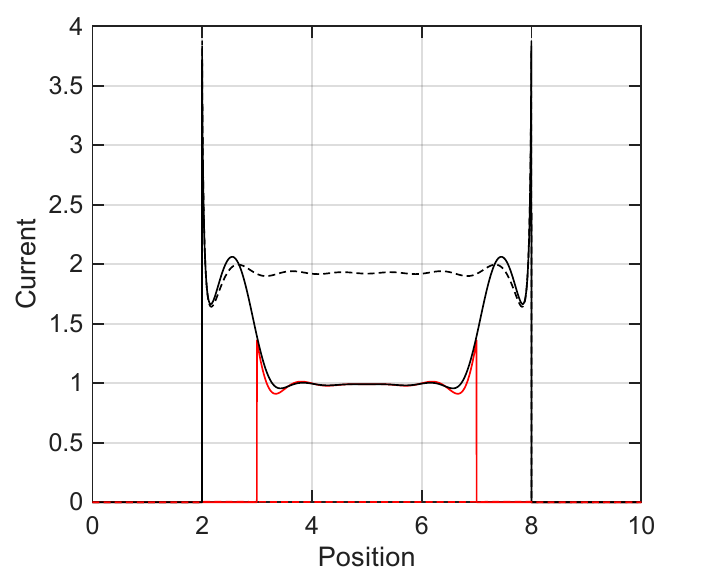}
          \caption{Angle 90 degrees}
            \label{fig_curr_grd_3}
     \end{subfigure}
          \hfill
     \begin{subfigure}[b]{0.45\textwidth}
         \centering
        \includegraphics[trim = 1cm 1cm 8cm 19cm, clip,width=70mm ]{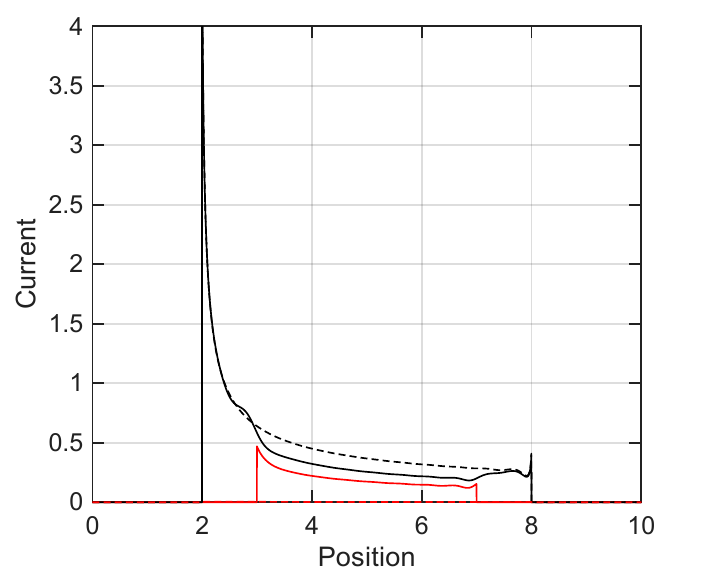}
                   \caption{Angle 180 degrees}
            \label{fig_curr_grd_4}
     \end{subfigure}
        \caption{Currents in the films as the angle of incidence of the plane wave is changed. The model is that same as that of Fig. \ref{fig_powabs_film_prog1_plot1}, but the sheet impedance of the absorbing film was held constant at 377 $\Omega$ in all cases. Current in the absorbing film (red solid); current in the ground plane (black solid); current in the ground plane when the absorbing film is removed (black dashed).}
        \label{fig_curr grnd_plot1}
\end{figure}
The response of the free-space film (red) appears to comprise four Lambertian forms: two large circles associated with the main body of the film; and two small circles associated with edge currents. The side responses correspond to a plane waves travelling over the surface and loosing energy within a wavelength of the edge. As the strip is made narrower, the relative sizes of the circles change, but they remain well defined until the strip becomes very narrow and the circles merge to create a single-mode incoming cylindrical wave.  When a ground plane is included (black), and the sheet impedance of the film increased to 377 $\Omega$, the beam is thrown in the forward direction, and the maximum response becomes that of the full width of the device: no power is reflected or transmitted. Interesting, the back lobe is supressed, but the proportionate width of the main beam remains unchanged.

Figure \ref{fig_curr grnd_plot1} shows the currents in the film (red solid) and ground plane (black solid) as the angle of incidence of the illuminating plane wave is changed. These plots correspond to the same arrangement as Fig. \ref{fig_powabs_film_prog1_plot1}. Various features are evident. The leading edge of the ground plane exhibits singular behaviour, which is also seen to some extent in the film and on the trailing edge of the ground plane. Ripples can be seen that currespond to the edge current producing a damped wave that travels over the surface. The dashed black line shows the current in the ground plane when the absorbing film is removed; when compared with the solid black line, the screening effect of the resistive strip is clear. 

\begin{figure}[H]
     \centering
     \begin{subfigure}[b]{0.45\textwidth}
         \centering
          \includegraphics[trim = 1cm 1cm 8cm 19cm, clip,width=70mm ]{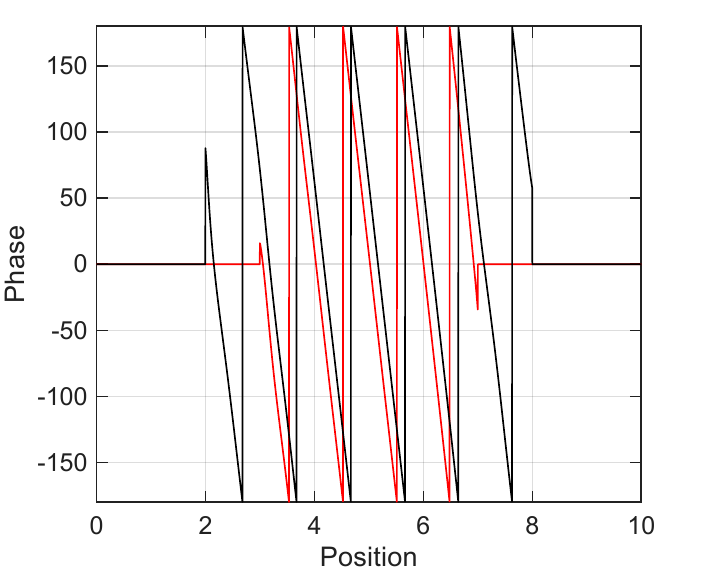}
         \caption{Angle 0 degrees}
         \label{fig_curr_grd_1}
     \end{subfigure}
     \hfill
     \begin{subfigure}[b]{0.45\textwidth}
         \centering
          \includegraphics[trim = 1cm 1cm 8cm 19cm, clip,width=70mm ]{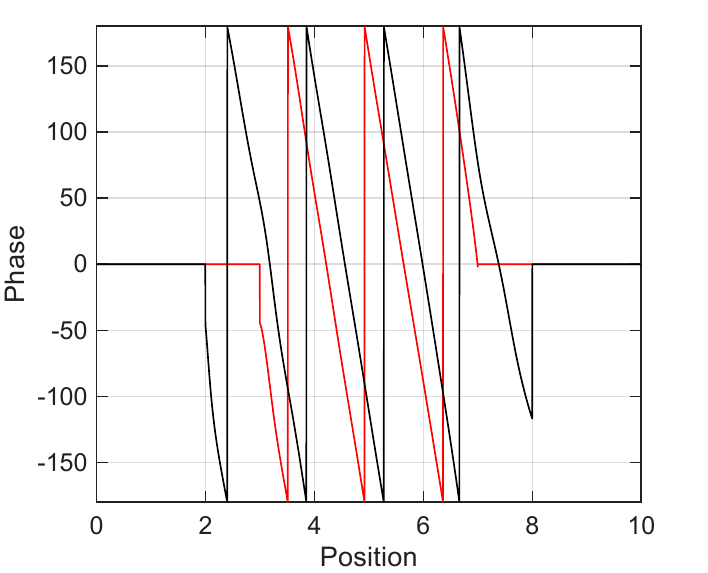}
           \caption{Angle 45 degrees}    
            \label{fig_curr_grd_2}
     \end{subfigure}
          \hfill
     \begin{subfigure}[b]{0.45\textwidth}
         \centering
         \includegraphics[trim = 1cm 1cm 8cm 19cm, clip,width=70mm ]{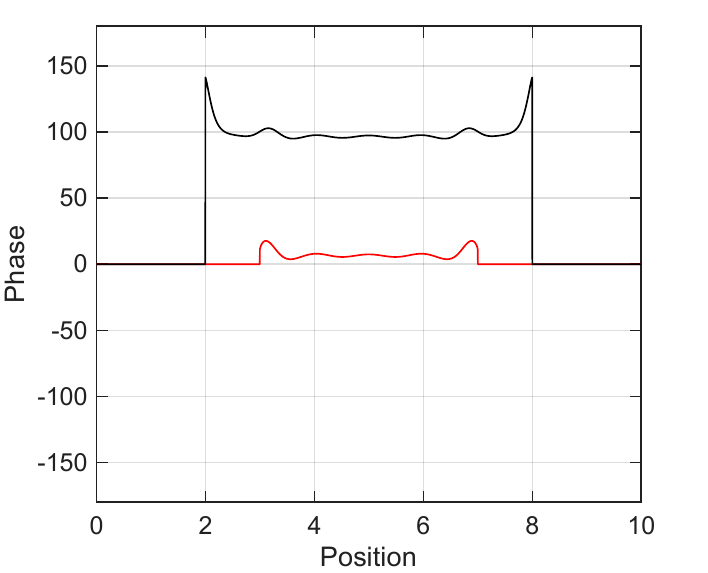}
          \caption{Angle 90 degrees}
            \label{fig_curr_grd_3}
     \end{subfigure}
          \hfill
     \begin{subfigure}[b]{0.45\textwidth}
         \centering
          \includegraphics[trim = 1cm 1cm 8cm 19cm, clip,width=70mm ]{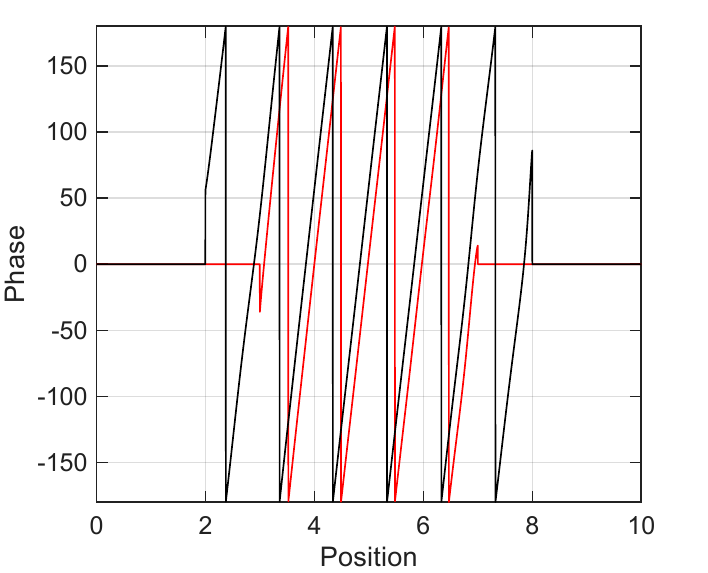}
                   \caption{Angle 180 degrees}
            \label{fig_curr_grd_4}
     \end{subfigure}
        \caption{Phase of currents in the films as the angle of incidence of the plane wave is changed. The model is that same as that of Fig. \ref{fig_powabs_film_prog1_plot1} but the sheet impedance of the absorbing film was held constant at 377 $\Omega$ in all cases. Current in the absorbing film (red solid). Current in the ground plane (black solid). }
        \label{fig_phase grnd_plot1}
\end{figure}

\begin{figure}[H]
     \centering
     \begin{subfigure}[b]{0.45\textwidth}
         \centering
         \includegraphics[trim = 1cm 1cm 8cm 18cm, clip,width=70mm ]{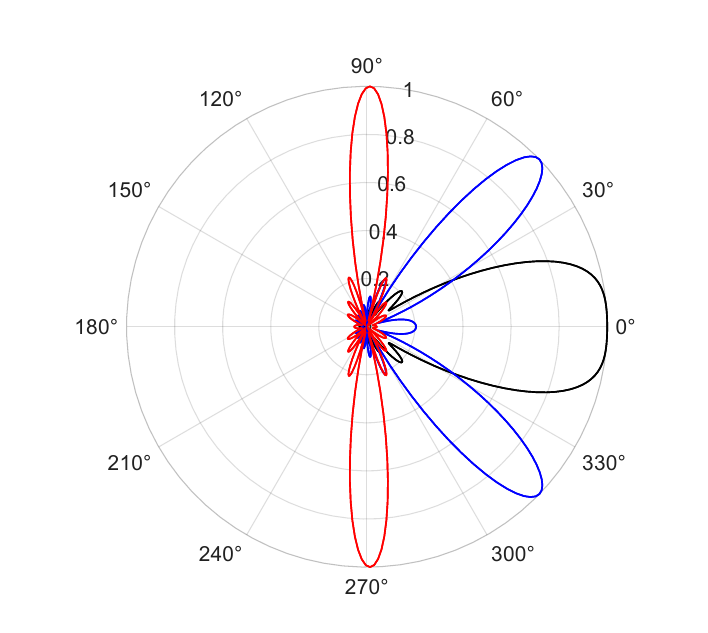}
         \caption{Thin resistive strip }
         \label{vis_1}
     \end{subfigure}
     \hfill
     \begin{subfigure}[b]{0.45\textwidth}
         \centering
          \includegraphics[trim = 1cm 1cm 8cm 18cm, clip,width=70mm ]{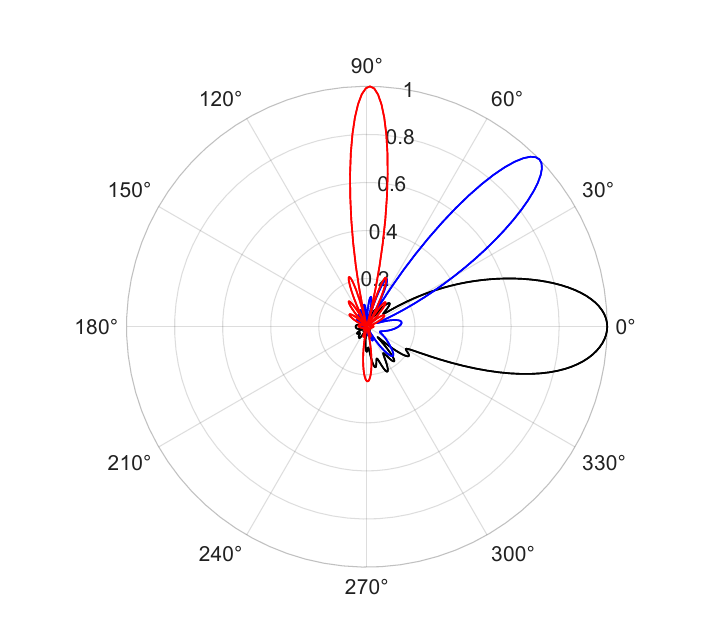}
           \caption{Thin resistive strip with ground plane}    
            \label{vis_2}
     \end{subfigure}
        \caption{Magnitude of the visibility functions of the power patterns shown in Fig. \ref{fig_powabs_film_prog1_plot1}. (a) Corresponds to the unbacked resistive film having sheet impedance 188 $\Omega$, and (b) corresponds to the backed resistive film having sheet impedance 377 $\Omega$. In each case, one of the sources was held at at a fixed reference angle of 0 degrees (black), 45 degrees (blue), 90 degrees (red), whilst the other source was swept over the full 360 degree range.}
        \label{fig_vis_plot}
\end{figure}

For face on illumination, the currents in the strip and ground plane are essentially the same, showing that a strip transmission-line type mode is being induced. The solid black line also shows that the ground-plane current is modified up to a distance of about $\lambda/2$ away from the edge, indicating that the effective width of the device can be greater than the physical width; suggesting that the individual modes may extend beyond the physical extent of the absorber. Figure \ref{fig_phase grnd_plot1} shows the phases of the currents corresponding to amplitudes of Fig. \ref{fig_curr grnd_plot1}. Plot \ref{fig_curr_grd_3} shows that for face-on illumination the current on the ground plane lags that of the strip by 90 degrees, as expected. The travelling wave forms induced by the other illuminations can be seen.
EAI traces out coherence areas, or angles in this case, of the SUT. Visibility functions of the resistive strip are shown in  Fig. \ref{fig_vis_plot}, where one source was held fixed at 0 degrees (black), 45 degrees  (blue) and 90 degrees (red), and the other source was swept around the full 360 degrees. These were calculated as follows. For each pair of source locations,  the absorbed power displays a fringe as the phase between the sources is varied,
\begin{equation}
\label{eqn_flm_2}
P = A + B \cos ( \theta + \phi_0)
\mbox{,}
\end{equation}
and so three power measurements $P_1 = P(\phi_0 = 0)$, $P_2 = P(\phi_0 = \pi)$, and $P_3 = P(\phi_0 = \pi/2)$ were made for every pair of source locations. 
The parameters $A,B$ and $\Phi_0$ were calculated through
\begin{eqnarray}
\label{eqn_flm_3}
A & = & \frac{P_{1}+P_{2}}{2} \\ \nonumber
B & = & \left[ \left( P_{2} - A \right)^{2} + \left( P_{3} - A \right)^{2} \right]^{1/2} 
\\ \nonumber
\mbox{tan} \theta_{o} & = & \frac{(P_{3}-A) }{(P_{2}-A)}
\mbox{,}
\end{eqnarray}
such that the magnitude of the fringe's visibility is given by $V = B/A$. Figure \ref{vis_1}  corresponds to the unbacked resistive film having sheet impedance 188 $\Omega$, and Fig. \ref{vis_2}  corresponds to the backed resistive film having sheet impedance 377 $\Omega$.

It can be shown that for a single sheet having a high impedance, low scattering, the visibility function, when viewed face on, is given by
\begin{equation}
\label{eqn_flm_4}
|V(\theta)| = \frac{\sin \left( \frac{W}{2 \lambda} \pi \sin \theta \right)}
{\left( \frac{W}{2 \lambda} \pi \sin \theta \right)}
\mbox{,}
\end{equation}
where $W/\lambda$ is the width in wavelengths, and $\theta$ is the angle relative to the surface normal. This underlying ${\rm sinc}(\theta)$ form can be seen in Fig. \ref{fig_vis_plot}. In the case of  the free-space strip,  Fig. \ref{vis_1}, the forward-facing lobes are correlated with the  backward-facing lobes, which occurs because for thin films, the current arising from front illumination shares the same volume as that arising from back illumination. The lobe associated with side-on illumination, however, is uncorrelated with the lobe associated with illumination from the other side. This occurs because the currents do not share the same volume because the field is attenuated as it propagates. The sidelobes are also wider than the front and back lobes due to the currents being more extended in the latter case. When the ground plane is present, Fig. \ref{vis_2}, the visibility functions are much the same, but now the back response is supressed. What is noticeable, particularly in the free-space case, is that the response associated with side-on illumination is not correlated appreaciably with that of face-on illumination.
\begin{figure}[h]
     \centering
      \includegraphics[trim = 1cm 1cm 8cm 19cm, clip,width=70mm ]{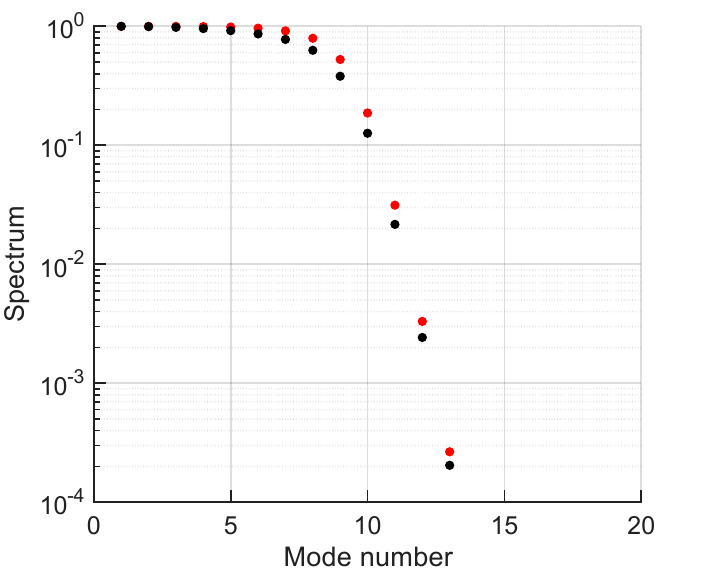}
        \caption{Normalised eigenvalue spectra of the free-space strip (red) and backed strip (black) recovered through EAI.}
        \label{fig_modes_plot1}
\end{figure}
EAI was used numerically to determine the spectra and modes of the free-space and metal-backed resistive strips described above, where
the off-diagonal elements of the detector response matrix were calculated through
\begin{align}
\label{eqn_flm_5}
D_{ii} & = \left\{ \frac{P_{1}+P_{2}}{4} \right\}_{ii} \\ \nonumber
D_{ij} & =  \frac{1}{2}  \left\{ B \left[ \cos(\theta) + i \sin(\theta) \right] \right\}_{ij} \\ \nonumber 
& = \frac{1}{4} \left\{ (A - P_2) + i  \frac{1}{4} (A - P_3) \right\}_{ij}.
\end{align}
The normalised spectra are shown in Fig. \ref{fig_modes_plot1}. The free-space film (red) has slightly larger eigenvalues compared with the metal-backed film (black), which occurs because the power in the side response is pushed into the forward response. 

\begin{figure}[H]
     \centering
     \begin{subfigure}[b]{0.45\textwidth}
         \centering
        \includegraphics[trim = 1cm 1cm 8cm 18cm, clip,width=70mm ]{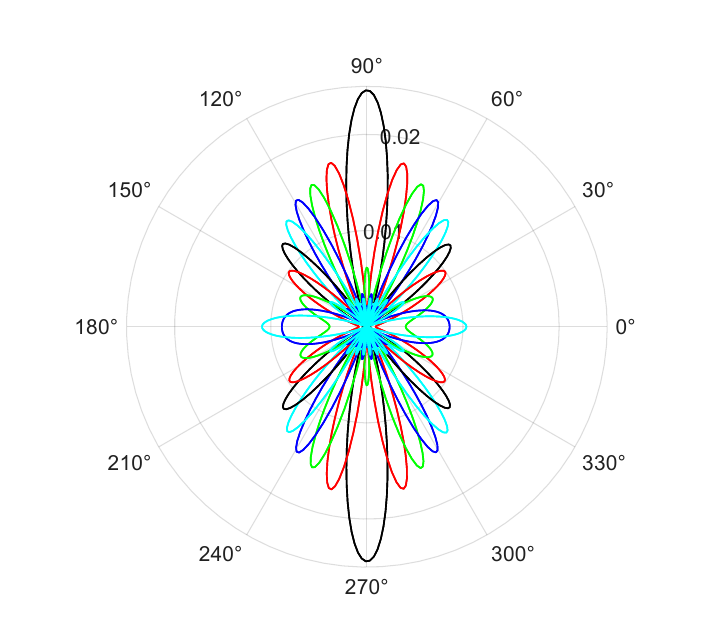}
         \caption{Thin resistive strip }
         \label{modes_plot1}
     \end{subfigure}
     \hfill
     \begin{subfigure}[b]{0.45\textwidth}
         \centering
         \includegraphics[trim = 1cm 1cm 8cm 18cm, clip,width=70mm ]{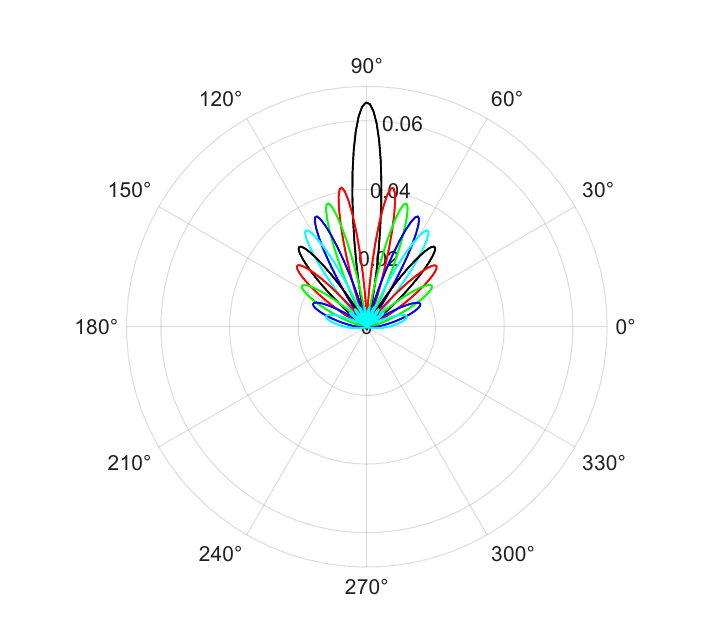}
           \caption{Thin resistive strip with ground plane}    
            \label{modes_plot2}
     \end{subfigure}
        \caption{Power patterns of the 10 lowest-order modes. }
        \label{fig_modes_plot3}
\end{figure}

The corresponding modes are shown in Fig. \ref{fig_modes_plot3} for the 10 largest eigenvalues. In each case, the modes fan out as the mode index is increased. Each maximum corresponds to having an integral number of half wavelengths across the film, as would be expected. It can be shown that peaks should occur at approximately
\begin{equation}
\label{eqn_flm_6}
\cos \theta = \frac{n \lambda}{2 W}.
\end{equation}
In the case of the free-space strip, Fig. \ref{modes_plot1}, modes 9 and 10 (blue and cyan) account for side response, whereas in the case of the free-space strip, Fig. \ref{modes_plot2}, they are pushed forward. The edge currents have their own independent modes. 
\begin{figure}[H]
     \centering
     \begin{subfigure}[b]{0.45\textwidth}
         \centering
        \includegraphics[trim = 1cm 1cm 8cm 18cm, clip,width=50mm ]{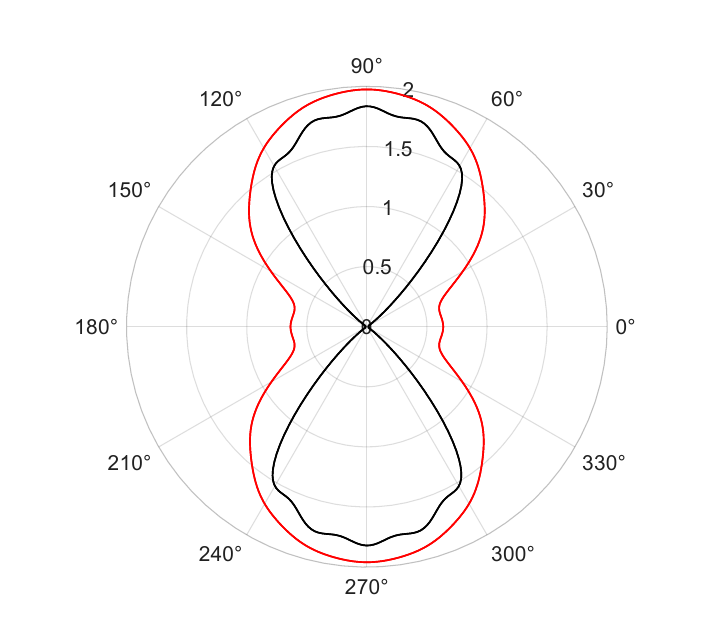}
         \caption{$N=$ 5 modes}
         \label{fig_modes_plot2_a}
     \end{subfigure}
     \hfill
     \begin{subfigure}[b]{0.45\textwidth}
         \centering
          \includegraphics[trim = 1cm 1cm 8cm 18cm, clip,width=50mm ]{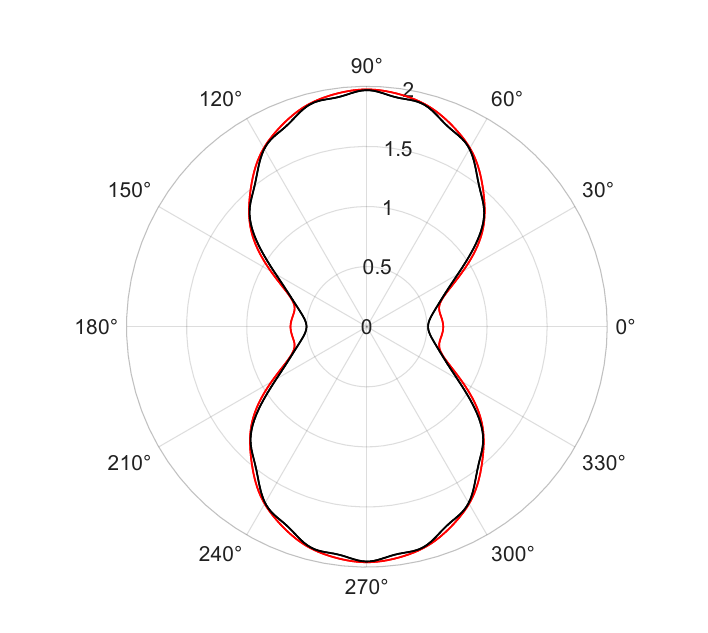}
           \caption{$N=9$ modes}    
            \label{fig_modes_plot2_b}
     \end{subfigure}
          \hfill
     \begin{subfigure}[b]{0.45\textwidth}
         \centering
         \includegraphics[trim = 1cm 1cm 8cm 18cm, clip,width=50mm ]{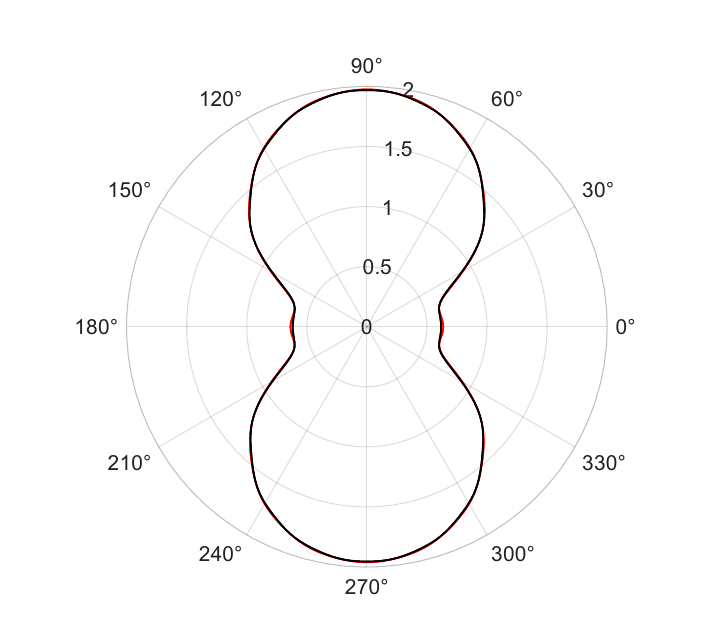}
          \caption{$N=10$ modes}
            \label{fig_modes_plot2_c}
     \end{subfigure}
          \hfill
     \begin{subfigure}[b]{0.45\textwidth}
         \centering
         \includegraphics[trim = 1cm 1cm 8cm 18cm, clip,width=50mm ]{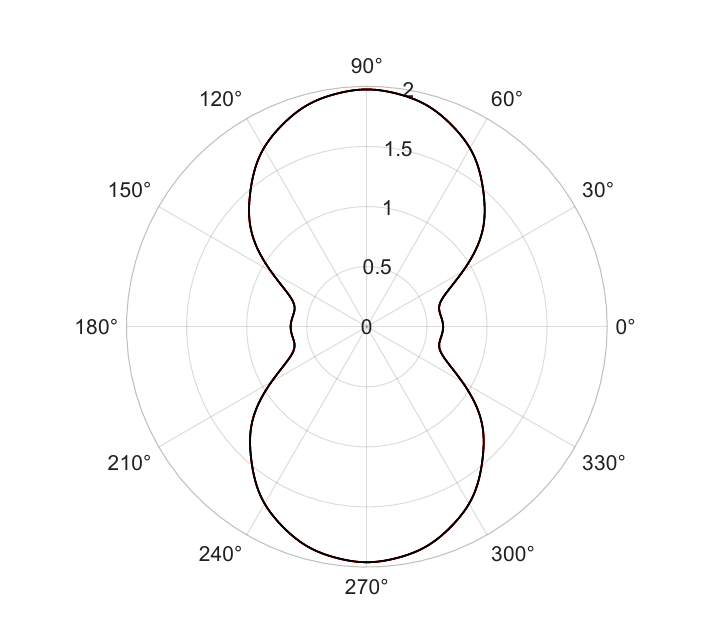}
                   \caption{$N=$ 11 modes}
            \label{fig_modes_plot2_d}
     \end{subfigure}
         \caption{Building up the power pattern of the free-space film from a weighted linear combination of modes. The four plots correspond to adding a total of $N=$ 5, 9, 10, 11 modes.  }
        \label{fig_modes_plot2_flm}
\end{figure}
\begin{figure}[H]
     \centering
     \begin{subfigure}[b]{0.45\textwidth}
         \centering
        \includegraphics[trim = 1cm 1cm 8cm 18cm, clip,width=50mm ]{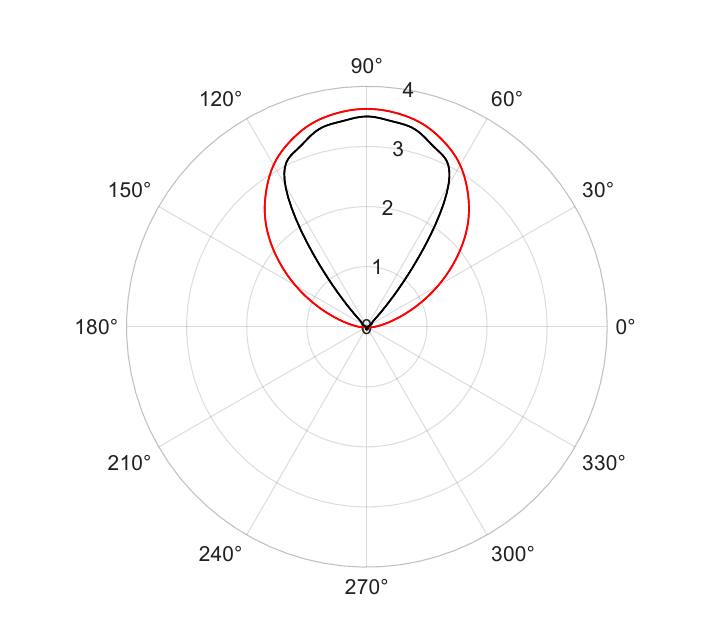}
         \caption{$N=$ 5 modes}
         \label{fig_modes_plot2_e}
     \end{subfigure}
     \hfill
     \begin{subfigure}[b]{0.45\textwidth}
         \centering
         \includegraphics[trim = 1cm 1cm 8cm 18cm, clip,width=50mm ]{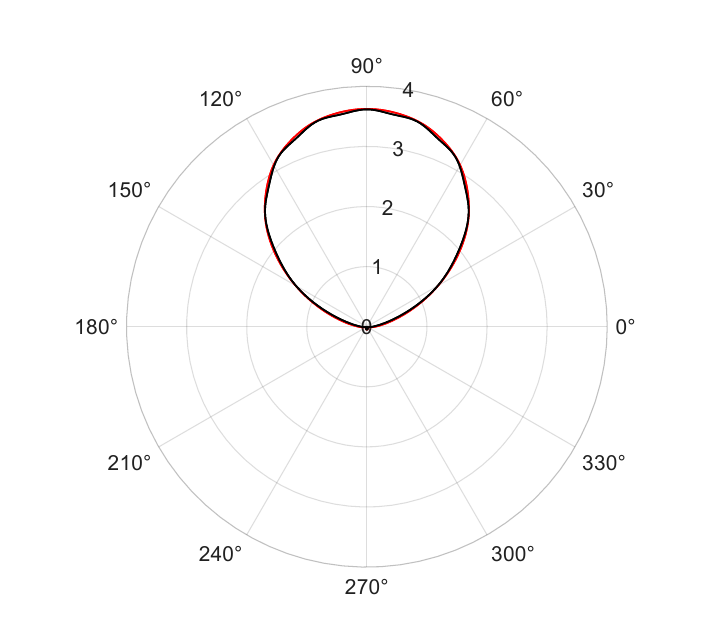}
           \caption{$N=$ 9 modes}    
            \label{fig_modes_plot2_f}
     \end{subfigure}
          \hfill
     \begin{subfigure}[b]{0.45\textwidth}
         \centering
          \includegraphics[trim = 1cm 1cm 8cm 18cm, clip,width=50mm ]{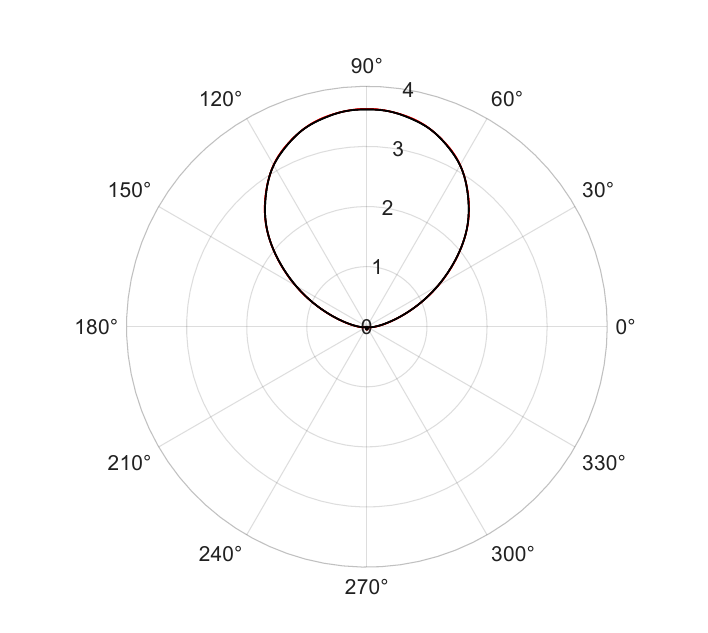}
          \caption{$N=$ 10 modes}
            \label{fig_modes_plot2_g}
     \end{subfigure}
          \hfill
     \begin{subfigure}[b]{0.45\textwidth}
         \centering
        \includegraphics[trim = 1cm 1cm 8cm 18cm, clip,width=50mm ]{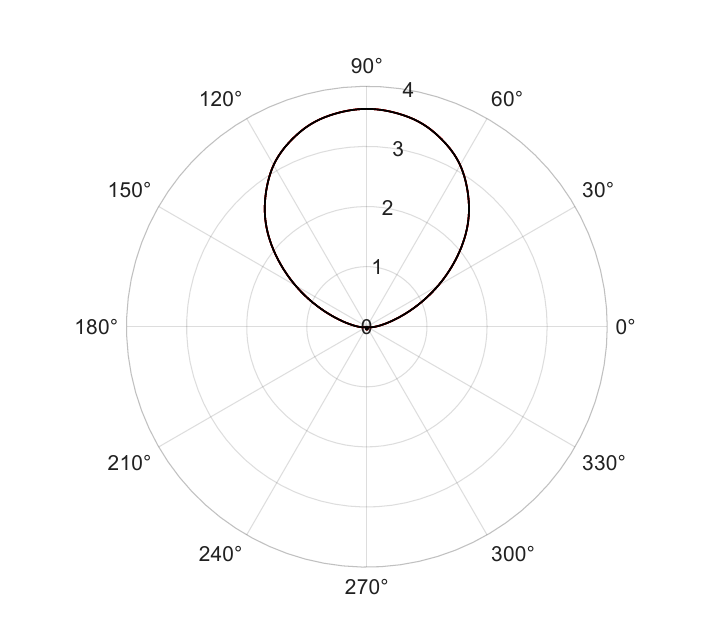}
                   \caption{$N=$ 11 modes}
            \label{fig_modes_plot2_h}
     \end{subfigure}
        \caption{Building up the power pattern of the backed film from a weighted linear combination of modes. The four plots correspond to adding a total of $N=$ 5, 9, 10, 11 modes. }
        \label{fig_modes_plot2_bck}
\end{figure}
Figures \ref{fig_modes_plot2_flm} and \ref{fig_modes_plot2_bck} illustrate the effects of summing the power patterns of the first $N$ modes, weighted by their corresponding eigenvalues, when $N=$ 5, 9, 10, 11. In each case, the power pattern of the device, taken from Fig.
\ref{fig_powabs_film_prog1_plot1}, is shown in red. The power pattern is completely accounted for by the number of degrees of freedom in the current distribution, which is 8 plus the two edge currents.  It can also be seen that in the case of the free-space film, modes 10 and 11, account for the side response.  

\begin{figure}[H]
     \centering
     \begin{subfigure}[b]{0.45\textwidth}
         \centering
        \includegraphics[trim = 1cm 1cm 8cm 19cm, clip,width=70mm ]{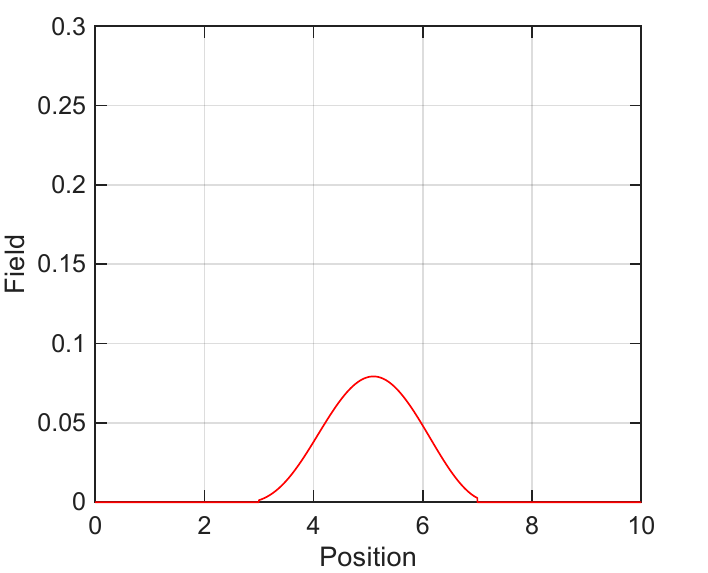}
         \caption{Mode 1}
         \label{fig_mode_curr_plot1_5}
     \end{subfigure}
     \hfill
     \begin{subfigure}[b]{0.45\textwidth}
         \centering
         \includegraphics[trim = 1cm 1cm 8cm 19cm, clip,width=65mm ]{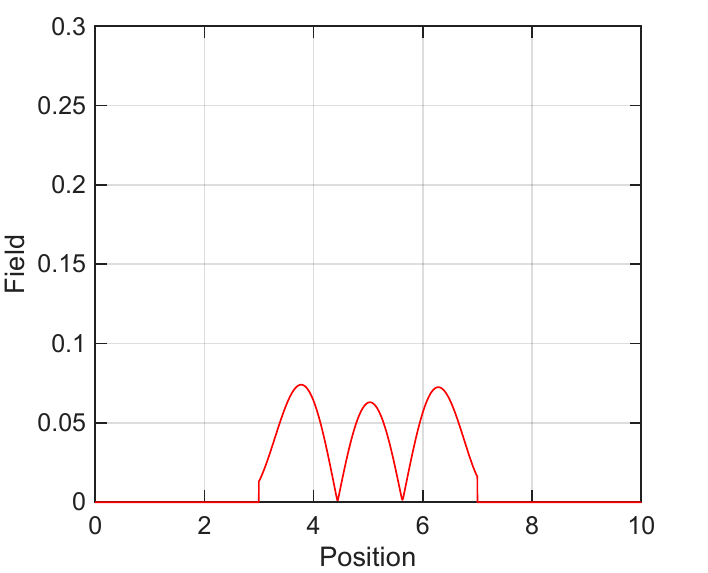}
           \caption{Mode 3 }    
            \label{fig_mode_curr_plot1_6}
     \end{subfigure}
          \hfill
     \begin{subfigure}[b]{0.45\textwidth}
         \centering
          \includegraphics[trim = 1cm 1cm 8cm 19cm, clip,width=65mm ]{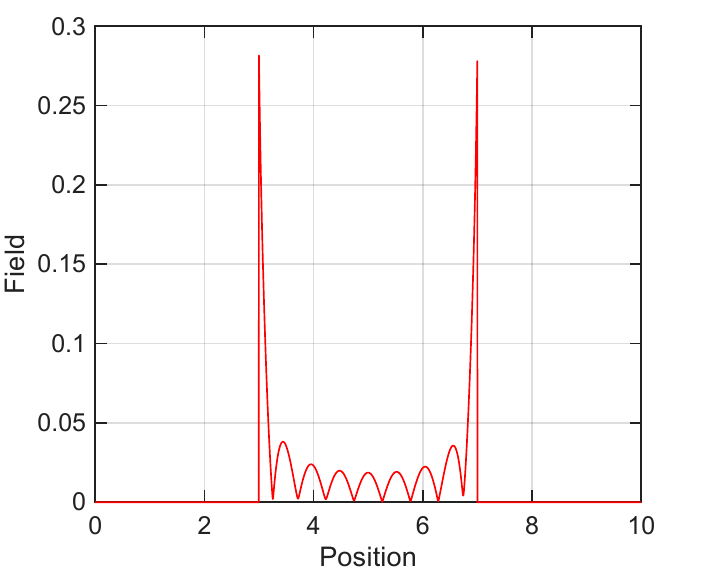}
          \caption{Mode 9}
            \label{fig_mode_curr_plot1_7}
     \end{subfigure}
          \hfill
     \begin{subfigure}[b]{0.45\textwidth}
         \centering
         \includegraphics[trim = 1cm 1cm 8cm 19cm, clip,width=65mm ]{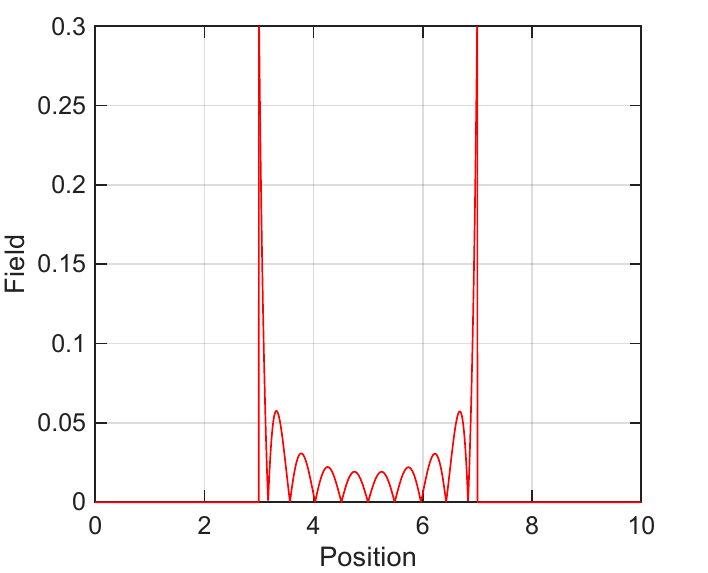}
                   \caption{Mode 10}
            \label{fig_mode_curr_plot1_8}
     \end{subfigure}
        \caption{Currents in the resistive strip corresponding to modes 1, 3, 9 and 10. }
        \label{fig_mode_curr_plot1_flm}
\end{figure}

\begin{figure}[H]
     \centering
     \begin{subfigure}[b]{0.45\textwidth}
         \centering
         \includegraphics[trim = 1cm 1cm 8cm 18cm, clip,width=65mm ]{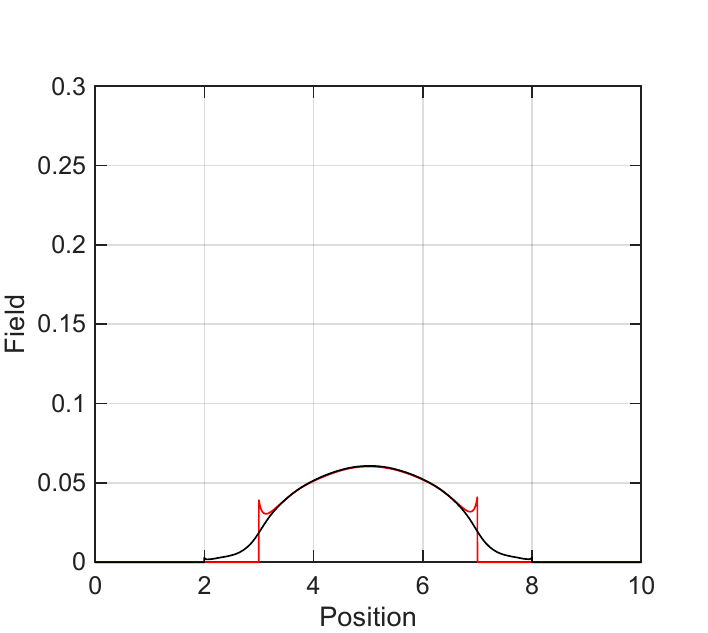}
         \caption{Mode 1}
         \label{fig_mode_curr_plot1_1}
     \end{subfigure}
     \hfill
     \begin{subfigure}[b]{0.45\textwidth}
         \centering
          \includegraphics[trim = 1cm 1cm 8cm 18cm, clip,width=65mm ]{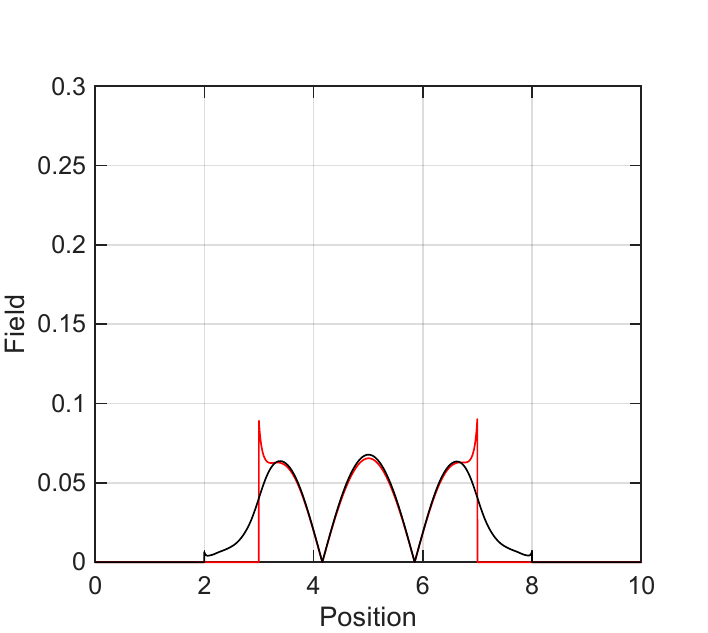}
           \caption{Mode 3}    
            \label{fig_mode_curr_plot1_2}
     \end{subfigure}
          \hfill
     \begin{subfigure}[b]{0.45\textwidth}
         \centering
         \includegraphics[trim = 1cm 1cm 8cm 18cm, clip,width=65mm ]{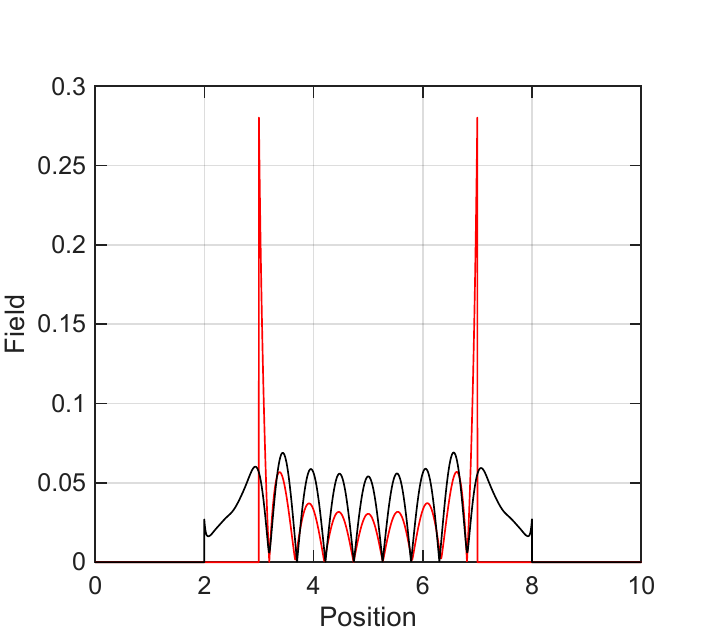}
          \caption{Mode 9 }
            \label{fig_mode_curr_plot1_3}
     \end{subfigure}
          \hfill
     \begin{subfigure}[b]{0.45\textwidth}
         \centering
          \includegraphics[trim = 1cm 1cm 8cm 18cm, clip,width=70mm ]{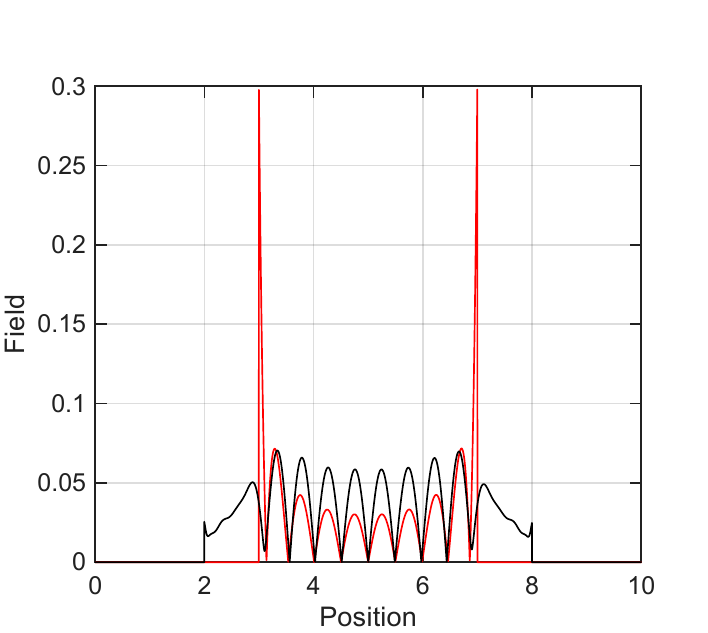}
                   \caption{Mode 10}
            \label{fig_mode_curr_plot1_4}
     \end{subfigure}
        \caption{Currents in the resistive strip (red) and ground plane (black) corresponding to modes 1, 3, 9 and 10. }
        \label{fig_mode_curr_plot1_bck}
\end{figure}

Once the discretised modes are known, it is possible to reconstruct the continuous forms of the modes, but in this case because a densely packed set of illumination angles was used, the illuminating fields are not orthogonal. The fields where therefore reconstructed using the dual set, and these fields where subsequently used to calculate the current distributions associated with the individual modes. Figure \ref{fig_mode_curr_plot1_flm} shows the current induced in the strip when the unbacked device was illuminated by modes 1, 3, 9 and 10. The lowest-order modes are approximately truncated Guassian functions, where the number of cycles corresponds to the angular positions of the lobes, as indicated by (75). It is likely, although this has not been verified, that these are Prolate Spheroidal Wavefunctions, which are the eigenfuctions of a propagator that is truncated in both the spatial and Fourier domains.  The currents associated with mode 9 and above are concentrated within about a wavelength of the edge, for the reasons described previously. In fact, as the order is increased above 9, the modes correspond to heavily truncated surface waves, which cannot be excited by a far-field source.
 Figure \ref{fig_mode_curr_plot1_bck} shows a similar set of plots when the backing plate was included. The behaviour is very similar to that of Fig.  \ref{fig_mode_curr_plot1_flm}, but now the current in the film starts to show enhanced singularities at the edges. These ultimately lead to cross talk when arrays of pixels are assembled.
%


\section{Noise and Fluctuations}
\label{sec_stray}

Noise arises in EAI experiments from two principle sources: (i) The SUT is bathed in a thermal radiation field, which makes a contribution to the measured power. This contribution, often called `straylight', can be small but is always present at some level. The detected power can be calculated using (\ref{eqn_7}) if the correlation tensor of the thermal field is known. It acts as an additive offset as long as the temperature of the field is stable. However, even if the temperature is constant, the field still fluctuates, and so the recorded power fluctuates. (ii) Noise can also appear at the output because of thermal fluctuations in the device itself, and because the readout electronics adds noise. In general terms, these effects are collectively characterised by Noise Equivalent Power (NEP), but in well-designed ultra-low-noise far-infrared detectors background noise dominates.  Before discussing how the fluctuations in the recorded power affect the responsivities and forms of the recovered modes, it is beneficial to consider the nature of the various noise contributions.

Equation (\ref{eqn_7}) can be used to calculate the power absorbed from a thermal background, where the integral over $\omega$ is determined by the bandwidth of the device $B$ on the `input side'.  Sometimes,  a device can absorb power over a large bandwidth and through many optical modes, and (\ref{eqn_7}) can lead to an offset that drifts with time. In low-noise sensors, unstable thermal loading can be problematic, and many techniques have been designed to mitigate its effects.

The spatial coherence tensor $\dyad{E} ({\bf r},{\bf r}',\omega)$ can be calculated straightforwardly for specific optical configurations. When the SUT is placed in a perfectly absorbing enclosure having uniform temperature $T$, the coherence tensor is given by Planck's law 
\begin{align}
\label{eqn_stray_1}
\dyad{E} ({\bf r},{\bf r}',\omega) & = \frac{1}{2 \pi} \frac{\hbar \omega}{ e^{\hbar \omega/kT} -1 } \dyad{I} \delta ({\bf r} - \bf{r}'), 
\end{align}
where the unit dyad $\dyad{I}$ indicates that the radiation is unpolarised. Strictly speaking, the coherence tensor has a finite coherence length and cross-polar correlations on the scale of a wavelength, which can be properly accounted for by using a k-domain representation of the background, but for the purposes of this discussion, (\ref{eqn_stray_1}) is adequate.

Substituting $(\ref{eqn_stray_1})$ in $(\ref{eqn_7})$ 
gives
\begin{align}
\label{eqn_stray_3}
P  & =  \frac{1}{2 \pi} \int_B {\rm d} \omega   \frac{\hbar \omega}{ e^{\hbar \omega/kT} -1 } \int_{\cal V}  \dyad{\bm \chi} ({\bf r},{\bf r},\omega)  \cdot \cdot \, \dyad{I} \, d^3{\bf r},
\end{align}
which is the background on the signal produced by the sources.  $\dyad{\bm \chi} ({\bf r},{\bf r},\omega)  \cdot \cdot \, \dyad{I}$ indicates the trace over an orthogonal set of polarisations, and overall $(\ref{eqn_stray_3})$ indicates that the device absorbes Planckian power through every degree of freedom available, taking into account the absoption efficiency of each mode. Equation (\ref{eqn_stray_3}) can also be written
\begin{align}
\label{eqn_stray_3b}
P  & =  \frac{1}{2 \pi} \int_B {\rm d} \omega   \frac{\hbar \omega}{ e^{\hbar \omega/kT} -1 } N_{\rm eff} (\omega),
\end{align}
where $N_{\rm eff} (\omega)$ is an {\em effective number of modes}. It can easily be shown that if the response tensor is written in diagonal form, 
$\dyad{\bm \chi} ({\bf r},{\bf r}',\omega) = \sum_n \lambda_n (\omega) {\bf u}_n({\bf r}',\omega) {\bf u}_n({\bf r},\omega)^\ast$, then through (\ref{eqn_stray_3})
\begin{align}
\label{eqn_stray_3c}
N_{\rm eff} (\omega) = \sum_n \lambda_n (\omega).
\end{align}
This analysis indicates that care is needed when designing an EAI measurement system for ultra-low-noise devices to ensure that background loading does not change as the coherent sources are moved.

Background loading is associated with thermal fluctuation noise. Suppose that two different devices, $a$ and $b$, say two pixels in an imaging array, are exposed to a common incoming partially coherent radiation field.  It can be shown \cite{Sak1}, using a Poisson mixture technique and the Gaussian moment theorem, that the covariance $C[P_{a},P_{b}]$ of the outputs of the detectors due to an incident thermal background is
\begin{align}
\label{eqn_stray_4}
& C[P_{a},P_{b}]  = \\ \nonumber
& \frac{1}{\tau} \int_B {\rm d} \omega \int_{\cal V} {\rm d}^{3} {\bf r}_{1}  \int_{\cal V} {\rm d}^{3}  {\bf r}_{2}
 \int_{\cal V} {\rm d}^{3} {\bf r}_{3}  \int_{\cal V} {\rm d}^{3} {\bf r}_{4} \,
\dyad{\bm \chi}_{a} ({\bf r}_{1},{\bf r}_{2},\omega) \cdot \dyad{E} ({\bf r}_{2},{\bf r}_{3},\omega)
\cdot \cdot \, \dyad{\bm\chi}_{b} ({\bf r}_{3},{\bf r}_{4},\omega) \cdot \dyad{E} ({\bf r}_{1},{\bf r}_{4},\omega) \\ \nonumber
& + \frac{\delta_{a b}}{\tau} \int_B {\rm d} \omega \, \hbar \omega \,    \int_{\cal V} {\rm d}^{3} {\bf r}_{1}  \int_{\cal V} {\rm d}^{3} {\bf r}_{2} \,  \dyad{\bm \chi}_{a} ({\bf r}_{1},{\bf r}_{2},\omega) \cdot \cdot \, \dyad{E} ({\bf r}_{1},{\bf r}_{2},\omega),
\end{align}
where  $\dyad{\bm \chi}_{a} ({\bf r}_1,{\bf r}_2,\omega)$ and $\dyad{\bm \chi}_{b} ({\bf r}_3,{\bf r}_4,\omega)$ are the response tensors of devices $a$ and $b$ respectively. $\tau$ is the effective post-detection integration time, introduced by the filtering on the output. If $B_o$ is the bandwidth of the readout system, then $\tau \approx 1/2 B_o$.  Equation (\ref{eqn_stray_4}) shows that the fluctuations in the outputs of any single detector, and the correlations between the outputs of pairs of detectors, can be calculated once the detectors' response tensors are known; adding additional value to EAI. Identical expressions hold when a reference surface is used, apart from the domain of integration. The modal approach described here is well suited to understanding straylight and radiation noise in pixels that couple poorly to the high-transmission modes of the preceding optical system. Indeed, (\ref{eqn_stray_1}) and (\ref{eqn_stray_4}) are valuable for optimising the sizes, spacings and layouts of the pixels in imaging arrays for efficiency and information recovery. 

The first term in (\ref{eqn_stray_4}) describes classical fluctuations, whereas the second term accounts for photon shot noise. When $a=b$, $ C[P_{a},P_{a}]$ is the variance of the output of detector $a$, and in this case the noise at the output is determined by both thermal and photon noise. When $a \neq b$, $\delta_{a b}$ indicates that the photon noise in different detectors is uncorrelated. For the purpose of EAI, we shall only be concerned with the fluctuations in the ouput of a single detector, not in the correlations.  Equation (\ref{eqn_stray_4}) may appear involved, but when $\dyad{\bm \chi} ({\bf r},{\bf r}',\omega)$ and  $\dyad{E} ({\bf r},{\bf r}',\omega)$  are sampled for numerical modelling the classical noise reduces to the trace of a product of four matrices, and photon noise reduces to the trace of a product of two matrices. 

In the case of a uniform, incoherent background, substituting (\ref{eqn_stray_1}) in (\ref{eqn_stray_4}) gives
\begin{align}
\label{eqn_stray_5}
C[P_{a},P_{b}]  & = \frac{1}{2 \pi \tau} \int_B {\rm d} \omega \frac{(\hbar \omega)^2}{( e^{\hbar \omega/kT} -1)^2 } \int_{\cal V} {\rm d}^{3} {\bf r}_{1}  \int_{\cal V} {\rm d}^{3}  {\bf r}_{2}  \, \dyad{\bm \chi}_{a} ({\bf r}_{1},{\bf r}_{2},\omega)   \cdot \cdot \, \dyad{\bm \chi }_{b} ({\bf r}_{2},{\bf r}_{1},\omega)  \\ \nonumber
& + \frac{\delta_{a b}}{2 \pi \tau} \int_B {\rm d} \omega  \frac{(\hbar \omega)^2}{ e^{\hbar \omega/kT} -1 } \int_{\cal V} {\rm d}^{3} {\bf r}_{1}   \,  \dyad{\bm \chi}_{a} ({\bf r}_{1},{\bf r}_{1},\omega) \cdot \cdot \, \dyad{I}.
\end{align}
If the response tensors are orthogonal, which would occur if they are physically distinct, then both terms evaluate to zero unless $a=b$. In other words, there are no correlations between the fluctuations of the outputs of different devices if they are embedded in an isothermal blackbody field. The variance of a single device becomes
\begin{align}
\label{eqn_stray_5b}
C[P_{a},P_{c}]  = \frac{1}{2 \pi \tau} \int_B {\rm d} \omega  \left( \frac{\hbar \omega}{ e^{\hbar \omega/kT} -1 } \right)^2\sum_n \lambda_n^2 (\omega) + \frac{\delta_{a b}}{2 \pi \tau} \int_B {\rm d} \omega \,  \frac{(\hbar \omega)^2}{ e^{\hbar \omega/kT} -1 } \sum_n \lambda_n(\omega).
\end{align} 
which reproduces equation 7 of \cite{With5}, but generalised to the case where the eigenvalues are not all unity. It can also be shown, using concepts from radiometry, that  $N_{eff} = A_e \Omega_e / \lambda^2$  where $A_e$ is the effective area of the device and $\Omega_e$ is the effective solid angle of the reception pattern. The power patterns and modal spectrum simulated in Section \ref{sec_flm} are consistent with this expression.

The last term in (\ref{eqn_stray_5}) can be compared with (\ref{eqn_stray_3}) to show that the variance of the photon noise can be written in terms of the absorbed power, which follows because the variance of a Poisson process is given by its mean. In the context of optical modelling, (\ref{eqn_stray_5}) can be used to calculate the fluctuations in the outputs of single detectors once the response tensors have been measured. The correlations between the outputs of two detectors can be calculated solely on the basis of the response tensors of each: no other information is needed. Even if the outputs of two detectors are added, which might be done in certain EAI experiments to investigate cross talk, the covariances can be used to calculate the fluctuations in the combined signal. 

An important consideration is that the fluctuations due to the different processes in (\ref{eqn_stray_5}) have different statistical forms: (i) The classical fluctuation noise associated with the thermal background is fourth-order in the field, and corresponds to the variance of a Rayleigh distribution. (ii) The photon noise relates to the variance of a Poisson process. (iii) The readout noise is assumed to be Gaussian, because it might correspond, for example, to a thermal voltage or current. Despite the difference between (i) and (ii) the variances add in quadrature, according to (\ref{eqn_stray_5}), and so adding the variance associated with the readout noise (iii) to give an overall variance  $\epsilon^2$ seems justified. Adding the variances in this way is commonplace when, for example, defining an overall NEP. 

Phase noise in the coherent illuminating sources represents another source of noise. It can often be assumed that the phase noise in the two arms is correlated because they are derived from a single source: assuming that the differential path length between the two sources is smaller that then coherence length of the sources, which for quasi temporally coherent sources is straightforward to achieve. Although not described explicitly, this noise is straightforward to incorporate, because the SUT measures power, and the complex fringe visibility comes from two power measurements. The role of a reference detector in reducing instability has already been addressed in Section \ref{sec_ref}.


\section{Error Analysis}
\label{sec_add}

Every EAI measurement yields a real-valued quantity that is proportional to power, and added noise appears as a fluctuation about this mean. Suppose that the flucuations have variance $\epsilon^2$. For each pair of EAI source locations,  the in and out-of-phase components of the resulting fringe are measured, and the complex-valued elements of  ${\bf \mathsf D}$ established, (\ref{eqn_eai_9}).  If the components are measured sequentially and sufficiently slowly, it can be assumed that the fluctuations in the two measured quadrature components are uncorrelated.  The variance of the sum or difference of two independent stochastic processes is the sum of their individual variances, and so we will assume that the overall distribution is Gaussian. Under this assumption,  each element of the measured response tensor is a complex analytic signal, $z = (x + i y)$, where $x$ and $y$ are each zero-mean stochastic Gaussian processes having variance $\sigma^2 = 2 \epsilon^2$. It follows that $z$ has a completely random phase and a Rayleigh amplitude distribution. Ordinarily, the complex analytic signal $z$ is defined with an additional factor of 2, but here we simply create a complex number from each pair of measurements. The noise may change from one spatial arrangement of the sources to the next, and so the variances of the elements of  ${\bf \mathsf D}$ may differ, but for brevity we will assume that they are the same: this assumption is easy to relax. 

Suppose that ${\bf \mathsf D}$ is the actual response matrix, but when the elements are measured, noise is introduced. For a single, complete set of experimental data 
\begin{align}
\label{eqn_add_1}
{\bf \mathsf H} & = {\bf \mathsf D} + {\bf \mathsf N},
\end{align}
where ${\bf \mathsf N}$ is a random matrix of complex analytic signals. Once output noise has been added, the measured response ${\bf \mathsf H}$ is no longer hermitian to within the accuracy imposed by the noise. Caution is required because non-hermitian matrices do not necessarily admit simple eigenvector representations; although this could be addressed by using Singular Value Decomposition (SVD). Selecting the hermitian part of the response removes the noise in the antihermitian part, which is similar to removing the quadrature component of some harmonic waveform to prevent noise in the quadrature component degrading the measurement.  

Assuming that all aspects of the measurement are ideal, apart from the noise,
\begin{align}
\label{eqn_add_2}
{\bf \mathsf H}^H & = {\bf \mathsf D} + {\bf \mathsf N}^H,
\end{align}
where the hermitian part of the measured response is given by
\begin{align}
\label{eqn_add_3}
{\bf \mathsf H}^H & = \frac{1}{2} \left[ {\bf \mathsf H} + {\bf \mathsf H}^\dagger \right],
\end{align}
${\bf \mathsf N}^H$ contains correlations, unlike ${\bf \mathsf N}$, because the off-axis terms form conjugate pairs. The moments of the elements of ${\bf \mathsf N}^H$ are calculated in Appendix \ref{appendixB}.
Because the measured DRF is now hermitian, it is possible to write 
\begin{align}
\label{eqn_add_4}
{\bf \mathsf H}^H & = \sum_n E_n {\bf \mathsf u}_n {\bf \mathsf u}_n^{\dagger} \\ \nonumber
& \mbox{as well as} \\ \nonumber
{\bf \mathsf D} & = \sum_n E_n^0 {\bf \mathsf u}^0_n {\bf \mathsf u}_n^{0 \dagger},
\end{align}
and the aim is to calculate the eigenvalues $E_n$ and eigenvectors ${\bf \mathsf u}_n$ of ${\bf \mathsf H}^H$ in terms of the eigenvalues $E_n^0$ and eigenvectors ${\bf \mathsf u}^0_n$ of  ${\bf \mathsf D}$.

The significance of measurement errors depends on the signal to noise ratio (SNR), but in what follows, it will often be convenient to use the noise-to-signal ratio (NSR). The magnitude of the noise is characterised by $\epsilon^2$, but the magnitude of the signal depends on the strengths of the sources, and so how can an SNR be defined? Various approaches are possible: (i) The most straightforward is to normalise the measured response matrix to the maximum power recorded as one source is moved over the sample positions, which is essentially the maximum value on the diagonal of ${\bf \mathsf H}^H$. Usually, this occurs when the source is on-axis. This information is available, and scales the  recorded power in a way that enables the SNR of a particular experiment to be calculated.  (ii) In simulations, it is possible to use the largest eigenvalue of the noiseless response matrix, which is the spectral norm $ \| {\bf \mathsf H}^H \|_2$. The largest diagonal element of the response matrix only places a lower bound on the spectral norm, and so the spectral norm is preferable. The spectral norm also has the advantage that  all of the eigenvalues lie in the range $0 \le \lambda_n \le 1$, and so aside from any overall output scaling or input inefficiency, the thermodynamic behaviour described in Section \ref{sec_stray} is correctly represented. (iii) The Frobenius norm, $ \|  {\bf \mathsf H}^H  \|_F = \sqrt{{\rm Tr} [ {\bf \mathsf H}^{H \dagger} {\bf \mathsf H}^{H}]}$, is also valuable when defining SNR, particularly when the source is partially coherent. 

In all cases, the normalising factor will be denoted $D_m$, such that 
\begin{align}
\label{eqn_add_6}
\frac{1}{D_m}  {\bf \mathsf H}^H & = \frac{1}{D_m} {\bf \mathsf D} + \frac{1}{D_m} { {\bf \mathsf N}^H} \\ \nonumber
\tilde{{\bf \mathsf H}}^H & = \tilde{\bf \mathsf D} + \frac{\sigma}{D_m} \tilde{\bf \mathsf N}^H.
\end{align}
$\tilde{{\bf \mathsf H}}$ and $\tilde{\bf \mathsf D}$ are the measured and actual response matrices normalised to $D_m$, and $\tilde{\bf \mathsf N}$ is the noise matrix normalised to $\sigma$. Each element of  $\tilde{\bf \mathsf N}$ is a complex random variable having unit variance.  If the NSR is small $\sigma / D_m  \ll 1$,   $\tilde{\bf \mathsf N}$ can be regarded as a perturbation of  $\tilde{\bf \mathsf D}$ with perturbation parameter $\sigma/D_m$. This approach will taken here, but it does not say anything about the absolute efficiency of the detector. To characterise absolute efficiency, one could compare the detected power with the power available from the source: the power that would be absorbed by a fully incoherent detector having the same physical footprint. In an imaging array, a reference detector can also be used to define an SNR against which all other pixels are measured: Section \ref{sec_ref}.

If the eigenvalues and eigenvectors of $\tilde{\bf \mathsf D}$ are known, which at the outset they are not, what are the statistical properties of  the eigenvalues and eigenvectors of $\tilde{{\bf \mathsf H}}^H$ in terms of the properties of  $\tilde{\bf \mathsf N}^H$. In other words, how does the noise manifest itself in changing the calculated eigenvalues and eigenvectors? This can be achieved through non-degenerate perturbation theory; degeneracy will be discussed also.


\subsection{Eigenvalue Shifts}
\label{sec_eigs}

The act of perturbing an operator is nonlinear in the eigenvalues and eigenvectors. In other words, the eigenvectors and eigenvalues of $\tilde{{\bf \mathsf H}}$ are not simply those of $\tilde{\bf \mathsf N}^H$ added to those of $\tilde{\bf \mathsf D}$. In what follows, all matrices, vectors, eigenvectors and eigenvalues correspond to those associated with the normalised operators $\tilde{{\bf \mathsf H}}^H$,  $\tilde{\bf \mathsf D}$ and $\tilde{\bf \mathsf N}^H$. 

The $n$'th eigenvalue of  $\tilde{{\bf \mathsf H}}^H$ in the presence of a perturbation can be written in terms of a 
power-series:
\begin{equation}
\label{eqn_eigs_1}
E_{n} (\lambda) =  E^{(0)}_{n} + \lambda E^{(1)}_{n} + \lambda^2 E^{(2)}_{n} + \lambda^3 E^{(3)}_{n} \cdots ,
\end{equation}
where $\lambda = \sigma/D_m$ is the perturbation parameter. In this section $\lambda$ is used, in line with usual convention, even though it is also used in earlier sections to denote eigenvalues. $E^{(0)}_{n}$ is the $n$'th eigenvalue in the absence of the perturbation, ($\lambda = 0$). $\lambda E^{(1)}_{n}$ is linear in the NSR,  $\lambda^2 E^{(2)}_{n}$ is quadratic, etc. In general, both $ \langle E_{n} (\lambda) \rangle $ and  $ \langle E_{n}^2(\lambda) \rangle - \langle E_{n} (\lambda) \rangle^2$, the mean and the variance, will be affected by noise. 

The expection value of the shift of the $n$'th eigenvalue is
\begin{equation}
\label{eqn_eigs_2}
\langle  \delta E_n  \rangle  = \lambda \langle E^{(1)}_{n} \rangle + \lambda^2  \langle E^{(2)}_{n} \rangle  +\lambda^3  \langle  E^{(3)}_{n} \rangle \cdots ,
\end{equation}
but as shown in Appendix \ref{appendixB}, the expectation values of all odd-ordered perturbation terms are zero, and therefore
\begin{equation}
\label{eqn_eigs_3}
\langle  \delta E_n  \rangle =  \lambda^2  \langle E^{(2)}_{n} \rangle +  \mathcal{O}(\lambda^4).  
\end{equation}

Appendix \ref{appendixC}, shows that, when a small deterministic perturbation $\hat{H}^{1}$ is applied to an unperturbed operator $\hat{H}^{0}$, the shift in the $n$'th eigenvalue is given by
\begin{align}
\label{eqn_eigs_4}
\delta E_{n} (\lambda) & \approx \lambda {\rm Tr} \left[\hat{H}^{1} \hat{P}_n \right]  - \lambda^2 {\rm Tr} \left[   \hat{H}^{1} \hat{S}_n  \hat{H}^{1} \hat{P}_{n} \right] \\ \nonumber
& + \lambda^3 {\rm Tr} \left[  \hat{H}^{1}  \hat{S}_n \hat{H}^{1} \hat{S}_n \hat{H}^{1} \hat{P}_n    \right]  - \lambda^3 {\rm Tr} \left[ \hat{H}^{1} \hat{S}_n^2 \hat{H}^{1}   \hat{P}_n  \hat{H}^{1} \hat{P}_n \right] +{\cal O}(\lambda^4).
\end{align}
where $\hat{P}_n$ and $\hat{S}_n$ are projectors and resolvents respectively, defined in (\ref{eqn_eigs_6}) and (\ref{eqn_eigs_7}) below.
In this context, `deterministic' means a small excursion of the perturbation operator $\hat{H}^{1}$ such as the deviation associated with a single realisation of a stochastic process. 

Because the expectation values of odd-ordered perturbation terms are zero, (\ref{eqn_eigs_3}) gives
\begin{align}
\label{eqn_eigs_5}
\langle \delta E_n  \rangle & = - \left( \frac{\sigma}{D_m} \right)^2 {\rm Tr} \left[ \langle \tilde{\bf \mathsf N}^H \tilde{\bf \mathsf S}_n \tilde{\bf \mathsf N}^H {\bf \mathsf P}_n \rangle \right]+  \mathcal{O}(\lambda^4) \\ \nonumber
&=  -  \left( \frac{\sigma}{D_m} \right)^2  \sum_{ijkl}   S_{n,ij} P_{n,kl}  \langle N^H_{jk}  N^H_{li} \rangle \\ \nonumber
&=  -  \left( \frac{\sigma}{D_m} \right)^2 \sum_{ijkl}  S_{n,ij} P_{n,kl}  \delta_{ij} \delta_{kl} \\ \nonumber
&=   -  \left( \frac{\sigma}{D_m} \right)^2 \ {\rm Tr} \left[  \tilde{\bf \mathsf S}_n \right]   {\rm Tr} \left[ \tilde{\bf \mathsf P}_n \right].
\end{align} 
where the Hilbert-space description of (\ref{eqn_eigs_4}) has been converted into matrices, and the expectation values of  Appendix \ref{appendixB}, used.

The projectors are defined by
\begin{align}
\label{eqn_eigs_6}
{\bf \mathsf P}_n & =  {\bf \mathsf u}^{(0)}_n  {\bf \mathsf  u}^{(0) \dagger}_n  \\ \nonumber
{\rm Tr} \left[ {\bf \mathsf P}_n \right] & = 1,
\end{align}
where ${\bf\mathsf  u}^{(0)}_m$ is the $m$'th eigenvector of the unperturbed response matrix, and the reduced resolvents are given by
\begin{align}
\label{eqn_eigs_7}
\tilde{\bf \mathsf S}_n & =  \sum_{m \neq n} \frac{{\bf \mathsf u}^{(0)}_m  {\bf \mathsf u}^{(0) \dagger}_m}{ E_{m}^{(0)} - E_{n}^{(0)}} \\ \nonumber
{\rm Tr} \left[\tilde{\bf \mathsf S}_n \right] & =  \sum_{m \neq n} \frac{1}{ E_{m}^{(0)} - E_{n}^{(0)}}.
\end{align}

In conclusion, readout noise leads to shifts in the recorded eigenvalues:
\begin{equation}
\label{eqn_eigs_8}
\langle \delta E_n  \rangle =  -   \left( \frac{\sigma}{D_m} \right)^2 {\rm Tr} \left[  \tilde{\bf \mathsf S}_n \right] =   \left( \frac{\sigma}{D_m} \right)^2 \sum_{m \neq n} \frac{1}{ E_{n}^{(0)}  -  E_{m}^{(0)}} + \mathcal{O} \left( \left( \frac{\sigma}{D_m} \right) ^4  \right).
\end{equation}
This expression has been normalised to $D_m$;  normalising again to $E_{n}^{(0)}$ yields the proportionate change:
\begin{align}
\label{eqn_eigs_9}
\frac{ \langle \delta E_n  \rangle }{E_{n}^{(0)}} & = \left( \frac{\sigma}{D_m E_{n}^{(0)}} \right)^2 \sum_{m \neq n} \frac{1}{1 - E_{m}^{(0)} / E_{n}^{(0)}},
\end{align}
where $D_m E_{n}^{(0)}$ is the eigenvalue that would have been calculated if the reponse matrix had not been normalised to $D_m$.
 
According to (\ref{eqn_eigs_8}), to ensure that an eigenvalue only shifts by a small amount, the difference between it and neighboring eigenvalues must be large compared with the variance of the noise. If the nearest-neighbour difference is of the order of the NSR, the shift  is of the order of NSR. Eigenvalues having neighbours that are closer than the NSR, can be shifted by appreciable amounts.  Crucially, the influence of a perturbation is not solely governed by the absolute values of the eigenvalues,  but also the difference between eigenvalues. 

Consider two isolated eigenvalues, when perturbed, the upper level will move up $\langle \delta E_n  \rangle > 0$, and the lower level will move down $\langle \delta E_n  \rangle < 0$; eigenvalues tends to repel. These shifts occur because a quasistatic fluctuation causes the levels to move, and repel, leading to shifts in the average values. The reduced resolvent is a measure of the sensitivity of an experiment to measurement noise. If near degeneracy occurs, eigenvalues split, which becomes more pronounced as the NSR is increased. For the spectra considered here, the largest and smallest eigenvalues tend to be degenerate on the scale of the noise, and so these will shift the most. Physical symmetries also lead to degeneracy, and these degeneracies will also broken by measurement noise.


\subsection{Eigenvalue Uncertainty}
\label{sec_eigu}

The variance of a measured eigenvalue is  
\begin{align}
\label{eqn_eigu_1}
\sigma^2_n &= \langle [ \delta E_n - \langle \delta E_n  \rangle ][ \delta E_n - \langle \delta E_n  \rangle] \rangle \\ \nonumber
& = \langle (\delta E_n )^2 \rangle -  \langle \delta E_n  \rangle^2.
\end{align}

The second term is already known, and the first term can be found from the perturbation expansion, 
\begin{equation}
\label{eqn_eigu_2}
\delta E_{n} (\lambda) = \lambda E^{(1)}_{n} + \lambda^2 E^{(2)}_{n} + \lambda^3 E^{(3)}_{n} \cdots ,
\end{equation}
giving
\begin{equation}
\label{eqn_eigu_3}
\sigma_{n}^2  = \lambda^2 \langle ( E^{(1)}_{n} )^2  \rangle  + \lambda^4  \langle  ( E^{(2)}_{n}  )^2 \rangle  + \lambda^4 2 \langle  E^{(1)}_{n} E^{(3)}_{n} \rangle  - ( \langle \delta E_n  \rangle )^2 +\mathcal{O}(\lambda^6).
\end{equation}
because all odd-order terms are zero. The first term on the RHS, which is a second-order term, evaluates to
\begin{align}
\label{eqn_eigu_4}
\langle ( E^{(1)}_{n} )^2  \rangle & = \left( \frac{\sigma}{D_m} \right)^2  \langle \left( {\rm Tr} \left[\tilde{\bf \mathsf N}^H {\bf \mathsf P}_n \right] \right)^2 \rangle\\ \nonumber
& = \left( \frac{\sigma}{D_m} \right)^2 \sum_{ijkl} \langle N^H_{ij}  N^H_{kl} \rangle P_{n, ji} P_{n, lk} \\ \nonumber
& =   \left( \frac{\sigma}{D_m} \right)^2 {\rm Tr} \left[  {\bf \mathsf P}_n  {\bf \mathsf P}_n \right]  = \left( \frac{\sigma}{D_m} \right)^2
\end{align}
which is a straightforward result. 

To second order, the variance on eigenvalue $n$ is
\begin{align}
\label{eqn_eigu_5}
\left( \frac{\sigma_n}{D_m} \right)^2 & = \left( \frac{\sigma}{D_m} \right)^2 -    \left[\left( \frac{\sigma}{D_m} \right)^2 \sum_{m \neq n} \frac{1}{E_{n}^{(0)} - E_{m}^{(0)} }\right]^2  \approx \left( \frac{\sigma}{D_m} \right)^2,
\end{align}
where the last inequality is true in the case of widely separated eigenvalues. The variance of each eigenvalue is simply given by the noise on each measurement, and indeed the shift in each eigenvalue is $\mathcal{O}((\sigma/D_m)^2)$, and the small correction term on (\ref{eqn_eigu_5}) is $\mathcal{O}((\sigma/D_m)^4)$.

Consider the second term on the RHS of (\ref{eqn_eigu_3}), which is the fourth-order term:
\begin{align}
\label{eqn_eigu_6}
\langle ( E^{(2)}_{n} )^2  \rangle & =  \left( \frac{\sigma}{D_m} \right)^4 \langle  \left( {\rm Tr} \left[  
\tilde{\bf \mathsf S}_n \tilde{\bf \mathsf N}^H {\bf \mathsf P}_n \tilde{\bf \mathsf N}^H \right] \right)^2 \rangle \\ \nonumber
& =  \left( \frac{\sigma}{D_m} \right)^4  \sum_{ijkl} \sum_{rstu} S_{n,ij}  P_{n,kl}  S_{n,rs}   P_{n,tu}  \langle N^H_{n,jk} N^H_{li} N^H_{n,st} N^H_{ur} \rangle \\ \nonumber
& =  \left( \frac{\sigma}{D_m} \right)^4  \left\{ \left({\rm Tr} [   \tilde{\bf \mathsf S}_n] \right)^2  +  {\rm Tr} [  \tilde{\bf \mathsf S}_n^2 ] \right\}.
\end{align}

The third term on the right of (\ref{eqn_eigu_3}), which is also fourth order, follows in a similar way:
\begin{align}
\label{eqn_eigu_7}
2 \langle  E^{(1)}_{n} E^{(3)}_{n} \rangle & =  2 \left( \frac{\sigma}{D_m} \right)^4 \langle {\rm Tr} \left[ \tilde{\bf \mathsf N}^H {\bf \mathsf P}_n \right]  {\rm Tr} \left[   \tilde{\bf \mathsf N}^H  \tilde{\bf \mathsf S}_n  \tilde{\bf \mathsf N}^H  \tilde{\bf \mathsf S}_n 
\tilde{\bf \mathsf N}^H {\bf \mathsf P}_n \right] \rangle \\ \nonumber
&  -  2 \left( \frac{\sigma}{D_m} \right)^4 \langle {\rm Tr} \left[\tilde{\bf \mathsf N}^H {\bf \mathsf P}_n \right] {\rm Tr} \left[  \tilde{\bf \mathsf N}^H  \tilde{\bf \mathsf S}_n^2 \tilde{\bf \mathsf N}^H  {\bf \mathsf P}_n \tilde{\bf \mathsf N}^H {\bf \mathsf P}_n \right] \rangle.
\end{align}
Again, consider the two terms. The first term in (\ref{eqn_eigu_7}) evaluates to zero because  $\tilde{\bf \mathsf S}_n {\bf \mathsf P}_n = 0$, and so only the second term remains
\begin{align}
\label{eqn_eigu_9}
& -  \left( \frac{\sigma}{D_m} \right)^4 \langle {\rm Tr} \left[\tilde{\bf \mathsf N}^H {\bf \mathsf P}_n \right] {\rm Tr} \left[  \tilde{\bf \mathsf N}^H  \tilde{\bf \mathsf S}_n^2 \tilde{\bf \mathsf N}^H  {\bf \mathsf P}_n \tilde{\bf \mathsf N}^H {\bf \mathsf P}_n \right] \rangle  = \\ \nonumber
& = -  \left( \frac{\sigma}{D_m} \right)^4 \sum_{ijklrstuv} P_{n,ji} S_{n,lr} S_{n,rs}  P_{n,tu} P_{n,vk} \langle N^H_{n,ij}  N^H_{n,kl} N^H_{n,st} N^H_{n,uv}  \rangle \\ \nonumber
& = - \left( \frac{\sigma}{D_m} \right)^4 \left\{ {\rm Tr} [ {\bf \mathsf P}_n^2 \tilde{\bf \mathsf S}_n^2]  {\rm Tr} [ {\bf \mathsf P}_n ] + {\rm Tr} [ {\bf \mathsf P}_n^2 \tilde{\bf \mathsf S}_n^2]  {\rm Tr} [ {\bf \mathsf P}_n ] + {\rm Tr} [ {\bf \mathsf P}_n^3 ]  {\rm Tr} [ \tilde{\bf \mathsf S}_n^2 ] \right\} \\ \nonumber
 & = - \left( \frac{\sigma}{D_m} \right)^4 \left\{{\rm Tr} [ \tilde{\bf \mathsf S}_n^2 ] \right\}
\end{align} 

Combining (\ref{eqn_eigu_4}), (\ref{eqn_eigu_6}), (\ref{eqn_eigu_7}), and (\ref{eqn_eigu_9}) gives the variance of the $n$'th eigenvalue as 
\begin{equation}
\label{eqn_eigu_10}
\left( \frac{\sigma_n}{D_m} \right)^2 = \left( \frac{\sigma}{D_m} \right)^2  \left\{ 1 - \left( \frac{\sigma}{D_m} \right)^2   {\rm Tr} [ 
\tilde{\bf \mathsf S}_n^2] \right\} .
\end{equation}

In addition to (\ref{eqn_eigs_6}) and (\ref{eqn_eigs_7}), it is useful to know 
\begin{align}
\label{eqn_eigu_11}
({\rm Tr} \left[ \tilde{\bf \mathsf S}_n \right] )^2 & =  \left( \sum_{m \neq n} \frac{1}{ E_{m}^{(0)} - E_{n}^{(0)}} \right)^2,
\end{align}
and
\begin{align}
\label{eqn_eigu_12}
{\rm Tr} \left[  \tilde{\bf \mathsf S}_n^2 \right] & = \sum_{m \neq n} \frac{1}{ ( E_{m}^{(0)} - E_{n}^{(0)} )^2}.
\end{align}
Once again, the variances are most affected by nearby eigenvalues. 


\subsection{Eigenmode Shifts}
\label{sec_emd}

Equation (\ref{eqn_appc_46}) of Appendix \ref{appendixC} shows that a deterministic perturbation changes the forms of the eigenvectors to
\begin{equation}
\label{eqn_emd_1}
| \phi_{n} \rangle (\lambda) = \left[ \hat{P}_n + \lambda {\hat W}_n^{(1)} + \lambda^2 {\hat W}_n^{(2)} + .... \right] | \phi^{(0)}_{n} \rangle,
\end{equation}
where
\begin{align}
\label{eqn_emd_2}
\hat{W}_n^{(1)} & =  - \hat{S}_n  \hat{H}^{1} \hat{P}_n \\ \nonumber
 \hat{W}_n^{(2)} & = \left(  \hat{S}_n  \hat{H}^{1} \right)^2 \hat{P}_n  - \left( \hat{S}_n \right)^2 \left( \hat{H}^{1} \hat{P}_n \right)^2 - \frac{1}{2} \hat{P}_n \hat{H}^{1} \left( \hat{S}_n \right)^2 \hat{H}^{1} \hat{P}_n.
\end{align}

The expectation value of a perturbed mode becomes
\begin{equation}
\label{eqn_emd_3}
\langle | \phi_{n} \rangle (\lambda) \rangle  = \langle \left[ \hat{P}_n + \lambda {\hat W}_n^{(1)} + \lambda^2 {\hat W}_n^{(2)} + .... \right] | \phi^{(0)}_{n} \rangle \rangle.
\end{equation}

Expressing this Hilbert-space representation in matrix form, and using the moments in \ref{appendixB}, it is clear that there are no first-order corrections:
\begin{equation}
\label{eqn_emd_4}
\langle  \tilde{\bf \mathsf W}_n^{(1)} \rangle =  - \tilde{\bf \mathsf S}_n \langle \tilde {\bf \mathsf N}^{H} \rangle {\bf \mathsf P}_n = 0.
\end{equation}
Considering the first term of (\ref{eqn_emd_2}), it is necessary to know
\begin{align}
\label{eqn_emd_5}
[ \langle  \tilde{\bf \mathsf N}^{H}  \tilde{\bf \mathsf S}_n  \tilde{\bf \mathsf N}^{H} \rangle ]_{il} \equiv  {\rm Tr}[\tilde{\bf \mathsf S}_n] \delta_{il},
\end{align}
and then
\begin{align}
\label{eqn_emd_6}
\lambda^2 [ \tilde{\bf \mathsf S}_n  \langle  \tilde{\bf \mathsf N}^{H}  \tilde{\bf \mathsf S}_n  \tilde{\bf \mathsf N}^{H} \rangle {\bf \mathsf P}_n {\bf u}^0_n ]_i & \equiv \left( \frac{\sigma}{D_m} \right)^2  \sum_{jkl} {S}_{n,i j} {\rm Tr}[\tilde{\bf \mathsf S}_n] \delta_{jk} {P}_{n,kl} u_{n,l} \\ \nonumber
& = \left( \frac{\sigma}{D_m} \right)^2  {\rm Tr}[\tilde{\bf \mathsf S}_n]  \tilde{\bf \mathsf S}_n  {\bf \mathsf P}_n {\bf \mathsf u}^0_n = 0.
 \end{align}
 
For the second term of (\ref{eqn_emd_2}) , we also need
\begin{align}
\label{eqn_emd_7}
[ \langle  \tilde{\bf \mathsf N}^{H}  {\bf \mathsf P}_n  \tilde{\bf \mathsf N}^{H} \rangle ]_{il} \equiv  \delta_{il},
\end{align}
giving
\begin{align}
\label{eqn_emd_8}
- \lambda^2 [ \tilde {\bf \mathsf S}_n  \tilde{\bf \mathsf S}_n  \langle  \tilde{\bf \mathsf N}^{H}  {\bf \mathsf P}_n  \tilde{\bf \mathsf N}^{H} \rangle  {\bf \mathsf P}_n {\bf \mathsf u}^0_n ]_i & = -  \left( \frac{\sigma}{D_m} \right)^2 \sum_{jkl}   {S}_{n,i j}  {S}_{n, jk}  \delta_{kl} {P}_{n,lm} u^0_{n m} \\ \nonumber 
& = - \left( \frac{\sigma}{D_m} \right)^2 \left( \tilde{\bf \mathsf S}_n \right)^2 {\bf \mathsf P}_n {\bf \mathsf u}^0_n = 0. 
\end{align}

Finally for the third term, 
\begin{align}
\label{eqn_emd_9}
[ \langle \tilde{\bf \mathsf N}^{H} \left( \tilde{\bf \mathsf S}_n \right)^2 \tilde{\bf \mathsf N}^{H} \rangle ] _{il} & =  {\rm Tr} [ \left( \tilde{\bf \mathsf S}_n \right)^2 ] \delta_{il},
\end{align}
and so
\begin{align}
\label{eqn_emd_10}
- \lambda^2 \frac{1}{2} [ {\bf \mathsf P}_n \langle \tilde{\bf \mathsf N}^{H} \left( \tilde{\bf \mathsf S}_n \right)^2 \tilde{\bf \mathsf N}^{H} \rangle {\bf \mathsf P}_n {\bf \mathsf u}^0_n ]_i  & \equiv - \frac{1}{2} 
\left( \frac{\sigma}{D_m} \right)^2 {\rm Tr} [ \left( \tilde{\bf \mathsf S}_n \right)^2 ]   \sum_{jkl} {P}_{n,ij}    \delta_{jk} {P}_{n,kl}   u^0_{n l} \\ \nonumber
& =  - \frac{1}{2} \left( \frac{\sigma}{D_m} \right)^2 {\rm Tr} [ \left( \tilde{\bf \mathsf S}_n \right)^2 ]  {\bf \mathsf u}^0_n.
\end{align}

Gathering these together, the change in the expectation value of mode $n$ to second order is  
\begin{equation}
\label{eqn_erm_11}
\langle \delta {\bf \mathsf u}_n \rangle   = - \frac{1}{2} \left( \frac{\sigma}{D_m} \right)^2    {\rm Tr} [ \left( \tilde{\bf \mathsf S}_n \right)^2 ]  {\bf \mathsf u}^0_n,
\end{equation}
which comprises a simple rescaling of the unperturbed vector. Then
\begin{equation}
\label{eqn_erm_12}
\langle {\bf \mathsf u}_n \rangle   =  \left\{ 1 - \frac{1}{2} \left( \frac{\sigma}{D_m} \right)^2    {\rm Tr} [ \left( \tilde{\bf \mathsf S}_n \right)^2 ] \right\}  {\bf \mathsf u}^0_n.
\end{equation}
Comparing with (\ref{eqn_eigu_10}) it can be seen that the eigenvector is scaled by a factor that is similar to that of the variance of the eigenvalue. This scaling occurs because the measured noisy modes have, by virtue of the perturbation method, been renormalised. 


\subsection{Eigenmode Covariance}
\label{sec_mvr}

The covariances between the spatial points of an eigenmode can be calculated as follows. According to Appendix \ref{appendixC},
\begin{align}
\label{eqn_mvr_1}
| \phi_{n} \rangle (\lambda) & = | \phi^{(0)}_{n} \rangle + \lambda | \phi^{(1)}_{n} \rangle + \lambda^{2} | \phi^{(2)}_{n} \rangle + \mathcal{O}(\lambda^3) \\ \nonumber
\left[ {\bf \mathsf u}_{n}  \right]_i & = \left[ {\bf \mathsf u}^{(0)}_{n}  \right]_i + \lambda \left[ {\bf \mathsf u}^{(1)}_{n}  \right]_i +  \mathcal{O}(\lambda^2) ,
\end{align}
where in the second line discretised eigenvectors have been introduced. Also, because the aim is to calculate the covariances between the elements of an eigenvector, the series has been terminated at first order, which gives the covariances to second order. The covariances are then accurate to third order because the third-order terms are zero.

The modes are complex-valued quantities, and so the variances and covariances are complex valued. The complex-valued covariances of interest are 
\begin{align}
\label{eqn_mvr_2}
(\tilde{\bf \mathsf K}_n)_{ij} & = \langle \delta {\bf \mathsf u}_{n,i}  \delta {\bf \mathsf u}_{n,j}^\ast \rangle \\ \nonumber
\boldsymbol{\kappa}_n & =  \left( \frac{\sigma}{D_m} \right)^2  \langle \left[  \tilde{\bf \mathsf S}_n  \tilde{\bf \mathsf N}^{H} {\bf \mathsf P}_n  {\bf \mathsf u}^0_n \right]   \left[  \tilde{\bf \mathsf S}_n  \tilde{\bf \mathsf N}^{H} {\bf \mathsf P}_n  {\bf \mathsf u}^0_n \right]^\dagger \rangle \\ \nonumber
& =  \left( \frac{\sigma}{D_m} \right)^2  \left[  \tilde{\bf \mathsf S}_n  \langle  \tilde{\bf \mathsf N}^{H}  {\bf \mathsf u}^0_n {\bf \mathsf u}^{0 \dagger}_n \tilde{\bf \mathsf N}^{H} \rangle \tilde{\bf \mathsf S}_n  \right],
\end{align}
but
\begin{align}
\label{eqn_mvr_3}
 \left( \frac{\sigma}{D_m} \right)^2 \langle  \tilde{\bf \mathsf N}^{H} {\bf \mathsf u}^0_n {\bf \mathsf u}^{0 \dagger}_n  \tilde{\bf \mathsf N}^{H} \rangle_{il} & = \left( \frac{\sigma}{D_m} \right)^2 {\rm Tr} [ {\bf \mathsf u}^0_n {\bf \mathsf u}^{0 \dagger}_n ] \delta_{il} =\left( \frac{\sigma}{D_m} \right)^2  \delta_{il}
\end{align}
and so
\begin{align}
\label{eqn_mvr_4}
\tilde{\bf \mathsf K}_n & =  \left( \frac{\sigma}{D_m} \right)^2 \left[  \tilde{\bf \mathsf S}_n  \langle  \tilde{\bf \mathsf N}^{H} {\bf \mathsf u}^0_n {\bf \mathsf u}^{0 \dagger}_n  \tilde{\bf \mathsf N}^{H} \rangle \tilde{\bf \mathsf S}_n  \right]  = \left( \frac{\sigma}{D_m} \right)^2   \tilde{\bf \mathsf S}_n^2.
\end{align}
In conclusion,
\begin{equation}
\label{eqn_mvr_5}
\tilde{\bf \mathsf K}_n  = \left( \frac{\sigma}{D_m} \right)^2 \tilde{\bf \mathsf S}_n^2 ,
\end{equation}
which is the Hermitian covariance matrix of the $n$'th noisy eigenvector. In some cases, Section \ref{sec_rer}, only the variances are needed, in which case
\begin{equation}
\label{eqn_mvr_6}
\tilde{\bf \mathsf \kappa} =  \left( \frac{\sigma}{D_m} \right)^2  {\rm Diag} \left[ \tilde{\bf \mathsf S}_n^2 \right].
\end{equation}
which is the point-by-point variance vector of the $n$'th noisy eigenvector. 


\subsection{Reconstruction Errors}
\label{sec_rer}

The shifts and variances of the eigenvalues and eigenvectors of the noisy response matrix are now known, but these are given in terms of the reduced resolvents $\tilde{\bf \mathsf S}_n$ of the unperturbed detector modes, which in an experiment are not known apriori.   According to (\ref{eqn_eigs_8}) and (\ref{eqn_erm_12}), however, the measured eigenvalues and eigenvectors are given by 
\begin{align}
\label{eqn_rer_1}
\langle E_n  \rangle & = E_n^0 +\mathcal{O}\left[ \left( \frac{\sigma}{D_m} \right)^2\right]  \\ \nonumber
\langle {\bf \mathsf u}_n \rangle & = {\bf \mathsf u}^0_n + \mathcal{O}\left[ \left( \frac{\sigma}{D_m} \right)^2\right],
\end{align}
and so in experiments it is sufficient to use the measured noisy eigenvalues and eigenvectors when calculating the resolvents, and the calculated shifts and variances will be correct to second order. In numerical simulations, the noiseless originals can be used, and indeed it is found that this approximation is correct.

The calculated shifts and variances are those pertaining to the discretised modes, and so what remains to be done is to calculate the shifts and variances of the reconstructed continuous modes. Following (\ref{eqn_eai_14})
\begin{align}
\label{eqn_rer_2}
{\bf u}_n({\bf r}) = \sum_{m}  u_{n,m} {\bf E}_m ({\bf r}) .  
\end{align}

From (\ref{eqn_rer_2}), the shifted reconstructed mode is
\begin{align}
\label{eqn_rer_3}
\langle {\bf u}_n({\bf r}) \rangle &  = \sum_{m}  \langle u_{n,m} \rangle {\bf E}_m ({\bf r})  \\ \nonumber
& =\sum_{m}  \langle u_{n,m}^0  \rangle {\bf E}_m ({\bf r}) +\sum_{m}  \langle \delta u_{n,m}  \rangle {\bf E}_m ({\bf r}) \\ \nonumber
& =  {\bf u}_n({\bf r}) + \delta {\bf u}_n({\bf r})
\end{align}
where  $\langle \delta u_{n,m} \rangle$ is given by  (\ref{eqn_erm_11}).  In conclusion
\begin{equation}
\label{eqn_rer_6}
\delta {\bf u}_n({\bf r}) = \sum_{m}  \langle \delta u_{n,m}  \rangle {\bf E}_m ({\bf r})
\end{equation}
is the shift in the reconstructed, continuous, eigenmodes.

The variances of the amplitudes of the reconstructed modes are
\begin{align}
\label{eqn_rer_7}
\langle | \delta {\bf u}_n({\bf r}) |^2 \rangle  & = \langle | \sum_{i} \delta  u_{n,i}  {\bf E}_i ({\bf r})  -  \sum_{j} \langle \delta u_{n,j} \rangle   {\bf E}_i ({\bf r}) \|^2 \rangle \\ \nonumber
 & =  \sum_{ii'} \langle \delta  u_{n,i}  \delta  u^{\ast}_{n,i'} \rangle {\bf E}_i ({\bf r}) \cdot  {\bf E}_{i'}^\ast ({\bf r}) + \sum_{jj'} \langle \delta u_{n,j} \rangle  \langle \delta u^{\ast}_{n,j'} \rangle  {\bf E}_j ({\bf r}) \cdot {\bf E}_{j'}^\ast ({\bf r}) 
 - 2 {\rm Re} \left[ \sum_{ij} \langle \delta u_{n,i} \rangle  \langle \delta u^{\ast}_{n,j} \rangle  {\bf E}_i ({\bf r}) \cdot {\bf E}_{j}^\ast ({\bf r}) \right] \\ \nonumber
  & =  \sum_{ij} \left[ \langle \delta  u_{n,i}  \delta  u^{\ast}_{n,j} \rangle - \langle \delta u_{n,i} \rangle  \langle \delta u^{\ast}_{n,j} \rangle  \right] {\bf E}_i ({\bf r}) \cdot {\bf E}_{j}^\ast ({\bf r}) 
\end{align}
where the first term indicates that the correlations between the variations in the coefficients of a single eigenvector $\tilde{\bf \mathsf K}_n$ must be known in order to calculate the variance of the reconstructed mode.

There are two possibilities: (i) the source fields do not overlap spatially; (ii) the sources fields do overlap, but are nevertheless mathematically orthogonal. In the case of (i)
\begin{equation}
\label{eqn_rer_8}
\langle | \delta {\bf u}_n({\bf r}) |^2 \rangle   =\sum_{i} \left[ \langle \delta  u_{n,i}  \delta  u^{\ast}_{n,i} \rangle - \langle \delta u_{n,i} \rangle  \langle \delta u^{\ast}_{n,i} \rangle  \right] | {\bf E}_i ({\bf r}) |^2, 
\end{equation}
and only the variances $\tilde{\mathsf \kappa}$ are needed. In this case, the variances of the discrete modes map in one-to-one correspondence onto the variances of the reconstructed continuous response tensor. In the case of (ii), if the modes overlap, but are mathematically orthogonal. $\langle | \delta {\bf u}_n({\bf r}) |^2 \rangle$ can then be integrated over ${\bf r}$ to give a scalar measure of the variance of the reconstructed mode, but if the spatial errors are needed, (\ref{eqn_rer_7}) must be used. 

In simulations, it is convenient to have a scalar measure of the degree to which the measured noisy modes  ${\bf \mathsf u}_n$ match the underlying noiseless modes ${\bf \mathsf u}^0_n$ . Given that ${\bf \mathsf u}_n$ and ${\bf\mathsf  u}^0_n$ are both normalised, their inner product can be regarded as the cosine of the angle between the two vectors, and so provides a suitable measure:
\begin{equation}
\label{eqn_rer_9}
\cos ( \alpha_n) = {\bf \mathsf u}_n^\dagger {\bf \mathsf u}^{0}_n .
\end{equation}
This same approach can be used to look for leakage between the measured noisy modes and the noiseless modes being sought. In the case of simulations, the error cosines can be calculated directed, or additionally when perturbation theory holds, using (\ref{eqn_erm_12}),
\begin{equation}
\label{eqn_rer_10}
\cos ( \alpha_n)    =  \left\{ 1 - \frac{1}{2} \left( \frac{\sigma}{D_m} \right)^2    {\rm Tr} [ \left( \tilde{\bf \mathsf S}_n \right)^2 ] \right\}.
\end{equation}
\begin{figure}[h]
     \centering
     \begin{subfigure}[b]{0.45\textwidth}
         \centering
         \includegraphics[trim = 1cm 1cm 8cm 19cm, clip,width=70mm ]{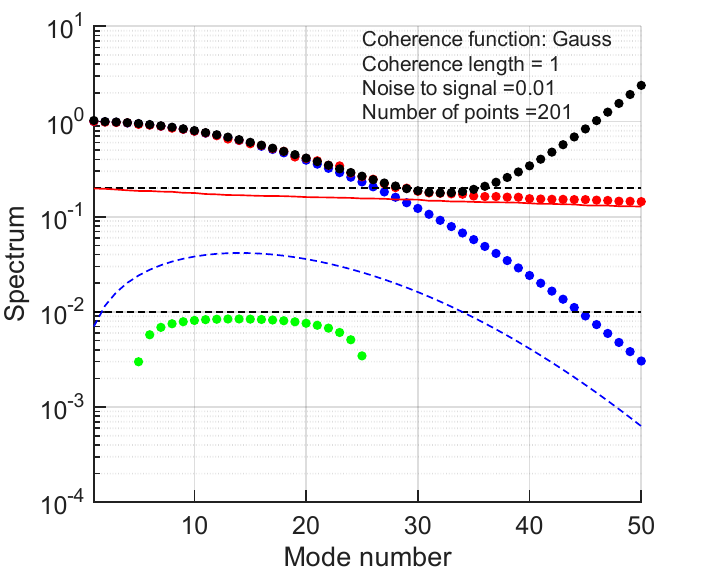}
         \caption{Coherence length $l=$ 1.}
         \label{fig_exp_1}
     \end{subfigure}
     \hfill
     \begin{subfigure}[b]{0.45\textwidth}
         \centering
          \includegraphics[trim = 1cm 1cm 8cm 19cm, clip,width=70mm ]{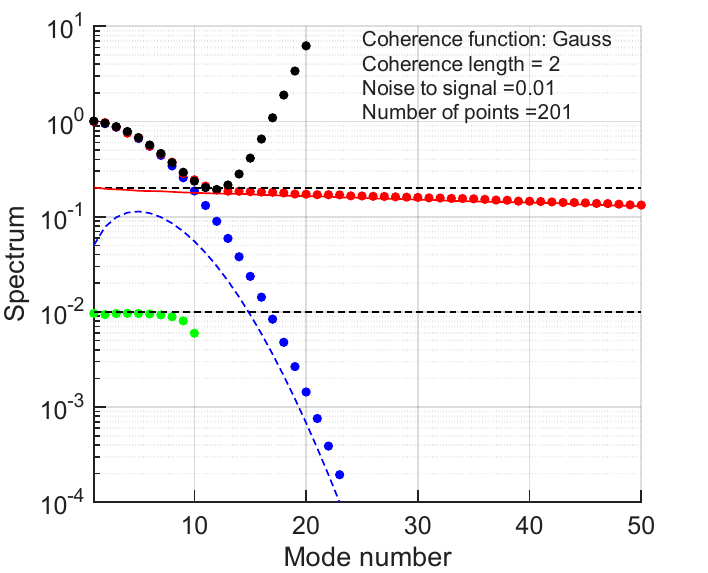}
            \caption{Coherence length  $l=$ 2.}
            \label{fig_exp_2}
     \end{subfigure}
          \centering
     \begin{subfigure}[b]{0.45\textwidth}
         \centering
         \includegraphics[trim = 1cm 1cm 8cm 19cm, clip,width=70mm ]{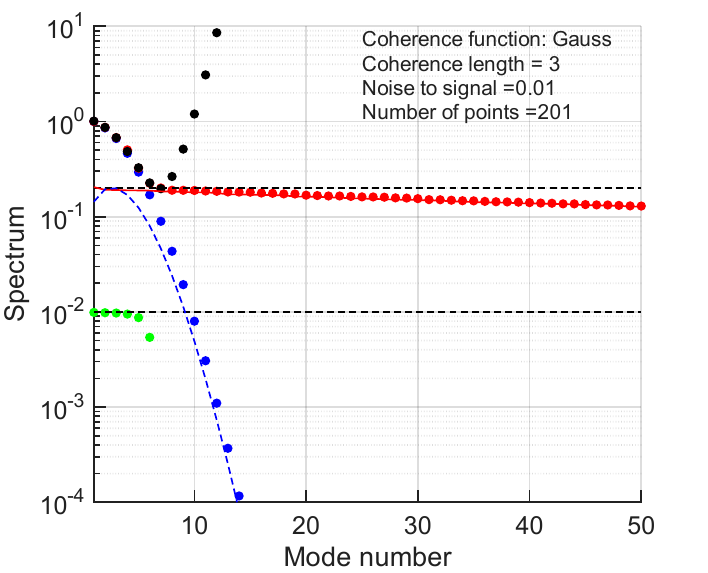}
        \caption{Coherence length  $l=$ 3.}
             \label{fig_exp_3}
     \end{subfigure}
     \hfill
     \begin{subfigure}[b]{0.45\textwidth}
         \centering
         \includegraphics[trim = 1cm 1cm 8cm 19cm, clip,width=70mm ]{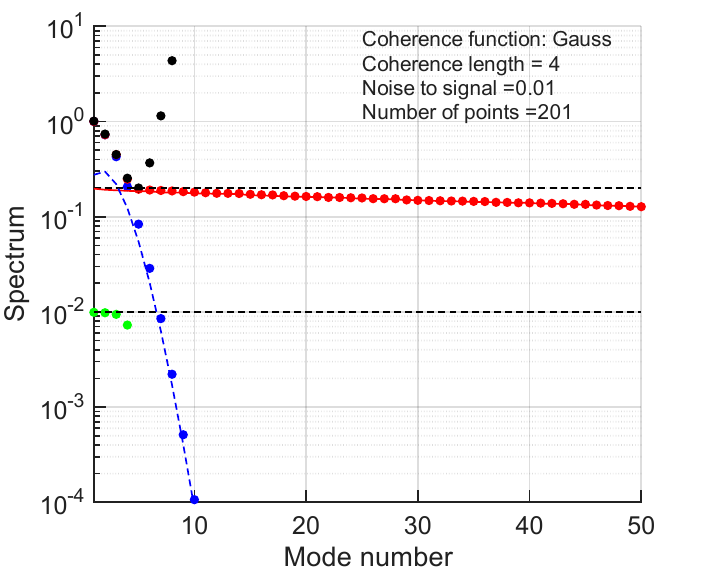}
 			\caption{Coherence length  $l=$ 4.}
              \label{fig_exp_4}
     \end{subfigure}
        \caption{Spectra of a Gaussian coherence function. The width was $w=$ 10.0, the NSR was $nts=$ 0.01, and $N=$ 201 sample points were used in all cases.  (a)-(d) are for coherence lengths of $l=$ 1, 2, 3 and 4. The various coloured plots are described in the text.}
        \label{fig_spec_1}
\end{figure}


\section{Illustrative Model}
\label{sec_sim}

The effects of noise on EAI are best appreciated through numerical simulations. Consider a 1D absorbing structure having width $w$, coherence length $l$. Two different coherence functions were used: a Gaussian
\begin{equation}
\label{eqn_sim_1}
D(r,r') =
\left\{
\begin{array}{cl}
 \mbox{exp} \left[- \frac{(r-r')^2}{l} \right] & \mbox{if} \,\, 0 \le r,r' \le w \\
0 & \mbox{otherwise}
\end{array}
\right. ,
\end{equation}
which has a smoothly tapering spectrum; and a sinc 
\begin{equation}
\label{eqn_sim_2}
D(r,r') =
\left\{
\begin{array}{cl}
 \mbox{sinc}( \frac{\pi( r-r')}{l}) & \mbox{if} \,\, 0 \le r,r' \le w \\
0 & \mbox{otherwise}
\end{array}
\right. ,
\end{equation}
which has a steeply falling spectrum.
In the simulations, the detectors were effectively probed using highly localised sources, such that $r$ and $r'$ were divided into $N$ intervals giving a discretised response function having $(N+1)^2$  elements. Noise was included, and a range of simulations carried out for different coherence lengths, NSR's and sample sizes.  In all cases, except where specified, the width was  $w=10$, the NSR was $nts=$ 0.01, and $N=$ 201 sample positions were used over the spatial domain. The normalisation factor $D_m$ was taken to be the largest eigenvalue of the noiseless response matrix: the spectral norm $\| D \|_2$. 


\subsection{Eigenvalue Spectra}

Figure \ref{fig_spec_1} shows the spectra of a set of simulations using Gaussian coherence functions having coherence lengths of $l=$ 1, 2, 3 and 4. The blue circles show the noiseless eigenvalue spectrum, and the red circles show the spectrum once measurement noise has been added. 
The number of appreciable eigenvalues is given approximately by the effective width-to-coherence-length ratio $w/l$. In fact it can be shown analytically that $w/l$ sits in the range
\begin{equation}
\label{eqn_sim_2}
N_{eff} \le \left( \frac{w}{l} \right) \le N_{max},
\end{equation}
where $N_{eff}$ is the effective number of modes described in Section \ref{sec_stray}, and $N_{max}$ is the maximum mode index needed to account to the majority for the response. This range is an indicator of the abruptness of the spectral cut off. High-throughput devices cut off more quickly, and so $w/l$ becomes a more precise indicator.

The appreciable eigenvalues of the noisy data are similar to those of the noiseless data; although the measured eigenvalues are slightly higher than the actual eigenvalues. The solid red line shows the spectrum of the hermitian part of the noise matrix alone, and although not shown, the spectrum of the noise in the antihermitian part is identical. The upper dashed black line indicates the threshold $\sqrt{2} nts \sqrt{N}$, which is implied by the Marchenko-Pastur distribution of a square matrix, and is indicative of the maximum eigenvalue of the noise matrix. Experimentally, the NSR can be measured, as an experiment proceeds, by turning the sources off as part of the calibration process, or it can found by fitting a line to the noisy part of the measured spectrum. The lower dashed black line shows the intrinsic noise level $nts$, which is the rms error on the recovered eigenvalues. In all cases, the spectra were calculated using Singular Value Decomposition (SVD), rather then an eigenvalue decomposition. In the case  of a positive definite square matrix, these are identical, but although the noise takes the form of a symmetric matrix, it is not positive definite in this simple case. This choice has little effect across the high-order eigenvalues of interest, but prevents the high-order noise modes yielding negative eigenvalues. Indeed, this is equivalent to diagonalising ${\bf \mathsf D} {\bf \mathsf D}^\dagger$  and taking the square root of the eigenvalues. This peculiarity is not present when simulating the measurement of a device fully, such as in the electromagnetic simulations of Section \ref{sec_flm}.

The black circles show the spectrum of the noisy response matrix predicted using perturbation theory based on the spectrum of the noiseless response. The dashed blue line shows the difference between neighboring eigenvalues of the noiseless response: $E_{n+1} - E_{n}$. The predicted spectrum, black circles, are very similar to the noisy spectrum, red circles, up to the point where the difference between neighboring eigenvalues, blue dashed line, becomes comparable with the noise, lower dashed black line. Importantly, and as to be expected, perturbation theory breaks down when the difference between neighboring eigenvalues becomes comparable with the noise, as described by the resolvent. The predicted eigenvalues are a good indicator of the actual noisy eigenvalues up to this limit. Once the absolute value of an eigenvalue becomes comparable with $\sqrt{2} nts \sqrt{N}$, the measured values are no longer a good indicator of the actual values. This is a severe constraint as more and more sample points are included because the limit scales as $\sqrt{N}$. The green circles show the rms error on the noisy eigenvalues predicted using perturbation theory based on the noiseless eigenvalues: legitimately, the noisy eigenvalues give a similar result. It can be seen that the rms error on the noisy measured eigenvalues is of order $nts$, 0.01 in this case, but actually improves higher up in the spectrum.

A key observation is that measurement errors appear, not when the absolute value of an eigenvalue approaches the NSR, thereby placing a large error on the recorded value, but when the difference between neighboring eigenvalues approach the noise level (dashed blue line of Fig. \ref{fig_exp_1}). This effect is apparent in the form of the resolvent, and is driven by the perturbation splitting eigenvalues that are degenerate on the scale of the noise. 
\begin{figure}[H]
     \begin{subfigure}[b]{0.45\textwidth}
         \centering
         \includegraphics[trim = 1cm 1cm 8cm 19cm, clip,width=70mm ]{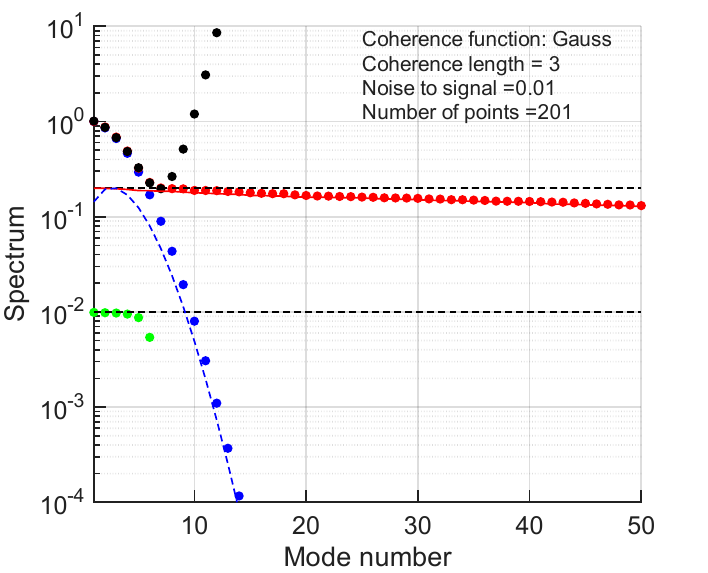}
        \caption{Number of sample points $N$=201}
             \label{fig_exp_sam_1}
     \end{subfigure}
     \hfill
     \begin{subfigure}[b]{0.45\textwidth}
         \centering
         \includegraphics[trim = 1cm 1cm 8cm 19cm, clip,width=70mm ]{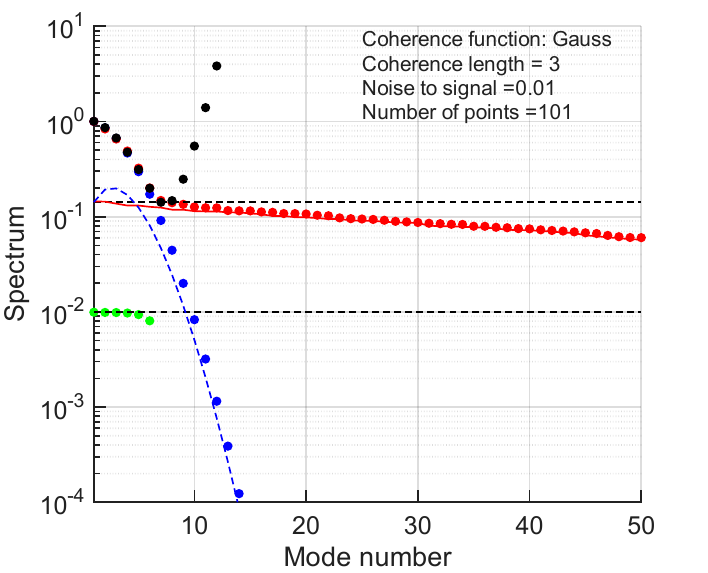}
        \caption{Number of sample points $N$=101}
             \label{fig_exp_sam_2}
     \end{subfigure}
     \hfill
     \centering
     \begin{subfigure}[b]{0.45\textwidth}
         \centering
		\includegraphics[trim = 1cm 1cm 8cm 19cm, clip,width=70mm ]{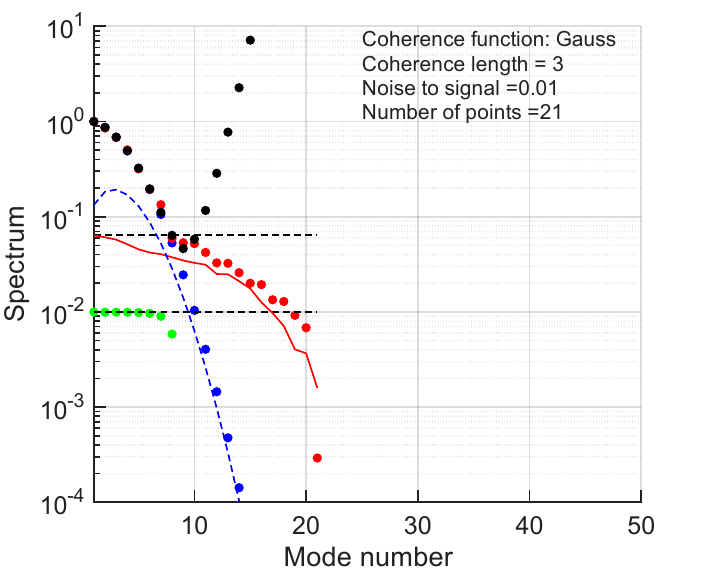}
         \caption{Number of sample points $N$=21}
         \label{fig_exp_sam_3}
     \end{subfigure}
     \hfill
     \begin{subfigure}[b]{0.45\textwidth}
         \centering
         \includegraphics[trim = 1cm 1cm 8cm 19cm, clip,width=70mm ]{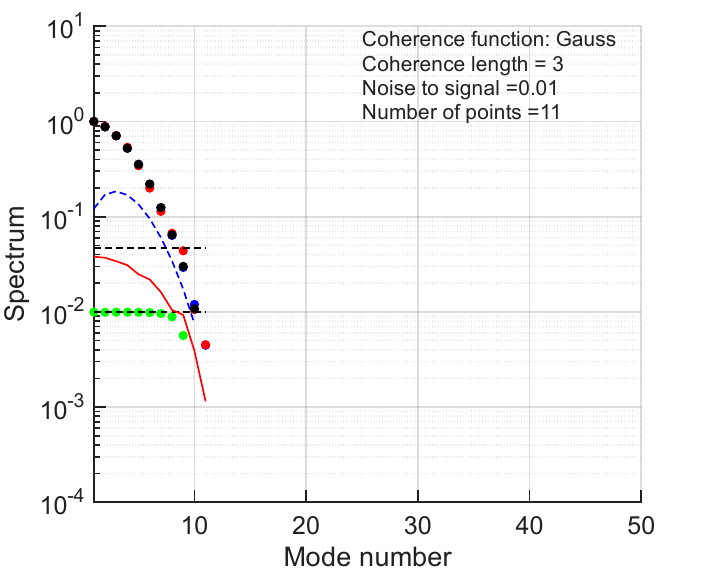}
         \caption{Number of sample points $N$=11}
          \label{fig_exp_sam_4}
     \end{subfigure}
\caption{Effect of reducing the number of sample points. All other parameters are the same as Figure \ref{fig_exp_3}. }
\label{fig_exp_9}
\end{figure}
The near-uniform splitting of the levels, on a log scale, is  evident in the red circles of all of the plots, and will be seen again shortly. Essentially, a noise excursion causes degeneracy to be broken, the eigenvalues to repel and pushed apart from each other as much as possible within the constraints of the problem. This effect is unfortunate because it causes a sea of near uniform eigenvalues to appear, which are at a level much higher than the intrinsic noise level itself: $ \sim \sqrt{2} nts \sqrt{N} $. The random errors on all of the eigenvalues are small, but the background, which scales as $\sqrt{N}$, causes eigenvalues of interest to be lost. 

This behaviour is illustrated further in Fig, \ref{fig_exp_9}, where the number of sample points was reduced from (a) $N=$ 201 to (b) 101, (c) 21, and (d) 11, keeping all other parameters the same as Fig. \ref{fig_exp_3}. Inspecting Fig. \ref{fig_exp_9}, it can be seen than the limit imposed by the noise falls appreciably to the point where 10 eigenvalues are clearly resolved. This drammatic effect illustrates the benefit of using the smallest number of samples possible whilst still recovering the number of appreciable degrees of freedom in the response. Even with this small number of sample points, the eigenvalues are well represented by the measurement. This effect occurs because the device is only receptive to a small number of degrees of freedom, and it is counterproductive to make measurements that only add noise. It is best to use just a few points spread over the field of view. As mentioned in Section  \ref{sec_sam}, analysis could proceed by adding in sample points incrementally during a measurement until all of the eigenvalues of interest have been revealed without increasing the noise floor, which has the benefit of reducing substantially the number of measurements needed, particularly when the number of degrees is not known apriori.

\begin{figure}[H]
     \centering
     \begin{subfigure}[b]{0.45\textwidth}
         \centering
       \includegraphics[trim = 1cm 1cm 8cm 19cm, clip,width=70mm ]{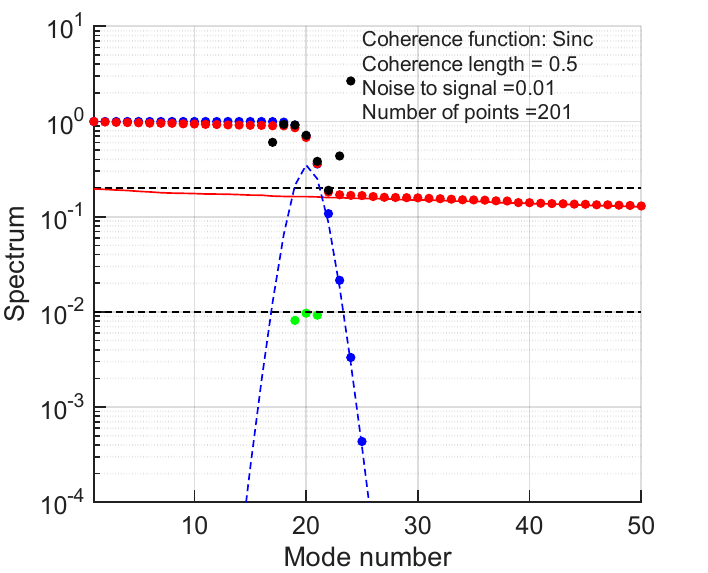}
         \caption{Coherence length of 0.5.}
         \label{fig_sin_1}
     \end{subfigure}
     \hfill
     \begin{subfigure}[b]{0.45\textwidth}
         \centering
         \includegraphics[trim = 1cm 1cm 8cm 19cm, clip,width=70mm ]{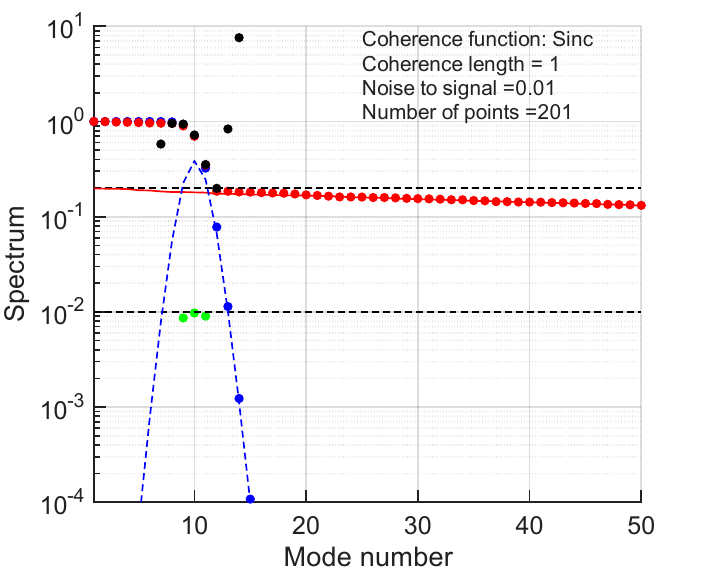}
            \caption{Coherence length of 1.0}
            \label{fig_sin_2}
     \end{subfigure}
          \centering
     \begin{subfigure}[b]{0.45\textwidth}
         \centering
        \includegraphics[trim = 1cm 1cm 8cm 19cm, clip,width=70mm ]{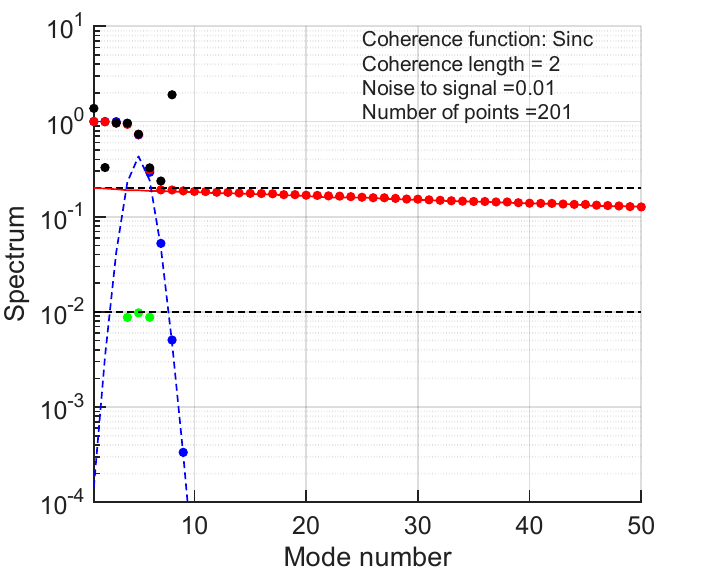}
        \caption{Coherence length of 2.0.}
             \label{fig_sin_3}
     \end{subfigure}
     \hfill
     \begin{subfigure}[b]{0.45\textwidth}
         \centering
        \includegraphics[trim = 1cm 1cm 8cm 19cm, clip,width=70mm ]{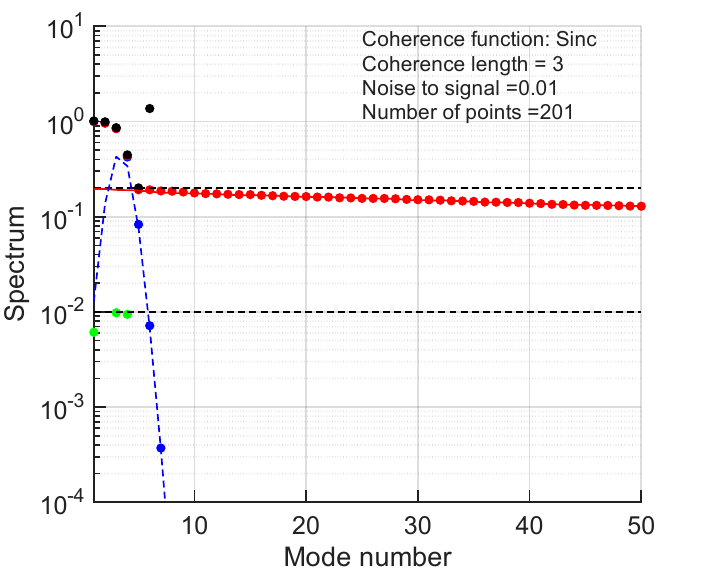}
 			\caption{Coherence length of 3.0.}
              \label{fig_sin_4}
     \end{subfigure}
        \caption{Spectra for a sinc coherence function. The width was 10.0, the NSR was 0.01, and 201 sample points were used in all cases. }
        \label{fig_spec_2}
\end{figure}

Figure \ref{fig_spec_2}  shows a similar set of spectra but now the coherence function was changed to a sinc, (\ref{eqn_sim_2}), and coherence lengths of (a) $l=$ 0.5, (b) 1, (c)  2, (d) 3 used. The main trends are the same as those of the Gaussian, (\ref{eqn_sim_1}),  but the spectrum falls off more abruptly. One additional observation, particularly in Fig. \ref{fig_sin_1}, is that the perturbation method fails on low-order modes, where degeneracy occurs, and the difference between neighboring eigenvalues is comparable with the intrinsic noise limit. These are well-resolved modes, and yet it is seen that the noise breaks the intrisinc degeneracy, causing splitting, and leads to a slope, red circles. Slopes of this kind are evident when high-throughput detectors are measured. The effect of reducing the number of samples, but keeping the NSR constant is the same as the Gaussian case. Thus, the overall trends are independent of the precise details of the coherence function.

\begin{figure}[H]
     \centering
     \begin{subfigure}[b]{0.45\textwidth}
         \centering
        \includegraphics[trim = 1cm 1cm 8cm 19cm, clip,width=70mm ]{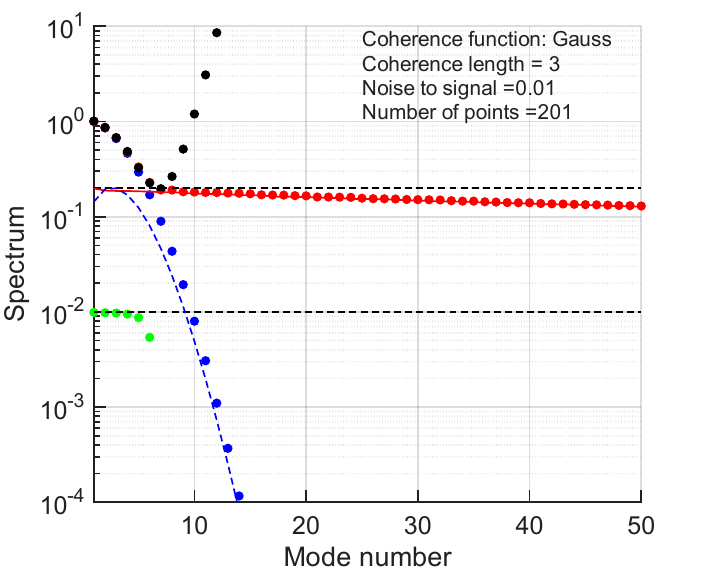}
         \caption{Noise to signal ratio $nts=$ 0.01.}
         \label{fig_spec_3a}
     \end{subfigure}
     \hfill
     \begin{subfigure}[b]{0.45\textwidth}
         \centering
        \includegraphics[trim = 1cm 1cm 8cm 19cm, clip,width=70mm ]{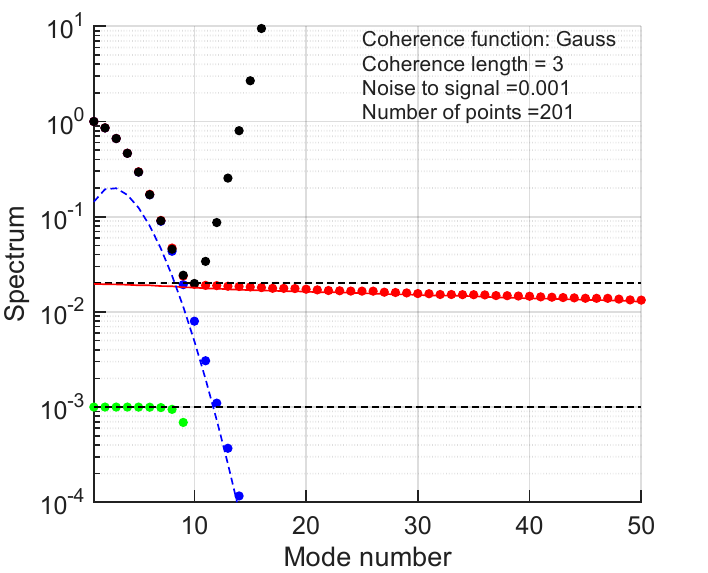}
            \caption{Noise to signal ratio  $nts=$ 0.001.}
            \label{fig_spec3b_2}
     \end{subfigure}
        \caption{Spectra of a Gaussian coherence function. The width was $w=$ 10.0, the NSR was $nts=$ 0.01 and 0.001, and $N=$ 201 sample points were used in all cases. }
        \label{fig_spec_3}
\end{figure}

Figure \ref{fig_spec_3} shows the effect of reducing the intrinsic noise level to $nts=$ 0.001 in the case of a Gaussian coherence fuction. All of the general comments made above remain applicable.
\begin{figure}[H]
     \centering
     \begin{subfigure}[b]{0.45\textwidth}
         \centering
        \includegraphics[trim = 1cm 1cm 8cm 19cm, clip,width=65mm ]{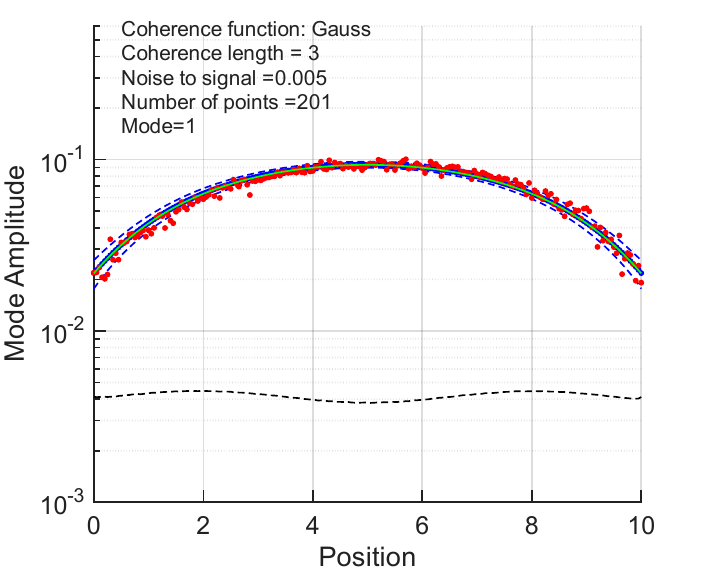}
         \caption{Mode 1.}
         \label{fig_exp_mde_1}
     \end{subfigure}
     \hfill
     \begin{subfigure}[b]{0.45\textwidth}
         \centering
        \includegraphics[trim = 1cm 1cm 8cm 19cm, clip,width=65mm ]{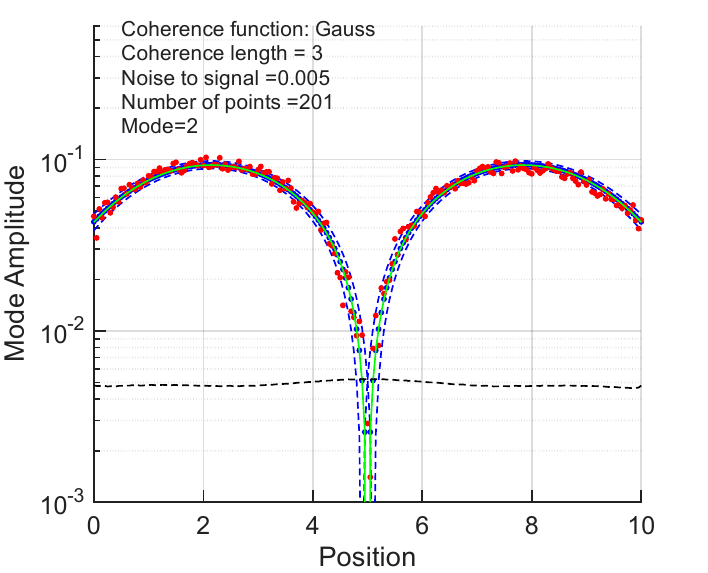}
           \caption{Mode 2.}    
            \label{fig_exp_mde_2}
     \end{subfigure}
          \hfill
     \begin{subfigure}[b]{0.45\textwidth}
         \centering
         \includegraphics[trim = 1cm 1cm 8cm 19cm, clip,width=65mm ]{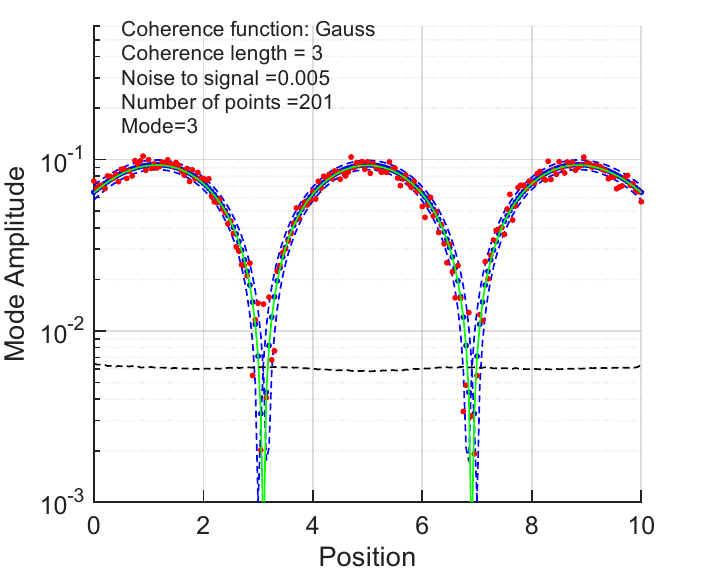}
          \caption{Mode 3.}
            \label{fig_exp_mde_3}
     \end{subfigure}
          \hfill
     \begin{subfigure}[b]{0.45\textwidth}
         \centering
        \includegraphics[trim = 1cm 1cm 8cm 19cm, clip,width=65mm ]{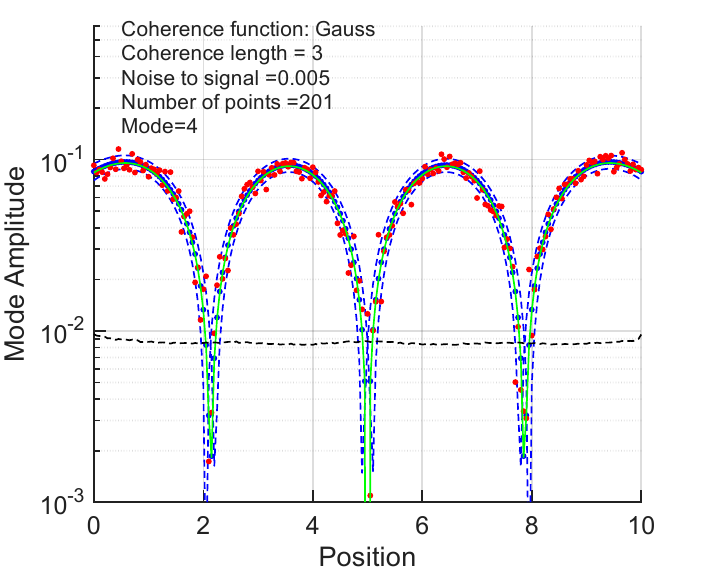}
                   \caption{Mode 4.}
            \label{fig_exp_mde_4}
     \end{subfigure}
        \caption{First 4 modes of a Gaussian coherence function. The width was $w=$ 10.0, the coherence length was $l=$ 3, the NSR was $nts=$ 5$\times 10^{-3}$, and $N=$ 201 sample points were used in all cases.}
        \label{fig_exp_modes_1}
\end{figure}
%


\subsection{Modal Forms}

Consider how noise affects the forms of the recovered modes.  In the following figures, blue circles show the amplitudes of the discretised modes of the noiseless response. Red circles show the amplitudes of the modes when noise was included. The solid green line shows the expected, shifted mode predicted on the basis of perturbation theory using the noisy resolvent. The dashed black line shows the spatial form of the variance predicted on the basis of perturbation theory, again using the noisy resolvent, and the dashed blue lines show the limits when the rms errors are applied to the modes of the noiseless response.
\begin{figure}[H]
     \centering
     \begin{subfigure}[b]{0.45\textwidth}
         \centering
         \includegraphics[trim = 1cm 1cm 8cm 19cm, clip,width=70mm ]{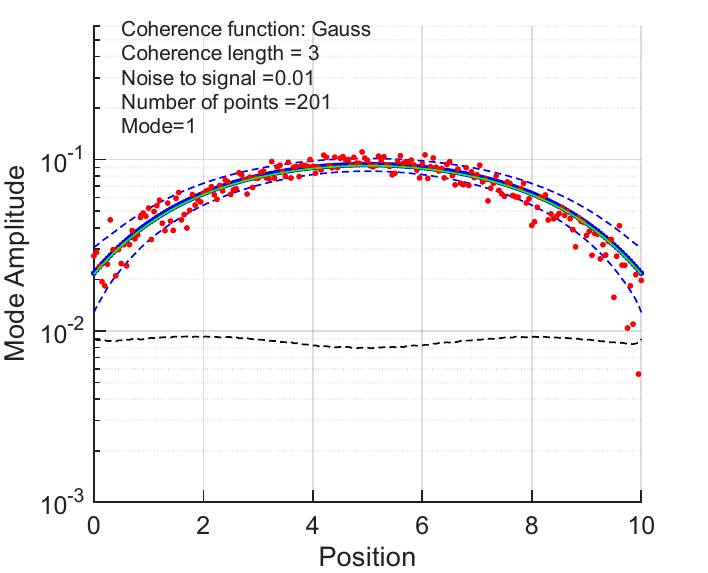}
         \caption{Mode 1.}
         \label{fig_exp_mde_1}
     \end{subfigure}
     \hfill
     \begin{subfigure}[b]{0.45\textwidth}
         \centering
          \includegraphics[trim = 1cm 1cm 8cm 19cm, clip,width=70mm ]{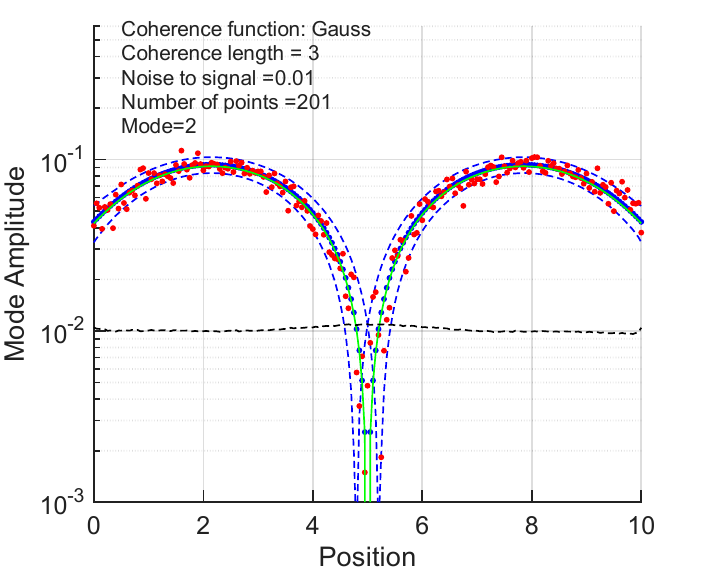}
           \caption{Mode 2.}    
            \label{fig_exp_mde_2}
     \end{subfigure}
          \hfill
     \begin{subfigure}[b]{0.45\textwidth}
         \centering
         \includegraphics[trim = 1cm 1cm 8cm 19cm, clip,width=70mm ]{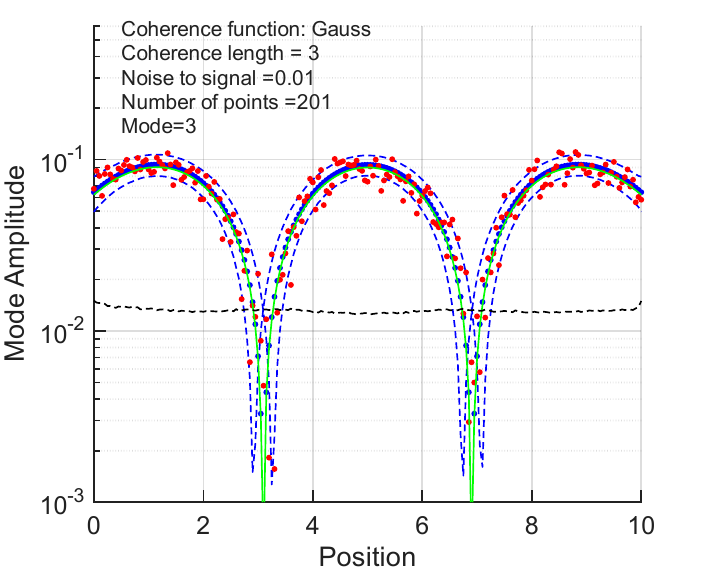}
          \caption{Mode 3.}
            \label{fig_exp_mde_3}
     \end{subfigure}
          \hfill
     \begin{subfigure}[b]{0.45\textwidth}
         \centering
          \includegraphics[trim = 1cm 1cm 8cm 19cm, clip,width=70mm ]{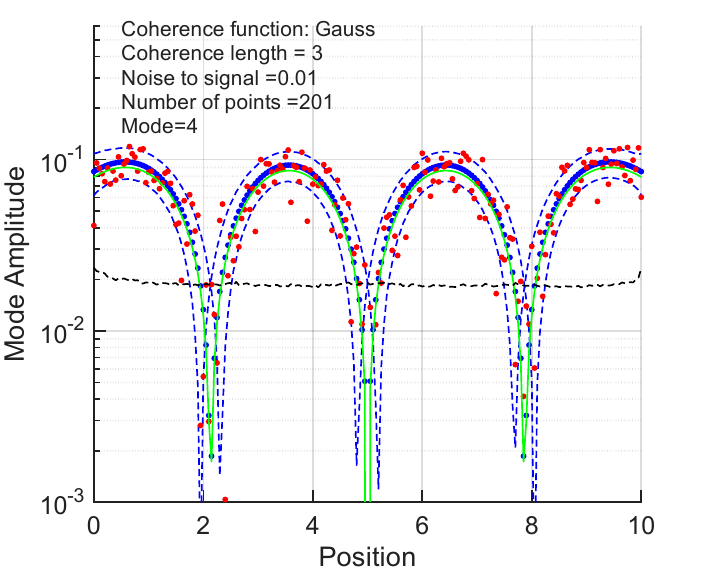}
                   \caption{Mode 4.}
            \label{fig_exp_mde_4}
     \end{subfigure}
               \hfill
     \begin{subfigure}[b]{0.45\textwidth}
         \centering
          \includegraphics[trim = 1cm 1cm 8cm 19cm, clip,width=70mm ]{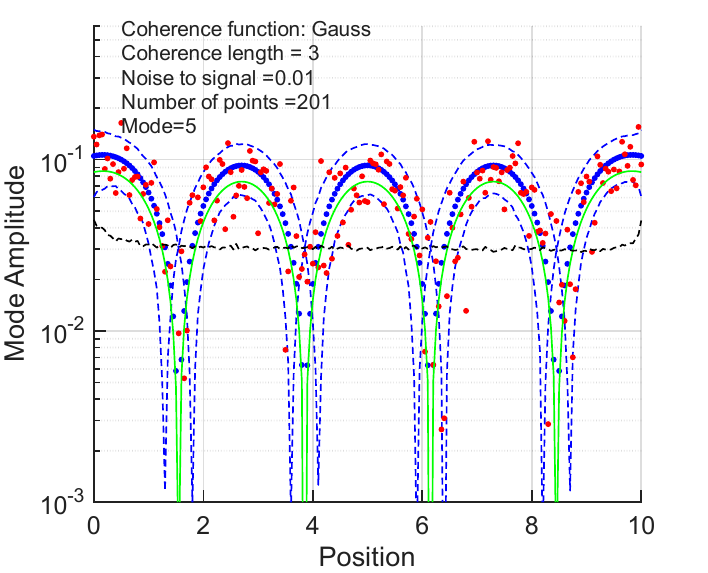}
          \caption{Mode 5.}
            \label{fig_exp_mde_5}
     \end{subfigure}
          \hfill
     \begin{subfigure}[b]{0.45\textwidth}
         \centering
           \includegraphics[trim = 1cm 1cm 8cm 19cm, clip,width=70mm ]{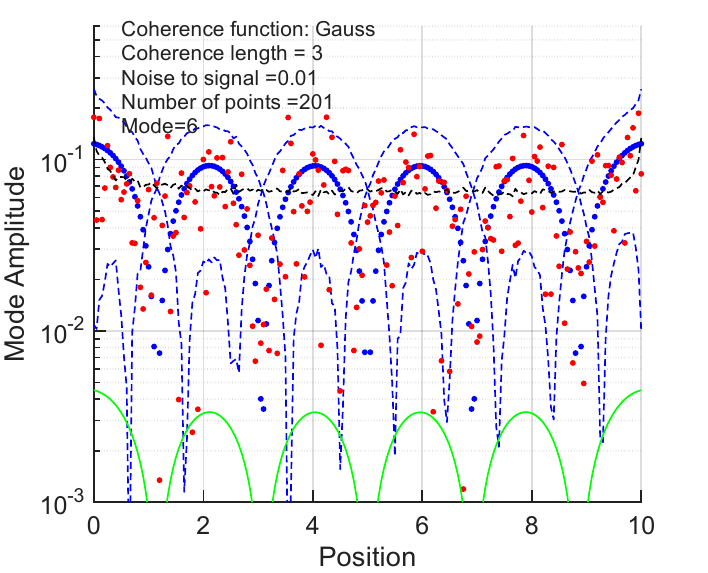}
                   \caption{Mode 6.}
            \label{fig_exp_mde_6}
     \end{subfigure}
        \caption{First 6 modes of a Gaussian coherence function. The width was $w=$ 10.0, the coherence length $l=$ 3, the noise to signal ratio $nts=$ 1$\times 10^{-2}$, and $N=$ 201 sample points were used in all cases.}
        \label{fig_exp_modes_2}
\end{figure}
Figure \ref{fig_exp_modes_1} shows the first 4 modes of a simulation where a Gaussian coherence function was used. These plots are `ideal' in the sense that the NSR was $nts =$ 5 $\times 10^{-3}$. In this low-noise case, a number of general features can be seen: (i) The forms of the modes are recovered well, and the zeros correctly located. (ii) The rms variations at each point of the measured noisy modes (red circles) are well represented by the limits based on perturbation theory (dashed blue line), which shows that the errors can be calculated from a single realisation of the the measured data, and so can be implemented as a part of the data analysis.  (iii) The rms variations (dashed black line) are just a few times the NSR, reasonably constant across the mode profile, and rise as the mode number increases. The errors are slightly larger when the rate of change of the modal form is greatest. This simulation was rerun for many stochastic realisations and the results consistent in all cases, with only the specific values of the noisy modes changing. This simulation indicates that the basic characteristics are as expected, and an EAI experiment with NSR = -30 dB would work well.
\begin{figure}[H]
     \centering
     \begin{subfigure}[b]{0.45\textwidth}
         \centering
         \includegraphics[trim = 1cm 1cm 8cm 19cm, clip,width=70mm ]{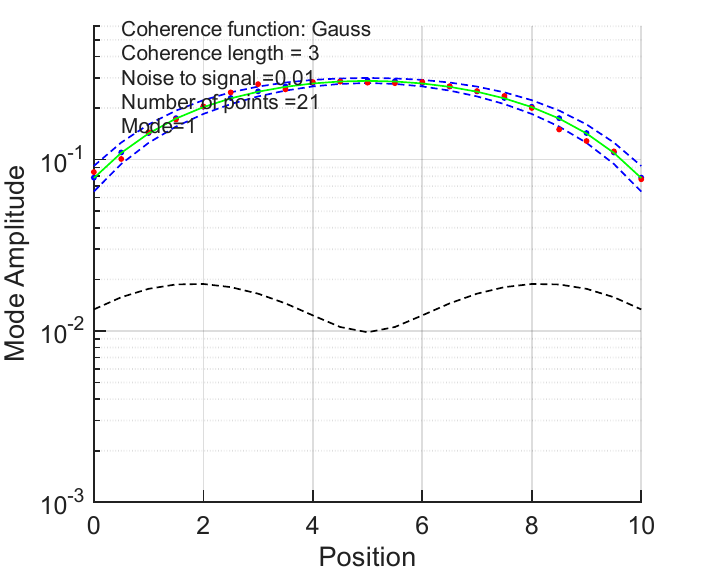}
         \caption{Mode 1.}
         \label{fig_exp_mde_1}
     \end{subfigure}
     \hfill
     \begin{subfigure}[b]{0.45\textwidth}
         \centering
                   \includegraphics[trim = 1cm 1cm 8cm 19cm, clip,width=70mm ]{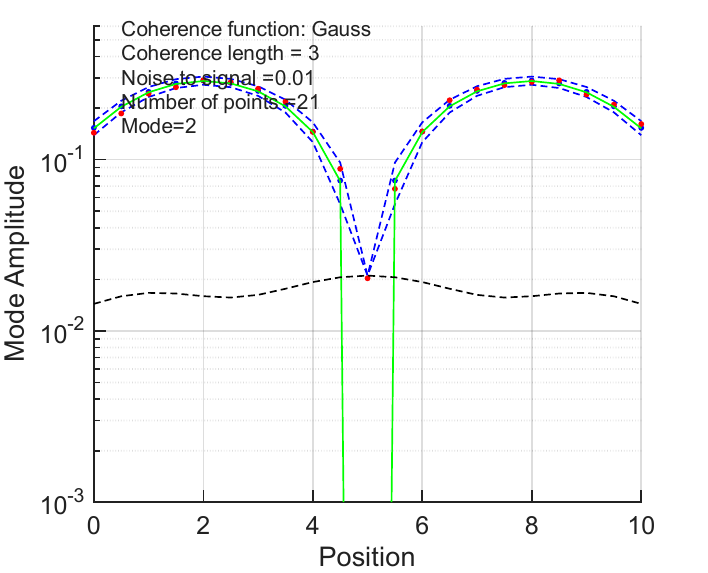}
           \caption{Mode 2.}    
            \label{fig_exp_mde_2}
     \end{subfigure}
          \hfill
     \begin{subfigure}[b]{0.45\textwidth}
         \centering
         \includegraphics[trim = 1cm 1cm 8cm 19cm, clip,width=70mm ]{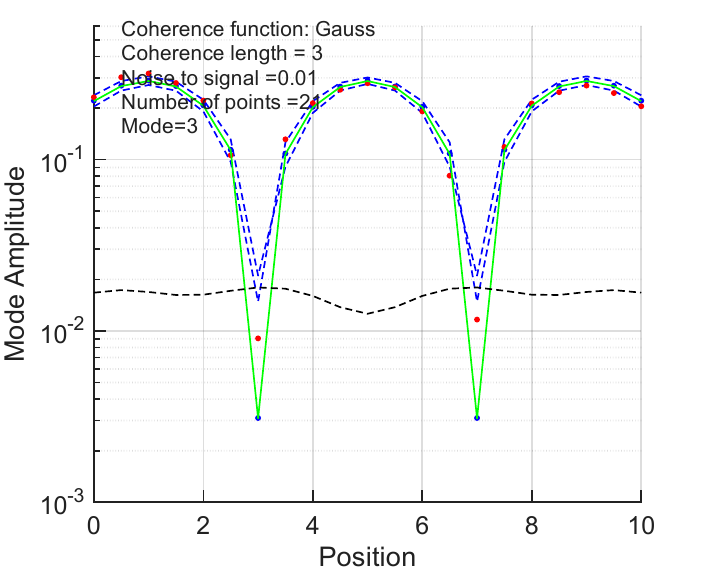}
          \caption{Mode 3.}
            \label{fig_exp_mde_3}
     \end{subfigure}
          \hfill
     \begin{subfigure}[b]{0.45\textwidth}
         \centering
         \includegraphics[trim = 1cm 1cm 8cm 19cm, clip,width=70mm ]{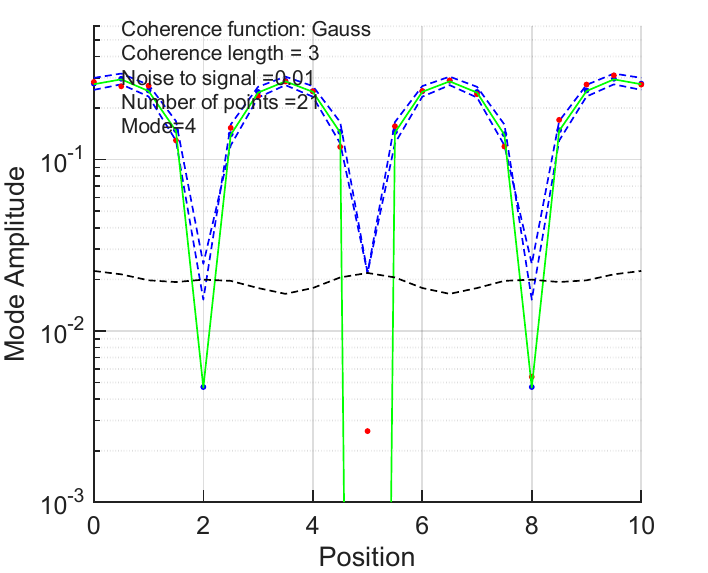}
                   \caption{Mode 4.}
            \label{fig_exp_mde_4}
     \end{subfigure}
               \hfill
     \begin{subfigure}[b]{0.45\textwidth}
         \centering
         \includegraphics[trim = 1cm 1cm 8cm 19cm, clip,width=70mm ]{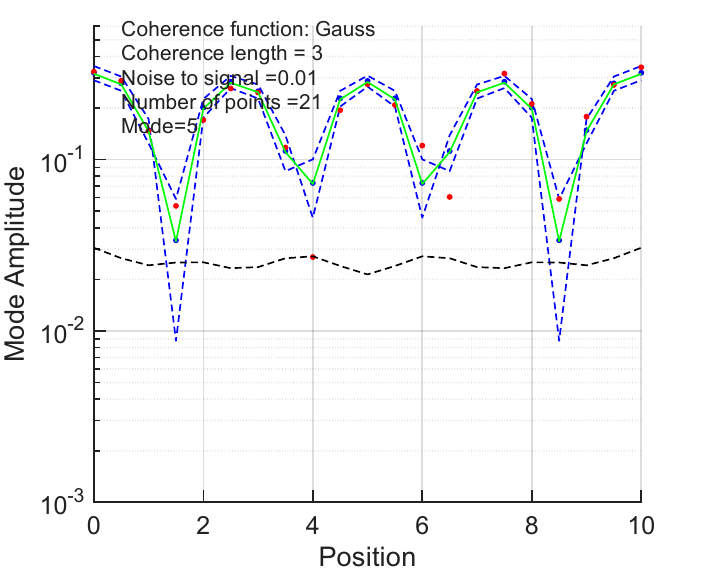}
          \caption{Mode 5.}
            \label{fig_exp_mde_5}
     \end{subfigure}
          \hfill
     \begin{subfigure}[b]{0.45\textwidth}
         \centering
         \includegraphics[trim = 1cm 1cm 8cm 19cm, clip,width=70mm ]{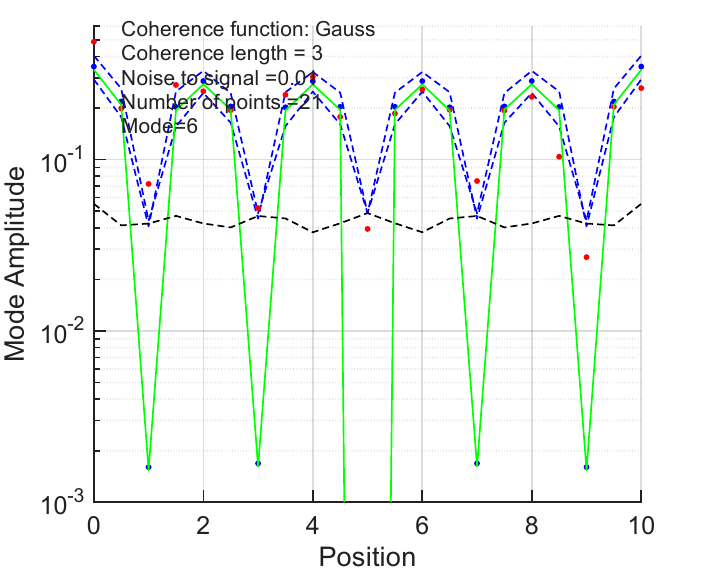}
                   \caption{Mode 6.}
            \label{fig_exp_mde_6}
     \end{subfigure}
        \caption{First 6 modes of a Gaussian coherence function. The width was $w=$ 10.0, the coherence length $l=$ 3, the NSR $nts=$ 1$\times 10^{-2}$, and $N=$ 21 sample points were used in all cases.}
        \label{fig_exp_modes_3}
\end{figure}

Figure \ref{fig_exp_modes_2} shows the first 6 modes of a simulation where the NSR was increased to $nts=$ 1$\times 10^{-2}$, which correspond to the spectrum shown in  Fig. \ref{fig_exp_3}. The 6'th mode is on the threshold where the spectrum dips down into the noise baseline, and the primary form can barely be seen; only the peaks are discrenable. The main features of Fig. \ref{fig_exp_modes_1} are maintained. It can be seen that the noise on the lowest-order mode is that of the the intrinsic $nts$, but increases with mode number until the 6'th mode, where they are determined by the baseline. Thus, not only are the spectra influenced by the baseline, but the modal forms are influenced in a similar way. Notice, on mode 1, that the distribution of the measured mode amplitudes (red circles) varies across the profile in a way that is predicted by perturbation theory. 

Figure \ref{fig_exp_modes_3} shows the first 6 modes of the same simulation as  Fig. \ref{fig_exp_modes_2}, but with a reduced number of sample ponts: from $N=$ 201 to 21, consistent with the number of degrees of freedom in the response. Comparing with Fig. \ref{fig_exp_modes_2}, the plot emphasises that the number of sample points should be kept to a minimum consistent with the degrees of freedom. Although there are fewer sample points across each profile, the basic form is still evident. 

\begin{figure}[H]
     \centering
     \begin{subfigure}[b]{0.45\textwidth}
         \centering
       \includegraphics[trim = 1cm 1cm 8cm 19cm, clip,width=70mm ]{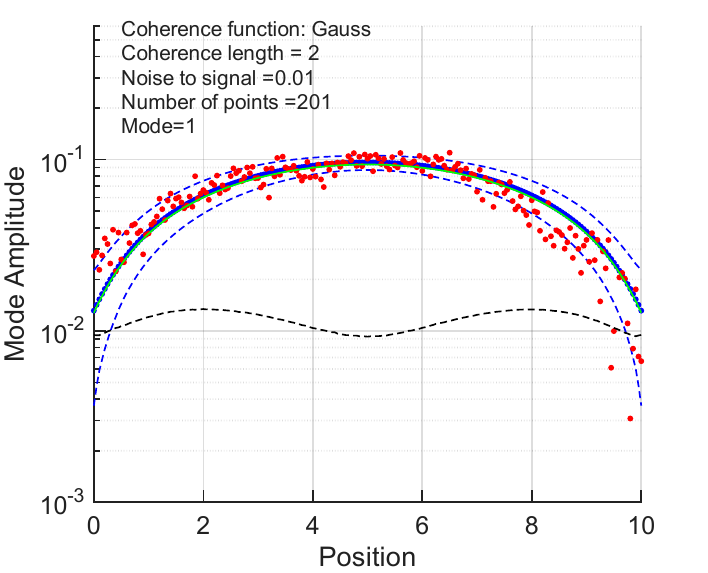}
         \label{fig_exp_mde_1}
     \end{subfigure}
     \hfill
     \begin{subfigure}[b]{0.45\textwidth}
         \centering
        \includegraphics[trim = 1cm 1cm 8cm 19cm, clip,width=70mm ]{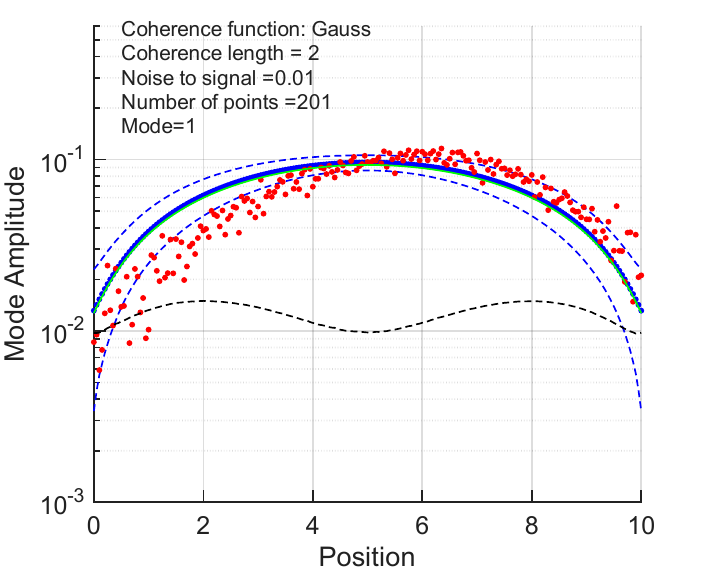}
            \label{fig_exp_mde_2}
     \end{subfigure}
          \hfill
     \begin{subfigure}[b]{0.45\textwidth}
         \centering
       \includegraphics[trim = 1cm 1cm 8cm 19cm, clip,width=70mm ]{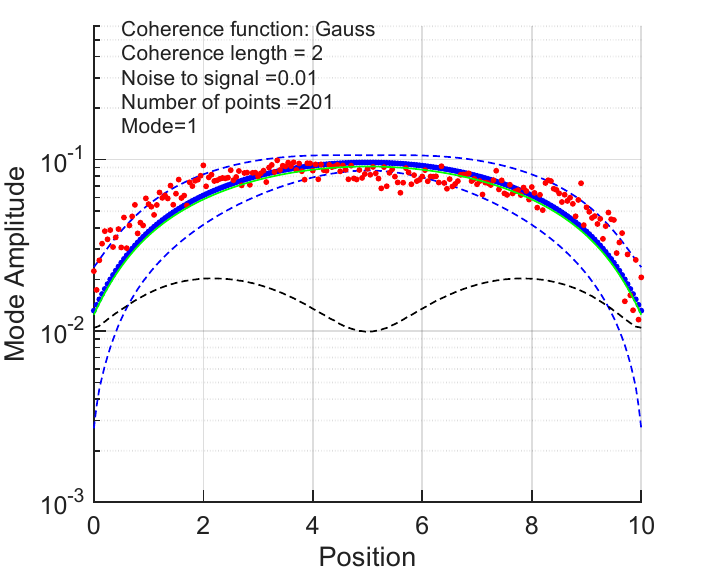}
            \label{fig_exp_mde_3}
     \end{subfigure}
          \hfill
     \begin{subfigure}[b]{0.45\textwidth}
         \centering
        \includegraphics[trim = 1cm 1cm 8cm 19cm, clip,width=70mm ]{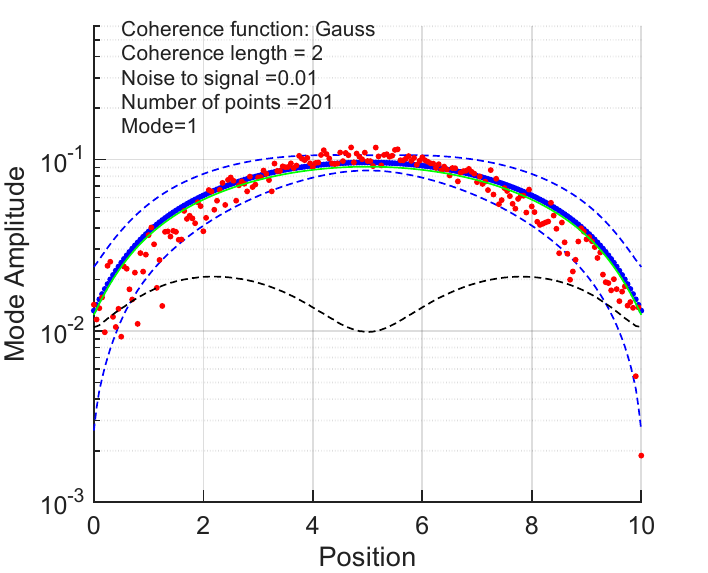}
            \label{fig_exp_mde_4}
     \end{subfigure}
        \caption{Lowest-order modes of a Gaussian coherence function. The width was $w=$ 10.0, the coherence length $l=$ 3, the NSR was  $nts=$ 1$\times 10^{-2}$, and $N=$ 21 sample points were used in all cases. The four plots correspond to rerunning the same simulation multiple times, but using different realisations of the underlying noise model.}
        \label{fig_exp_modes_4}
\end{figure}

\begin{figure}[H]
     \centering
     \begin{subfigure}[b]{0.45\textwidth}
         \centering
       \includegraphics[trim = 1cm 1cm 8cm 19cm, clip,width=65mm ]{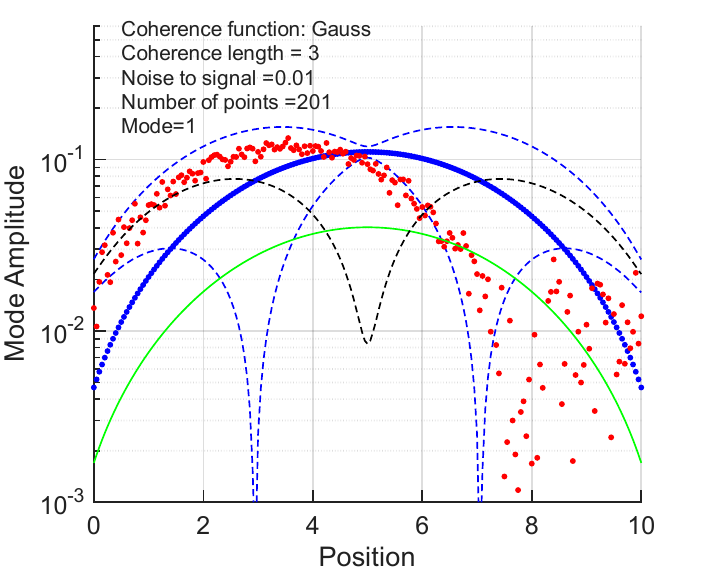}
         \label{fig_sin_mde_1}
     \end{subfigure}
     \hfill
     \begin{subfigure}[b]{0.45\textwidth}
         \centering
      \includegraphics[trim = 1cm 1cm 8cm 19cm, clip,width=65mm ]{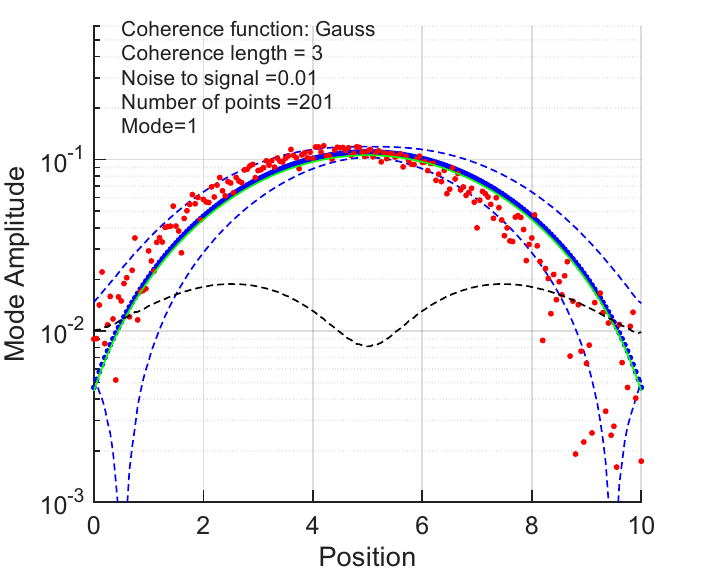}
            \label{fig_sin_mde_1}
     \end{subfigure}
          \hfill
     \begin{subfigure}[b]{0.45\textwidth}
         \centering
      \includegraphics[trim = 1cm 1cm 8cm 19cm, clip,width=65mm ]{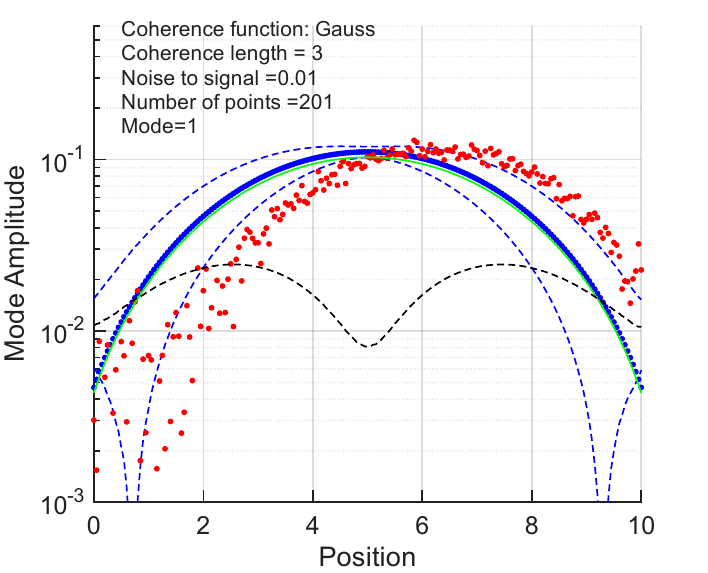}
            \label{fig_sin_mde_3}
     \end{subfigure}
          \hfill
     \begin{subfigure}[b]{0.45\textwidth}
         \centering
       \includegraphics[trim = 1cm 1cm 8cm 19cm, clip,width=65mm ]{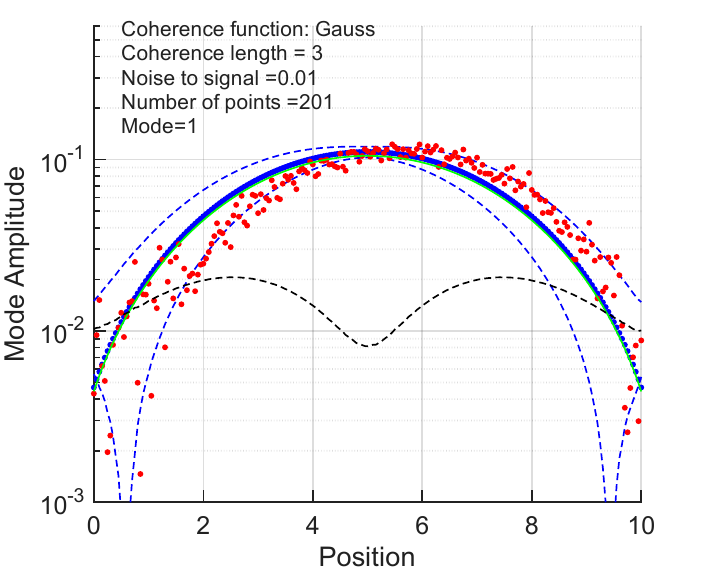}
            \label{fig_sin_mde_4}
     \end{subfigure}
        \caption{Lowest-order mode of a sinc coherence function. The width was $w=$ 10.0, the coherence length $l=$ 3, the NSR $nts=$ 1$\times 10^{-2}$, and $N=$ 201 sample points were used in all cases.}
        \label{fig_exp_modes_6}
\end{figure}

Figure \ref{fig_exp_modes_4} shows the lowest-order mode of a set of repeated simulations using a Gaussian coherence function having a coherence length of $l=$ 2. This set of plots was obtained by rerunning the same simulation using different realisations of the underlying Gaussian noise. With this coherence length, the spectrum has degeneracies on the scale size of the noise on lowest order modes. Rerunning  the same simulation causes the recovered mode to change form, but interestingly the form stays within the bounds calculated using perturbation theory. Higher-order, non-degenerate  modes are not affected in this way. When the coherence length is reduced further, such that the modes become increasingly degenerate, the effect becomes more pronounced. This behaviour is to be expected because the modal forms of the degenate modes are not unique, and be be rotated into other basis sets.  It is unfortunate, however, that readout noise causes apparent distortions, and this should be kept in mind when analysing data.
Figure \ref{fig_exp_modes_6} shows the lowest-order mode of a set of repeated simulations using a sinc coherence function having a coherence length of $l=$ 3. These plots correspond to the spectrum shown in Fig. \ref{fig_sin_4}. Again the plots are consistent with those described previously. As seen in Fig. \ref{fig_sin_4}, as the diference between the two noiseless eigenvalues approaches the noise level, the form of the recovered lowest-order mode becomes unstable, but moves around in a way that is bound by the errors. The forms of Figs \ref{fig_exp_modes_4} and \ref{fig_exp_modes_6} suggest that mode mixing occurs when the difference between the eigenvalues of states is comparable with the noise level.

\begin{figure}[H]
     \centering
     \begin{subfigure}[b]{0.45\textwidth}
         \centering
    \includegraphics[trim = 1cm 1cm 8cm 19cm, clip,width=65mm ]{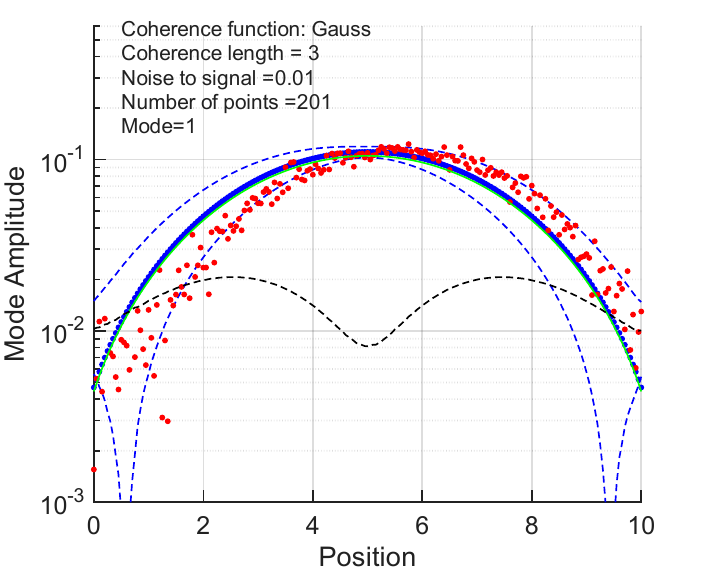}
         \caption{Mode 1}
         \label{fig_sin_mde_1}
     \end{subfigure}
     \hfill
     \begin{subfigure}[b]{0.45\textwidth}
         \centering
     \includegraphics[trim = 1cm 1cm 8cm 19cm, clip,width=65mm ]{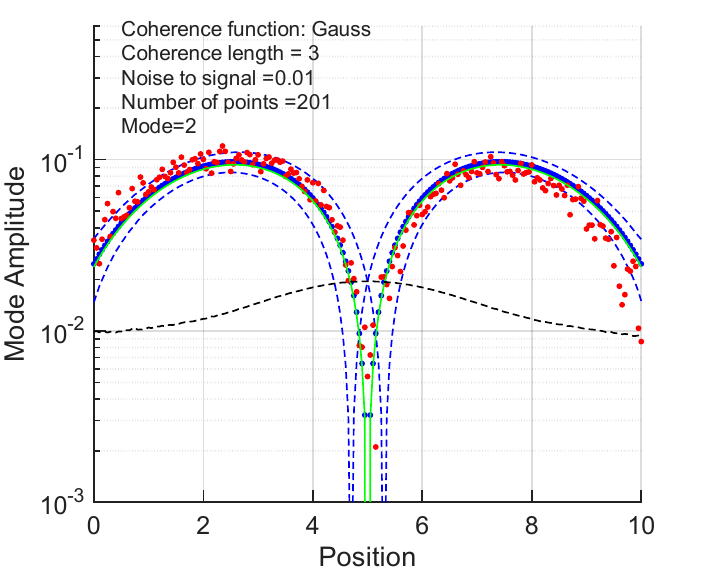}
            \caption{Mode 2}
            \label{fig_sin_mde_2}
     \end{subfigure}
          \hfill
     \begin{subfigure}[b]{0.45\textwidth}
         \centering
     \includegraphics[trim = 1cm 1cm 8cm 19cm, clip,width=65mm ]{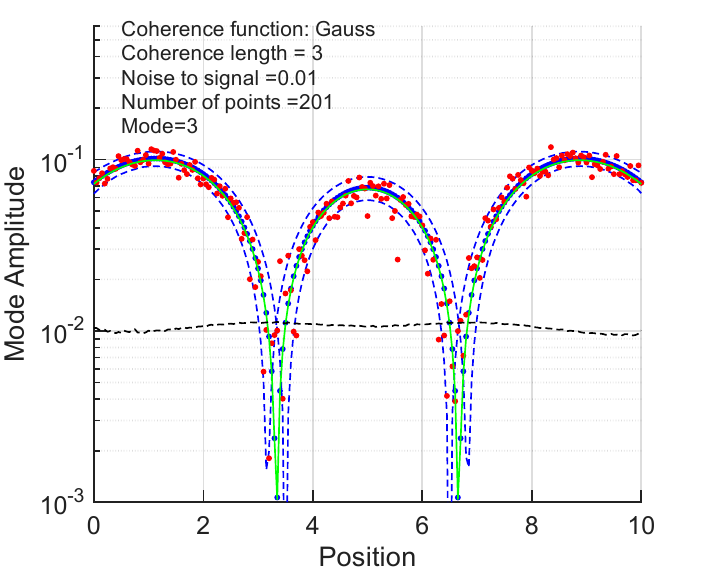}
           \caption{Mode 3}
            \label{fig_sin_mde_3}
     \end{subfigure}
          \hfill
     \begin{subfigure}[b]{0.45\textwidth}
         \centering
    \includegraphics[trim = 1cm 1cm 8cm 19cm, clip,width=65mm ]{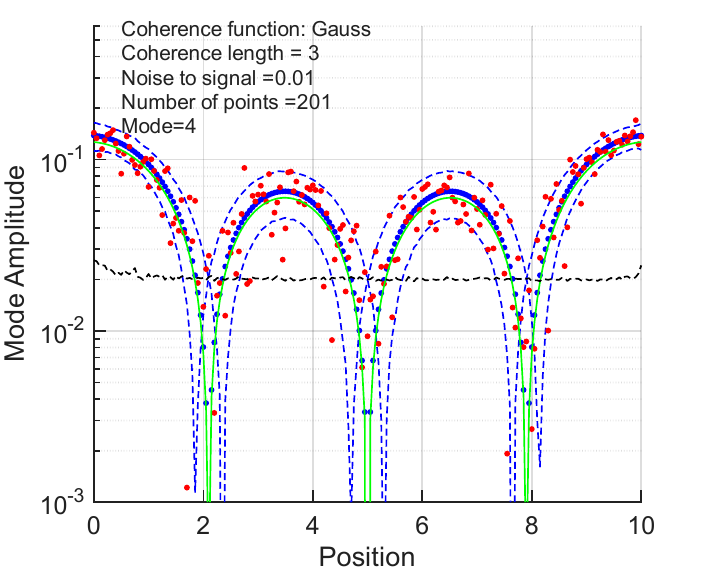}
           \caption{Mode 4}
            \label{fig_sin_mde_4}
     \end{subfigure}
        \caption{First four modes of a sinc coherence function. The width was $w=$ 10.0, the coherence length $l=$ 3, the NSR was $nts=$ 1$\times 10^{-2}$, and $N=$ 201 sample points were used in all cases.}
        \label{fig_exp_modes_5}
\end{figure}

Figure \ref{fig_exp_modes_5} shows the first four modes of a sinc coherence function. When this simulation was rerun multiple times, it was clear that mode mixing is indeed taking place between the two lowest-order modes, with different realisations giving distributions consistent with the predicted errors. Mode mixing is perhaps the main constraint on the NSR needed, although to some extent symmetry could be used to rotate the degenerate set back into the forms of the noiseless response. Nevertheless any visual inspection of a set of noisy data will shown this effect.
Although it is instructive to compare the modes of the noiseless response ${\bf \mathsf u}^0$ with those of the measured response  ${\bf \mathsf u}$ visually,  a more concise way of comparing them is needed. To that end it is instructive calculate the matrix ${\bf \mathsf U}^{0 \dagger} {\bf \mathsf U}_{n}$. This can be viewed in various ways: (a) Each column contains the coefficients of the measured modes in terms of the basis of the actual modes. In this sense the entries are expansion coefficients. (ii)  Each entry is an inner product between the noiseless and noisy modes, and as such can be regarded as the cosine of an angle giving the projection, (\ref{eqn_rer_9}). The diagonal elements give the projection of the noisy modes onto the their own noiseless modes, and  so ideally the diagonal entries should be unity for all of the modes having appreciable eigenvalues. Likewise the off-diagonal entries can be used to indicate overlaps. 

\begin{figure}[H]
     \centering
     \begin{subfigure}[b]{0.45\textwidth}
         \centering
         \includegraphics[trim = 1cm 1cm 8cm 19cm, clip,width=70mm ]{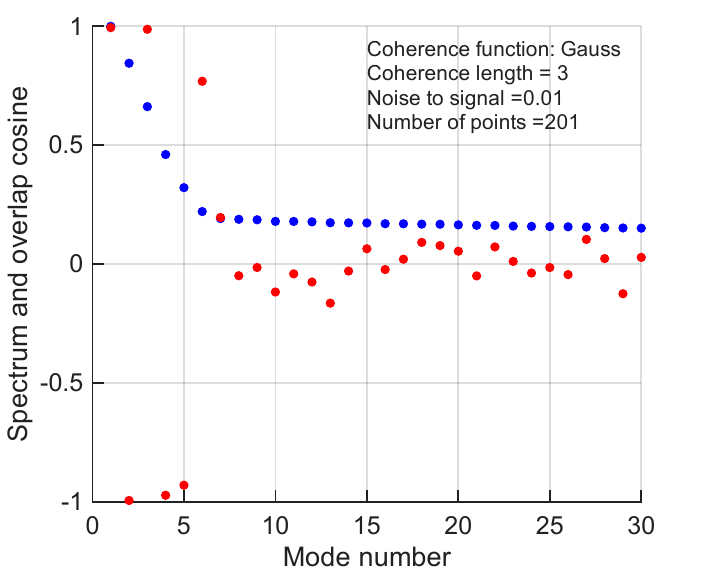}
         \caption{Number of samples 201 }
         \label{fig_mde_error_sub1}
     \end{subfigure}
     \hfill
     \begin{subfigure}[b]{0.45\textwidth}
         \centering
           \includegraphics[trim = 1cm 1cm 8cm 19cm, clip,width=70mm ]{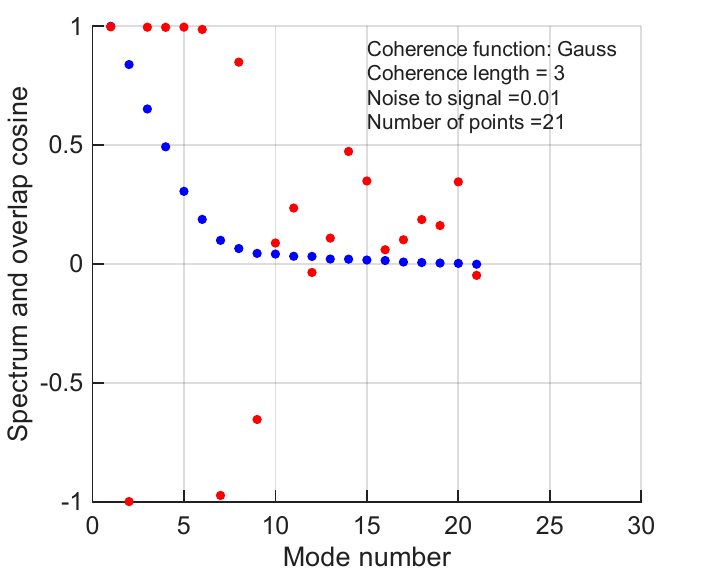}
           \caption{Number of samples 21}    
            \label{fig_mde_error_sub2}
     \end{subfigure}
        \caption{Spectrum (blue circles) and cosine error (red circles) for a Gaussian coherence function. The width was $w=$ 10.0, the coherence length $l=$ 3, the NSR was $nts=$ 1$\times 10^{-2}$.}
        \label{fig_mde_error_1}
\end{figure}

\begin{figure}[H]
     \centering
     \begin{subfigure}[b]{0.45\textwidth}
         \centering
        \includegraphics[trim = 1cm 1cm 8cm 19cm, clip,width=70mm ]{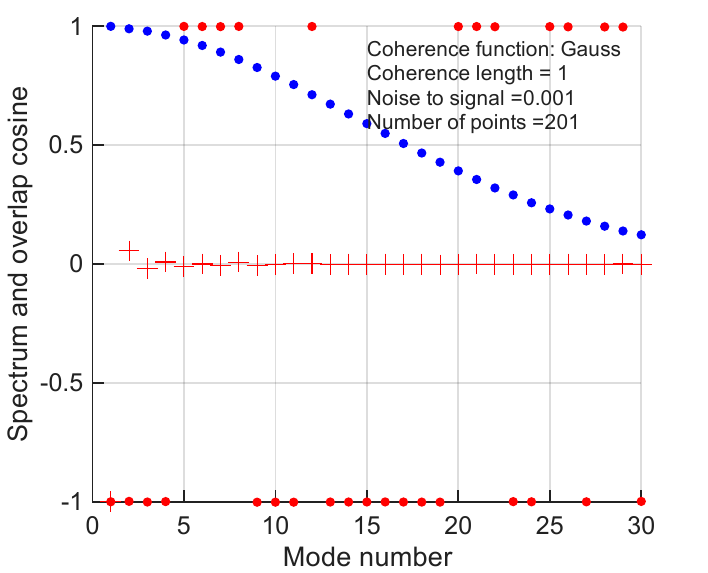}
         \caption{Noise to signal ratio $nts = 1 \times 10^{-3}$. }
         \label{fig_mde_error2_sub1}
     \end{subfigure}
     \hfill
     \begin{subfigure}[b]{0.45\textwidth}
         \centering
         \includegraphics[trim = 1cm 1cm 8cm 19cm, clip,width=70mm ]{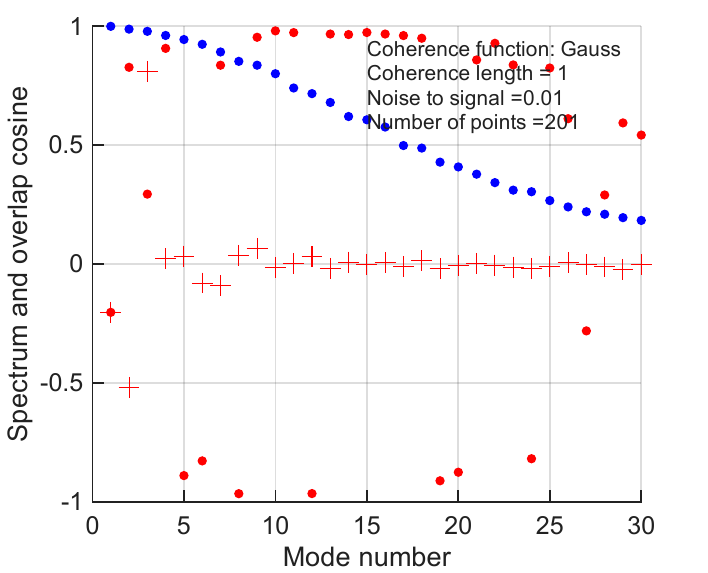}
           \caption{Noise to signal ratio $nts = 1 \times 10^{-2}$.}    
            \label{fig_mde_error2_sub2}
     \end{subfigure}
        \caption{Spectrum (blue circles) and self cosine error (red circles) and cross cosine error relative to the first mode (red crosses) for a Gaussian coherence function. The width was $w=$ 10.0, the coherence length$l=$ 1, and $N= 201$ samples points were used. The two plots show the effect of increasing the NSR.}
        \label{fig_mde_error_2}
\end{figure}

Figure \ref{fig_mde_error_sub1} shows the spectrum (blue circles) and error cosine (red circles) of a Gaussian coherence function having a coherence length of $l=3$ sampled with $N=201$ points. The low-order modes have essentially perfect forms, but the sign is flipped for some indicating a phase error. Because in these simulations the phase is either 1 or -1, this effect is caused by the noise flipping the sign between the two states. The higher order modes show the effect of noise. Figure \ref{fig_mde_error_sub2} shows the effect of reducing the number of sample points to 21, again indicating the benefit of mode fidelity of not oversampling the measurement. 
Figure \ref{fig_mde_error_2} shows the effect of reducing the coherence length to $l=$ 1.  Additionally, the red crosses show the error cosines between the lowest-order mode and all of the others. Figure \ref{fig_mde_error2_sub1} shows the effect of using $nts =$ 0.001: a near-perfect measurement.  Figure \ref{fig_mde_error2_sub2} shows the effect of increasing the NSR to $nts =$ 0.01. The lowest-order modes now show poor reproducibility; in fact the lowest-order noisy mode has very little coupling to the lowest-order noiseless mode; instead it couples appreciably to the second-order mode. This mode mixing varies from one statistical realisation to the next. Because, however, the mixed modes share the same eigenvalues they are not physically significant, and so to some extent can be rotated back into their`expected' forms. 
%
\subsection{Reconstruction}
\label{sec_rec}

The above simulations focus on the forms of the measured, discretised modes, but these are not the same as the continuous reconstructed modes. The simulations relate to measuring a device having a non-local response using a set of `point' sources. In this case, there is a trade off between the need to
keep the number of sample points to a minimum whilst capturing the degrees of freedom present, and the desire to well sample the spatial forms of the modes recovered. The primary problem is that the field patterns of the sources are poorly matched to the spatial forms of the modes; or equivalently, the sizes of the sources are not well matched to the coherence sizes of the device. 

A complementary approach, which uses this same generic model, is to say that a perfectly spatially incoherent device is scanned by sources having extended and shifted, Gaussian or sinc, field distributions $f_i(x)$. In this case, the matrix elements become
\begin{align}
\label{eqn_rec_1}
D_{ij} & = \int \int f_i^\ast (r) \delta(r - r')f_j(r') \, dr dr' \\ \nonumber
 & =  \int f_i^\ast (r) f_j(r) \, dx
\end{align} 
which in matrix form is
\begin{align}
\label{eqn_rec_2}
{\bf \mathsf D} & = {\bf \mathsf H}^\dagger {\bf \mathsf H}.
\end{align} 
The measured response can be regarded as that of ${\bf \mathsf D}$, rather than ${\bf \mathsf H}$ used for the simulations above. The full response can then be reconstructed as a superposition of the source modes. The basis functions, however, are not orthogonal, and so the duals must be used for the reconstruction, which can be included in (\ref {eqn_rec_1}) as a definition of the duals. It would be straightforward to rederive the various errors in this case, but we shall not do so here because in electromagnetic measurements, with sufficiently large step sizes, the source fields are orthogonal.

Looking at the forms of the modes in the above simulations, it seems that Fourier source field distributions might be best. In an electromagnetic measurement this happens when point sources are moved around a suitable set of angles in the far field: Section \ref{sec_flm}. In order to replicate the effects of noisy far-field measurements on the device being modelled in this section,  a set of sine and cosine source functions was used:
\begin{align}
\label{eqn_rec_3}
f_0 & = \frac{1}{\sqrt{w}} \\ \nonumber
f_i^{c} (r) & =
\left\{
\begin{array}{cl}
\sqrt{\frac{2}{w}} \cos (i 2 \pi r/w) & \mbox{for}  -1/2 \le r/w \le + 1/2 \\
0 & \mbox{otherwise}
\end{array}
\right. \\ \nonumber
f_i^{s} (r) & =
\left\{
\begin{array}{cl}
\sqrt{\frac{2}{w}} \sin (i 2 \pi r/w) & \mbox{for}  -1/2 \le r/w \le + 1/2 \\
0 & \mbox{otherwise}
\end{array}
\right. 
.
\end{align}
In comparison with the electromagnetic case, the constant term corresponds to the field of a far-field on-axis point source, and the paired sine and cosine terms correspond to the in and out of phase components described previously, but here we include them as separate basis functions rather than calculating the real and imaginary parts of the fringe. 

In order to carry out numerical simulations, the source fields in (\ref{eqn_rec_3}) were assembled into an $N \times M$ matrix, ${\bf \mathsf S}$, where the $M$ basis functions were sampled at $N$ points. Crucially, in EAI the spatial integrals are intrinsic to the method, but in our simulations the integrals were evaluated numerically using a large number of sample points $N \gg M$, and the calculating the $M \times M$ discretised response matrix:
\begin{align}
\label{eqn_rec_4}
{\bf \mathsf H}^{0'} & = (\Delta w)^2 {\bf \mathsf S}^\dagger {\bf \mathsf H}^0  {\bf \mathsf S},
\end{align} 
where ${\bf \mathsf H}^0$ is the $N \times N$ spatially sampled noiseless response, and $\Delta w = W/(N-1)$ is the sample length.
\begin{figure}[H]
     \centering
     \begin{subfigure}[b]{0.45\textwidth}
         \centering
  \includegraphics[trim = 1cm 1cm 8cm 19cm, clip,width=70mm ]{figure15a.pdf}
         \caption{Spectrum with point source sampling: 201 source positions.}
         \label{fig_spec_fourier_sub1} 
     \end{subfigure}
     \hfill
     \begin{subfigure}[b]{0.45\textwidth}
         \centering
              \includegraphics[trim = 1cm 1cm 8cm 19cm, clip,width=70mm ]{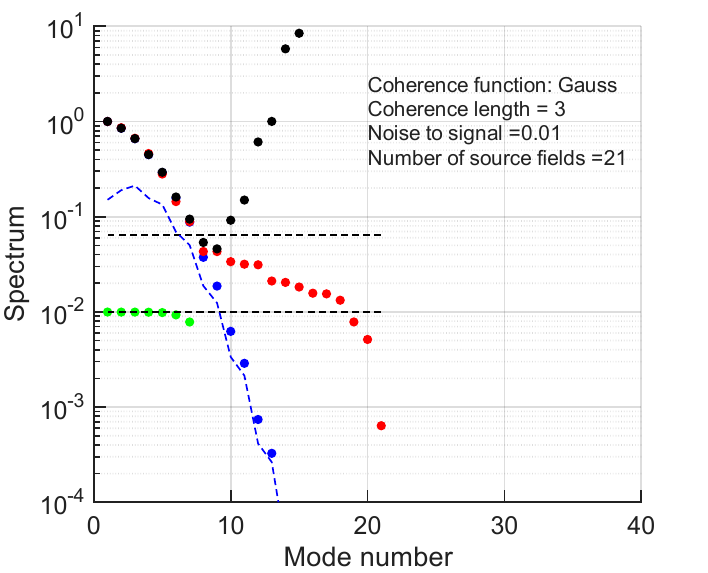}
           \caption{Spectrum with harmonic field sampling: 21 sources.}    
            \label{fig_spec_fourier_sub2}
     \end{subfigure}
        \caption{Spectrum (blue circles) of a Gaussian coherence function. The width was $w=$ 10.0, the coherence length $l=$ 3. The plot serves to indicate the benefits of using source field distributions that are reasonably well scaled to the forms of the individual modes.}
        \label{fig_spec_fourier_1}
\end{figure}
Figure \ref{fig_spec_fourier_1} shows the effect of `measuring' a device having  a Gaussian coherence function, $lc=3$, using the set of harmonic field distributions (\ref{eqn_rec_3}). Whereas $N=501$ points were used to calculate the matrix elements, amounting to near continuous spatial integrals, only 21 fields distributions were used to simulate the EAI experiment (1 uniform, and 10 each of sines and cosines). The original spectrum, using point-source sampling is shown in Figure \ref{fig_spec_fourier_sub1}, whereas the harmonic measurements are shown in Figure \ref{fig_spec_fourier_sub2}. The number of harmonic fields was chosen to capture the main degrees of freedom in the large scale structure of the response. Even though the number of EAI measurements has been reduced by a factor of 10, the primary eigenvalues remain unchanged. However, the noise floor falls significantly in-line with the the reduced amount of noise in the response matrix. The outcome is largely the same as simply reducing the number of points as in  Fig. \ref{fig_exp_sam_3}. The noise floor has fallen by $\sqrt{201/10}$ not $\sqrt{201/21}$, which occurs because previously each entry in ${\bf \mathsf D}$ required an in-phase and out-of phase-measurement, but here the shifting in the fringe from its cosine to sine form is simply regarded as another source field distribution. In other words, measurements with different phase offsets are simply treated as different illuminating fields. The crucial point is that the spectrum has been recovered with a much lower noise floor, revealing higher-order eigenvalues. Reconstruction can now be carried out using the continuous Fourier basis. The spatial reconstruction of the modes is not tied to the number of measurements needed, and because the illuminating fields are better matched to the large-scale structure of the response, a higher-dynamic range is achieved.
\begin{figure}[H]
     \centering
     \begin{subfigure}[b]{0.45\textwidth}
         \centering
     \includegraphics[trim = 1cm 1cm 8cm 19cm, clip,width=65mm ]{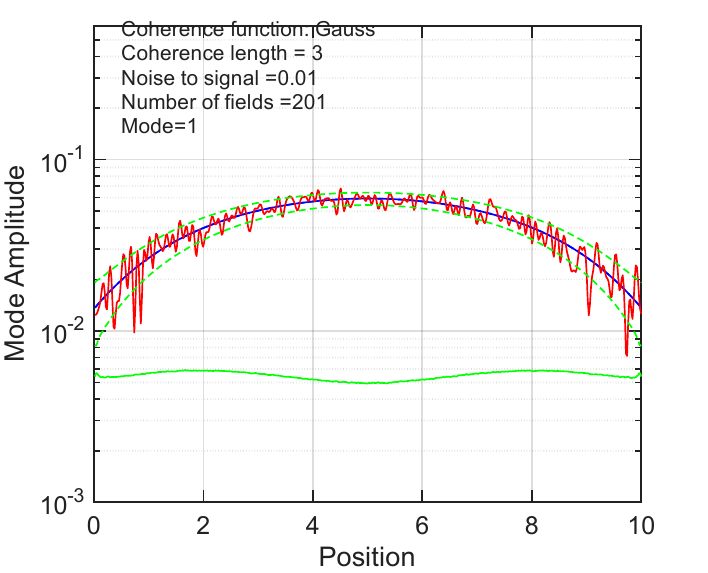}
         \caption{Mode 1}
         \label{fig_exp_mde_amp_sub1}
     \end{subfigure}
     \hfill
      \begin{subfigure}[b]{0.45\textwidth}
         \centering
     \includegraphics[trim = 1cm 1cm 8cm 19cm, clip,width=65mm ]{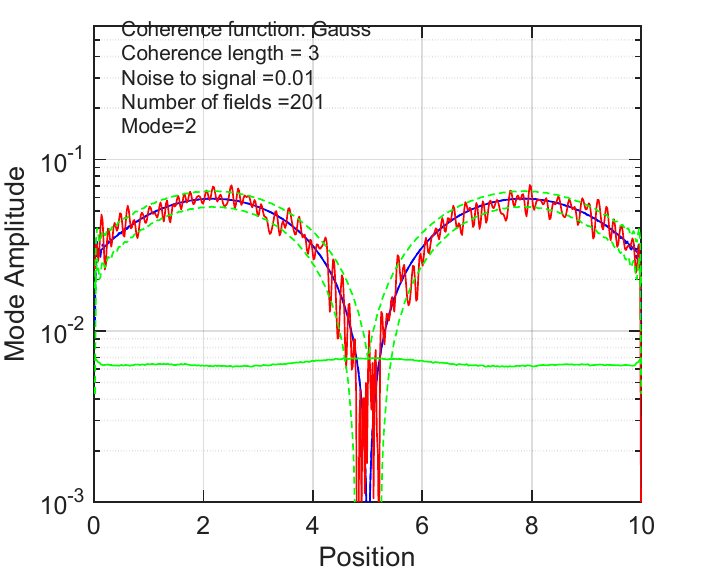}
         \caption{Mode 2}
         \label{fig_exp_mde_amp_sub2}
     \end{subfigure}
     \hfill
      \begin{subfigure}[b]{0.45\textwidth}
         \centering
     \includegraphics[trim = 1cm 1cm 8cm 19cm, clip,width=65mm ]{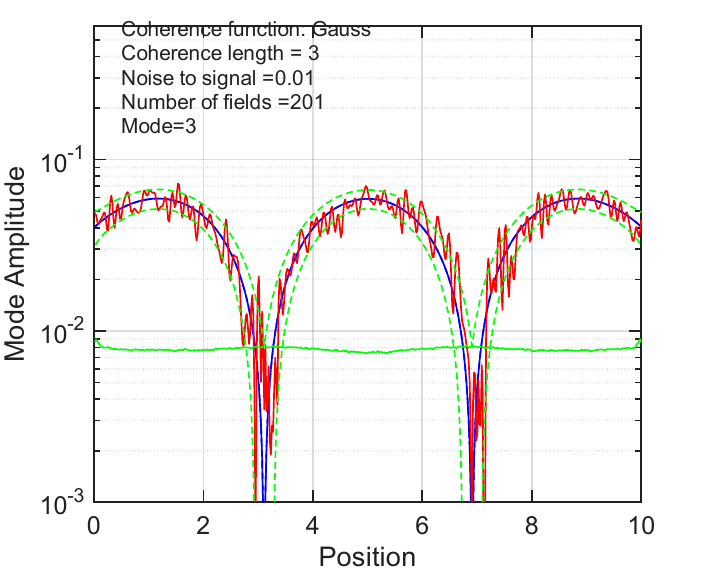}
         \caption{Mode 3}
         \label{fig_exp_mde_amp_sub3}
     \end{subfigure}
     \hfill
      \begin{subfigure}[b]{0.45\textwidth}
         \centering
      \includegraphics[trim = 1cm 1cm 8cm 19cm, clip,width=65mm ]{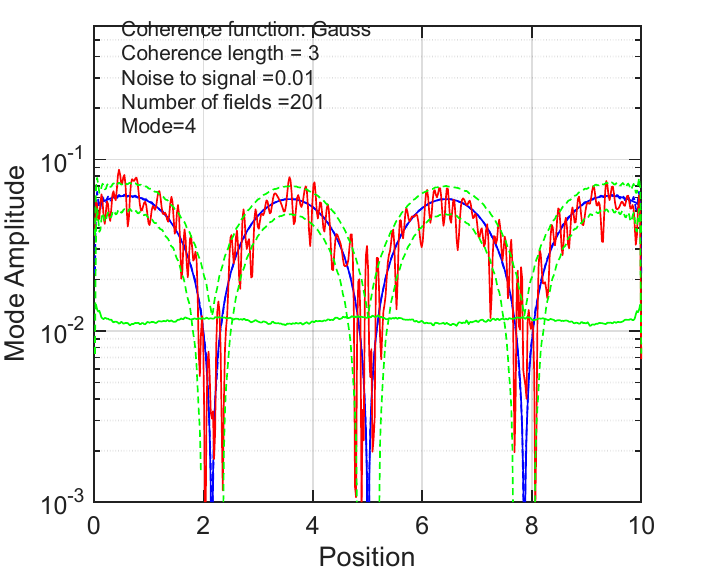}
         \caption{Mode 4}
         \label{fig_exp_mde_amp_sub4}
     \end{subfigure}
     \hfill
      \begin{subfigure}[b]{0.45\textwidth}
         \centering
     \includegraphics[trim = 1cm 1cm 8cm 19cm, clip,width=65mm ]{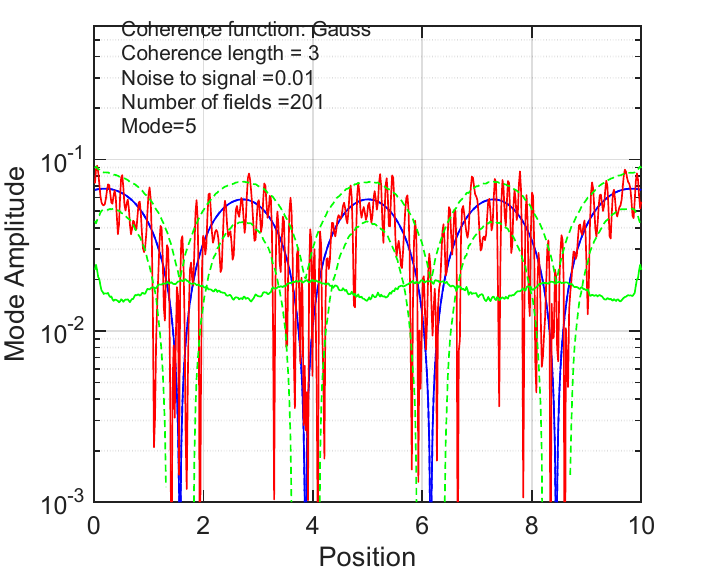}
         \caption{Mode 5}
         \label{fig_exp_mde_amp_sub5}
     \end{subfigure}
     \hfill
      \begin{subfigure}[b]{0.45\textwidth}
         \centering
      \includegraphics[trim = 1cm 1cm 8cm 19cm, clip,width=65mm ]{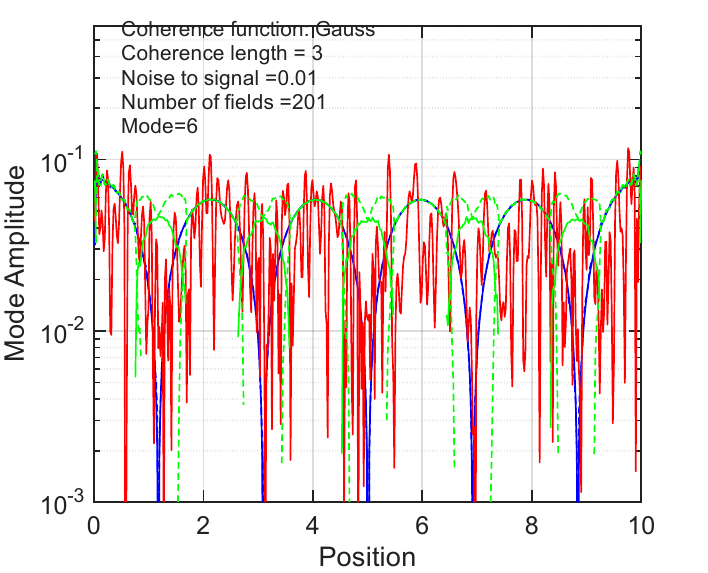}
         \caption{Mode 6}
         \label{fig_exp_mde_amp_sub6}
     \end{subfigure}
     \hfill     
        \caption{First 6 modes of a Gaussian coherence function. The width was $w=$ 10.0, the coherence length $l=$ 3, the NSR $nts=$ 1$\times 10^{-2}$, and $N=$ 201 source fields were used in all cases.}
        \label{fig_exp_modes_amp1}
\end{figure}
Figure \ref{fig_exp_modes_amp1} shows the first 6 reconstructed modes when 201 (1 uniform and 100 each of cosine and sine) source fields were used. The modes of the actual noiseless response are shown as solid blue. The noiseless modes reconstructed from the 201 source measurements are shown as dashed blue, and with this number of basis functions, the two are indistinguishable. The solid red line shows the reconstruction after noise has been added. A large number of source fields has been used to illustrate that the behaviour is essentially the same as when point sources are used. The solid line shows the error on each spatial point of the reconstruction, and the dashed green lines show the error applied to the noiseless recovered response.  The consistency with the point source case is clear. 

\begin{figure}[H]
     \centering
     \begin{subfigure}[b]{0.45\textwidth}
         \centering
        \includegraphics[trim = 1cm 1cm 8cm 19cm, clip,width=65mm ]{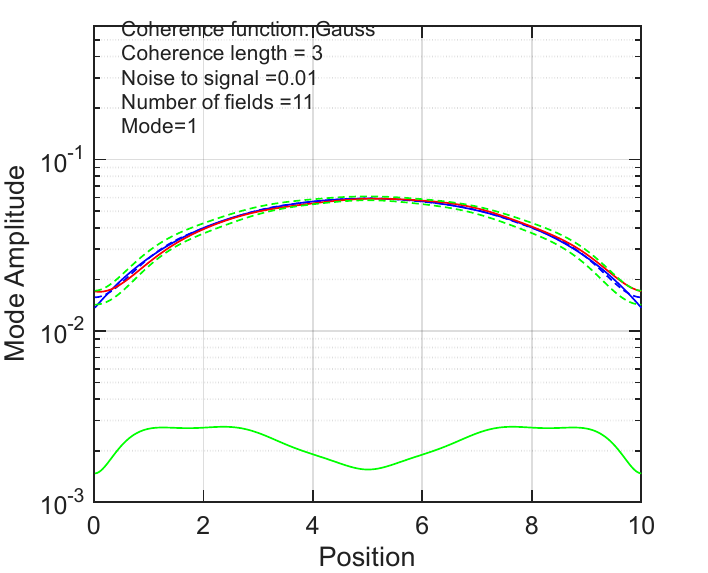}
         \caption{Mode 1}
         \label{fig_exp_mde_amp3_sub1}
     \end{subfigure}
     \hfill
      \begin{subfigure}[b]{0.45\textwidth}
         \centering
         \includegraphics[trim = 1cm 1cm 8cm 19cm, clip,width=65mm ]{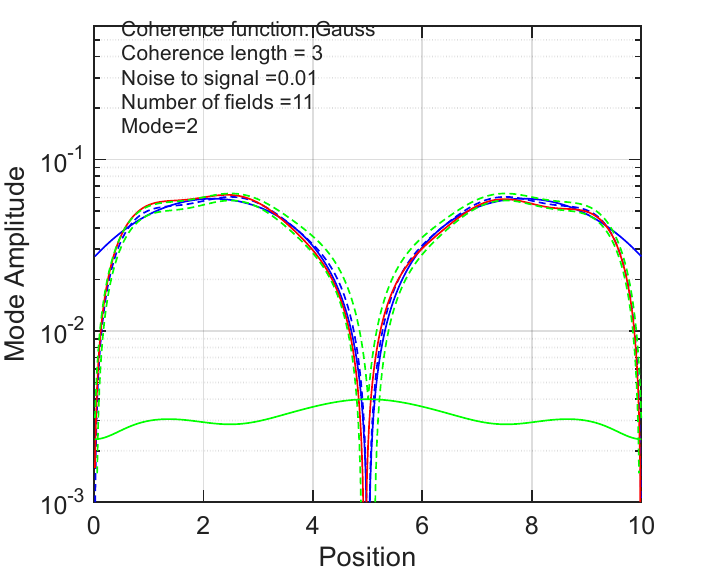}
         \caption{Mode 2}
         \label{fig_exp_mde_amp3_sub2}
     \end{subfigure}
     \hfill
      \begin{subfigure}[b]{0.45\textwidth}
         \centering
         \includegraphics[trim = 1cm 1cm 8cm 19cm, clip,width=65mm ]{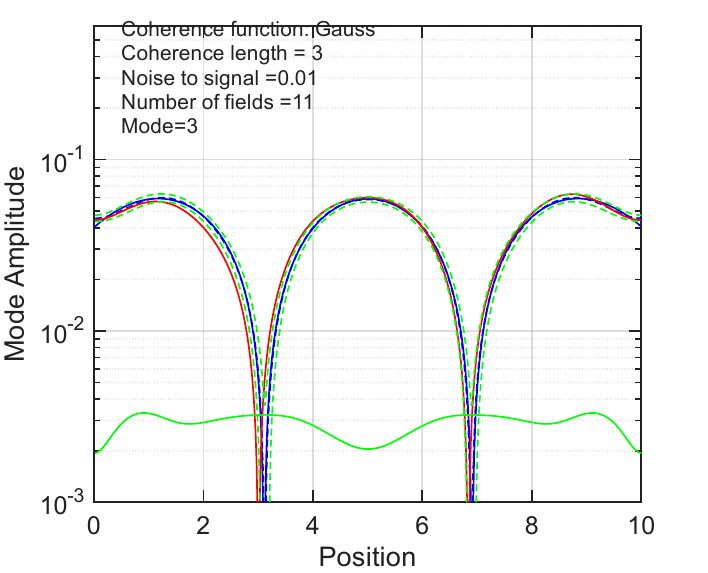}
         \caption{Mode 3}
         \label{fig_exp_mde_amp3_sub3}
     \end{subfigure}
     \hfill
      \begin{subfigure}[b]{0.45\textwidth}
         \centering
         \includegraphics[trim = 1cm 1cm 8cm 19cm, clip,width=65mm ]{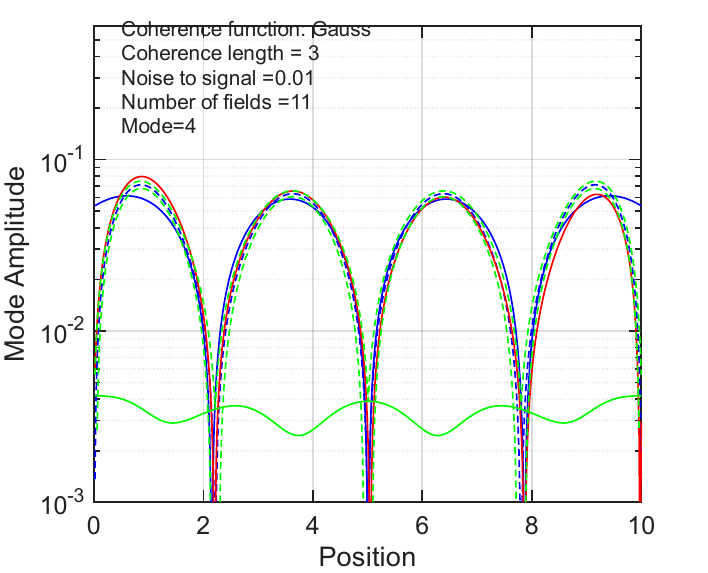}
         \caption{Mode 4}
         \label{fig_exp_mde_amp3_sub4}
     \end{subfigure}
     \hfill
      \begin{subfigure}[b]{0.45\textwidth}
         \centering
         \includegraphics[trim = 1cm 1cm 8cm 19cm, clip,width=65mm ]{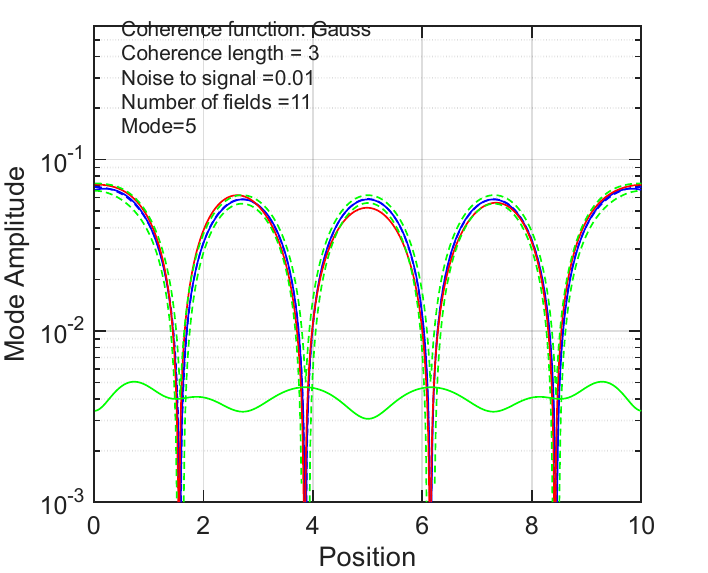}
         \caption{Mode 5}
         \label{fig_exp_mde_amp3_sub5}
     \end{subfigure}
     \hfill
      \begin{subfigure}[b]{0.45\textwidth}
         \centering
        \includegraphics[trim = 1cm 1cm 8cm 19cm, clip,width=65mm ]{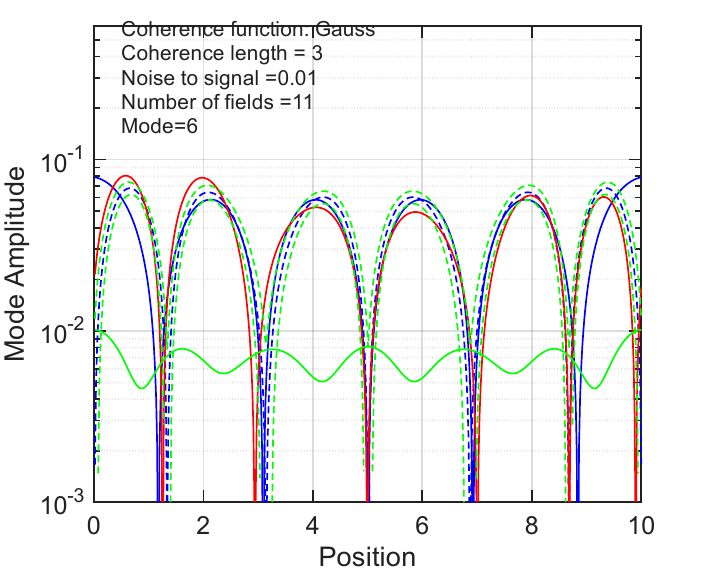}
         \caption{Mode 6}
         \label{fig_exp_mde_amp3_sub6}
     \end{subfigure}
     \hfill     
        \caption{First 6 modes of a Gaussian coherence function. The width was $w=$ 10.0, the coherence length $l=$ 3, the NSR $nts=$ 1$\times 10^{-2}$, and only $M=$ 11 source fields were used.}
        \label{fig_exp_modes_amp3}
\end{figure}

\begin{figure}[H]
     \centering
     \begin{subfigure}[b]{0.45\textwidth}
         \centering
         \includegraphics[trim = 1cm 1cm 8cm 19cm, clip,width=65mm ]{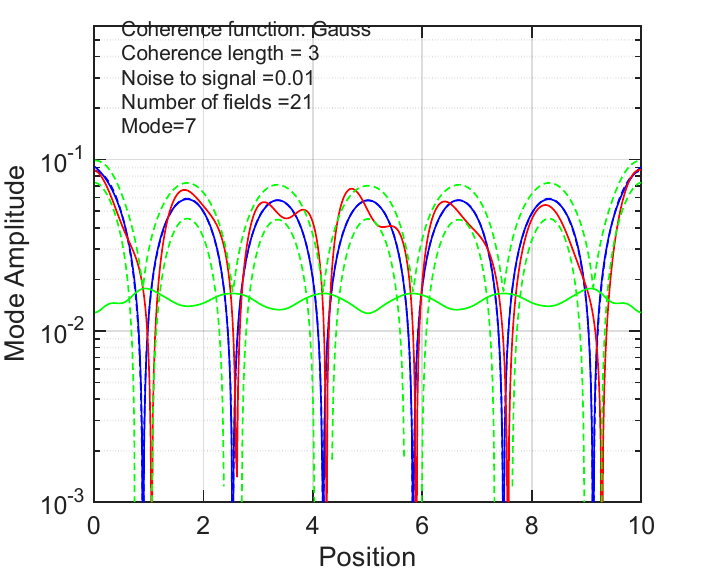}
         \caption{Noise to signal ratio 0.01}
         \label{fig_exp_mde_amp4_sub1}
     \end{subfigure}
     \hfill
      \begin{subfigure}[b]{0.45\textwidth}
         \centering
         \includegraphics[trim = 1cm 1cm 8cm 19cm, clip,width=65mm ]{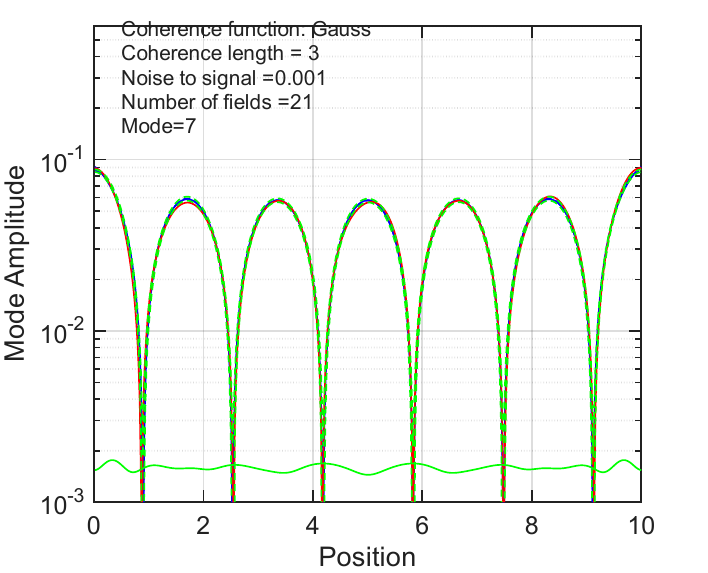}
          \caption{Noise to signal ratio 0.001}
         \label{fig_exp_mde_amp4_sub2}
     \end{subfigure}
     \hfill     
        \caption{Mode 7 of a Gaussian coherence function with two different noise levels.}
        \label{fig_exp_modes_amp4}
\end{figure}

Figure \ref{fig_exp_modes_amp3} is identical to Fig. \ref{fig_exp_modes_amp1}, but with the number of source fields reduced to 11 (one uniform and 5 each of sine and cosine). This number of source fields is sufficient to capture all of the significant degrees of freedom in the response, and yet provides a smoothed reconstruction of the response. It was found that the noisy modes change slightly with noise realisation, but stay bound to the predicted errors. At the edges, a Gibb's-like phenomena can be seen due the the detector having finite size, and yet using only a small number of source fields.  The higher-order modes are well represented, but by mode 6, where the noise is comparable with the spacing between the eigenvalues, some mode switching was seen. This switching occurs within the limits of the error bars, as discussed previously. Crucially, even the 6'th mode is now cleanly recovered, in comparison with the point-source case. Figure \ref{fig_exp_modes_amp4} shows mode 7 of the same simulation, but with two different noise levels $nts =$ 0.001 and 0.0001. The recovered modes and the actual modes show excellent agreement, verifying the integrity of the simulations as the noise is reduced.


\section{Conclusions}

Energy Absorption Interferometry (EAI) is a technique for measuring the complex-valued spatial and polarimetric forms of the individual degrees of freedom through which a structure can absorb energy. It also gives the responsivities of the modes, enabling the overall throughput of the structure to be determined. The individual complex-valued reception patterns measured are intimately related to the internal dissipative dynamical modes of the device being studied: such as electrical currents, spin waves, acoustic oscillations, etc. Each measurement is carried out at a single frequency using two coherent sources, allowing the spatial forms of modes to be tracked across spectral features, such as resonances, rather than inferring spatial forms from spectral features. EAI can be used theoretically, numerically and experimentally. It can be applied at any wavelength, and using different kinds of excitation, such as electromagnetic, elastic and acoustic. It can also be applied using different kinds of source simultaneously to yield the spatial forms of dynamical processes that couple to different kinds of field simultaneously. EAI can also be applied to measuring the coherence properties of thermally radiating and luminescent devices, because high-power sources can be used, avoiding the need to measure the correleations between the often weak emitted fields. In the specific context of far-infrared and optical detectors, EAI can be used to characterise the multimode behaviour of ultra-low-noise detectors, imaging arrays, and complete instruments, where it can be used to ensure that a system is maximally responsive to those partially coherent fields that carry signal whilst avoiding those that only carry noise. 

After reviewing the basic principles of the method, procedures have been described for calibrating and analysing data, and a list of potential errors and mitigations discussed. Noise analysis has been carried out for the first time, allowing the effects of additive noise to be understood, and giving an insight into the signal-to-noise level needed to guarantee a given level of precision in both the recovered spectrum and the spatial forms of the modes. 

It has been shown that oversampling is not only time consuming, but also leads to a significant degradation of the overall noise performance, because samples are included that carry no information about the SUT itself. The fidelity of the final data should be enhanced by increasing the integration time on each EAI measurement, rather than including sample points  that contribute no additional spatial information beyond that allowed by the response of the device. Strictly speaking, the number of sample points needed is determined by the number of degrees of freedom in the response, which is small for a few-mode device. However, this minimum number may have to be traded off against the need to smoothly reconstruct the forms of the individual modes. Ideally, the field patterns of the sources should be chosen to span and efficiently represent the forms of the modes over the reference surface chosen. For example, a far-field point-like source may efficiently represent the behaviour of a device over its surface, but only inefficiently represent the behaviour over a far field surface. If a source is needed that can efficiently represent behaviour in both the spatial and Fourier domains, an extended source is required, such as the field produced by a horn. For optimum performance, the suitability of a particular source field with respect to some particular class of device can be assessed using the theory of Frames. However, if the dynamic range of each measurement is intrinsically high, $>$20 dB, the need for optimisation can often be relaxed.

Additionally, incremental sampling can be applied, which means that the recovered spectrum can be systematically enhanced by adding more and more samples until all of the degrees of freedom have been found. Notions of separability may be beneficial when choosing optimum search strategies. Near-field, far-field, dual-surface, and phase-space scanning can be carried out; although it should be borne in mind that near-field scanning relates to whether evanescent modes are actually measured, whereas near-field and far-field reconstruction relates to the reference surface used for presenting the results. It has been observed that measurement noise couples and splits any near-degenerate modes of the ideal device. This mixing may have implications when using devices in real instruments, where noise is always present at some level, and where for high-throughput devices, the lowest-order modes are degenerate.

Various practical measures have been described for reducing errors. A reference detector is particularly valuable when measuring the individual and collective behaviours of pixels in short-wavelength imaging arrays, because it provides a common reference in terms of which the modes of the pixels can be described, such as the far-field phase slopes on the modes of displaced pixels. Full characterisation of arrays becomes possible. Also, when scanning the sources over a plane in the far field, possible ambiguities associated with phase wrapping are eliminated because the phases of the modes are referenced to those of the reference detector.

Although this paper has focused on using EAI for characterising the behaviour of partially-coherent far-infrared and optical low-noise detectors of the kind used in astronomy, many other techniques and applications have been identified, and it seems inevitable that EAI with become a valuable tool for characterising and studying the spatial-spectral forms of energy-absorbing devices, including detectors and energy-harvesting components. EAI can be used to determine the modal content of the partially coherent radiation fields emitted by active devices, such as LEDs, by measuring their absorption characteristics. Moreover, the extension of the method to allow the absorption characteristics of nonlinear devices to be determined under `bright-light' conditions could be particularly fruitful; for example, in astronomy it is essential to understand dynamic range when observing bright and weak sources in the same field of view.


{\bf Acknowledgements}

The authors would like to thank the many colleagues who have contributed, over the last 10 years, to the development of energy absorption interferometry (EAI) and its application to the design and characterisation of long-wavelength optical systems and instruments. We are particularly grateful to Stephen Yates (SRON), Ian Veenendaal (SRON), and Edgar Castillo Dominguez (SRON) for prototyping, developing and demonstrating EAI at FIR wavelengths using photo-mixer-based sources, and Shahab Dabironezare (TUD), and Daan Roos (TUD) for their work on using EAI for modelling, characterising, and verifying the optical performance of ultra-low-noise lens-absorber coupled Kinetic Inductance Detectors (MKIDs). This team is actively using EAI for developing and testing a new generation of ultra-low-noise  imaging arrays for the next generation of far-infrared space telescopes, and we are most grateful for our regular and ongoing discussions. Prof. Stafford Withington would like to thank the Space Research Organisation Netherlands, Groningen for being a strong supporter of a Dutch-UK collaboration on ultra-low-noise space-based instrumentation for well over 30 years.  



\appendix

\section{Power}
\label{appendix_A}

Consider multiplying two time-harmonic signals having the same frequency, where the product has the dimensions of power. Ordinary trigonometric reasoning can be used to write the instantaneous power in the form
\begin{align}
\label{eqn_apa_1}
P(t) &  = x_1 \cos (\omega t) x_2 \cos (\omega t + \phi) \\ \nonumber
& =  x_1  \cos (\omega t)  x_2 \cos (\omega t)  \cos ( \phi ) -  x_1  \cos (\omega t)  x_2 \sin (\omega t)  \sin ( \phi ) \\ \nonumber
& = x_1 x_2 \cos^2( \omega t) \cos ( \phi) - x_1 x_2 \frac{1}{2} \sin ( 2 \omega t) \sin (\phi), 
\end{align}
where $\phi$ is an arbitrary phase difference. The first term on the second line is the product of $x_1$ and the in-phase component of $x_2$, and the second term is the product of $x_1$ and the out-of-phase component of $x_2$. In the third line, if $\phi$ lies in the range $- \pi/2$ to   $+ \pi/2$, the first term is always positive, and describes the instantaneous rate at which energy is dissipated.  It has a cycle-averaged value of $ P = x_1 x_ 2 \cos (\phi) /2$. Using the language of electrical engineering $\cos (\phi)$ is the {\em power factor}.  If $\phi$ lies outside the range $- \pi/2$ to   $+ \pi/2$ dissipative energy flows in the other direction, and the load now seems to be acting as a source of energy; this situation occurs in electrical circuits when negative resistance is present. The second term however has a cycle average value of zero, regardless of $\phi$, and describes the instantaneous rate at which reactive energy flows. Its magnitude $ Q = x_1 x_ 2 \sin (\phi) /2$ gives the maximum instantaneous rate of the reactive flow, first in one direction and then the other.

For a finite duration $T$, the average power is given by
\begin{align}
\label{eqn_apa_2}
P &  =  x_1 x_2 \frac{1}{T} \int_{-T/2}^{T/2}  \left[ \cos^2( \omega t) \cos ( \phi) - \frac{1}{2} \sin ( 2 \omega t) \sin (\phi)\right]  \, dt, 
\end{align}
which evaluates to
\begin{align}
\label{eqn_apa_3}
P &  =  \frac{x_1 x_2}{2} \cos ( \phi) \left[ {\rm sinc} (\omega T) + 1 \right], 
\end{align}
because the reactive term is asymmetric about $t=0$. The peak reactive power, however, is  $x_1 x_2 \sin (\phi) /2$. For large $T$, or for samples of duration $1/2f$,  $P$
is just the average dissipated power.

Consider the product of two analytic signals
\begin{align}
\label{eqn_apa_4}
S & =\frac{1}{2} x_1 (t) x_2^\ast (t)  \\ \nonumber
& = | x_1 (\omega_0) || x_2 (\omega_0) |\left[  \cos(\phi(\omega_0)) - i  \sin(\phi(\omega_0)) \right]
\end{align}
$x_1(t) x^{a \ast}_2(t)/2$ is a complex-valued power: the real part gives the average rate at which dissipated energy flows, {\em real power}, and the imaginary part gives the maximum rate at which stored energy flows, {\em reactive power}. $P$ is the {\em apparent power}, and together they form a vector representation of the energy flow.  Accordingly, 
\begin{align}
\label{eqn_apa_5}
P & = \frac{1}{2} \mbox{Re} [ \langle x^a_1(t) x^{a \ast}_2(t)) \rangle],
\end{align}
and
\begin{align}
\label{eqn_apa_6}
Q & = \frac{1}{2} \mbox{Im} [ \langle x^a_1(t) x^{a \ast}_2(t)) \rangle],
\end{align}
is the reactive power. 


\section{Noise Moments}
\label{appendixB}

The elements of ${\bf N}$, labelled ${N}_{ij}$, are complex-valued zero-mean Gaussian random variables:
\begin{equation}
\label{eqn_appb_1}
P({\bf n}) = \frac{1}{\pi^N \mbox{det} ({\bf \Sigma})} \mbox{exp} \left[ - {\bf n}^{\dagger} {\bf \Sigma}^{-1} {\bf n} \right],
\end{equation}
where for notational convenience , the complex-valued elements of $\hat{\bf n}$ are the same as those of $\hat{\bf N}$ but written as a vector. ${\bf \Sigma} = \langle \hat{\bf n} \hat{\bf n}^\dagger \rangle$ is the covariance matrix, and the prefactor ensures that the distribution is normalised. $\mbox{`det'}$ indicates the determinant, and $N$ is the number of element in $\hat{\bf N}$.  Any off-diagonal correlations in the noise are handled by introducing suitable correlations into ${\bf \Sigma}$. 

Consider a stochastic complex variable $z$, where the real $u$ and imaginary $v$ parts are zero-mean Gaussian random variables: 
$z = u + i v$. The real and imaginary parts are uncorrelated  $ \langle u v \rangle = 0$ but have the same variance $ \langle u^2 \rangle =  \langle v^2 \rangle = \sigma^2$. It follows that $\langle z \rangle = 0$. Several second-order moments can be formed:
\begin{align}
\label{eqn_appb_4}
\langle z  z \rangle & = \langle (u + i v)(u + iv) \rangle = 0 \\ \nonumber
\langle z^\ast  z^\ast \rangle & = \langle (u - i v)(u - iv) \rangle = 0 \\ \nonumber
\langle z  z^\ast \rangle & = \langle (u + i v)(u - iv) \rangle = 2 \sigma^2.
\end{align}
$\langle z  z^\ast \rangle = \langle | z |^2 \rangle$ is the expectation value of the amplitude squared. The amplitude has a Rayleigh distribution, and so the mean and variance of the amplitude of the added noise are $\sigma \sqrt{\pi/2}$ and $\sigma^2 (4-\pi)/2 \approx \sigma^2/2$ respectively.

If $z_m$ and $z_n$ are drawn from a set of mutually uncorrelated complex random variables
\begin{align}
\label{eqn_appb_5}
\langle z_m  z_n \rangle & = \langle (u_m + i v_m)(u_n + iv_n) \rangle = 0 \hspace{2mm} \forall \, m,n\\ \nonumber
\langle z_m^\ast  z_j^\ast \rangle & = \langle (u_m - i v_m)(u_n - iv_n) \rangle = 0 \hspace{2mm} \forall \, m,n \\ \nonumber
\langle z_m  z_n^\ast \rangle & = \langle (u_m + i v_m)(u_n - iv_n) \rangle = 2 \sigma^2 \delta_{mn}.
\end{align}

Consider a random matrix $Z$ where each element $Z_{ij}$ is a zero-mean complex random process, and additionally assume that all of the elements are uncorrelated $\langle Z_{ij} Z_{mn} \rangle = 0 \, \, \forall \, i,j,m,n$. The matrix $Z$ can be decomposed into its hermitian and antihermitian parts:
\begin{align}
\label{eqn_appb_6}
Z & = Z^H + Z^A \\ \nonumber
Z^H & = \frac{1}{2} \left( Z + Z^\dagger \right) \\ \nonumber
Z^A & = \frac{1}{2} \left( Z - Z^\dagger \right),
\end{align}
where the elements are
\begin{align}
\label{eqn_appb_7}
Z^H_{ij} = \frac{1}{2} \left( Z_{ij} + Z_{ji}^\ast \right) \\ \nonumber
Z^A_{ij} = \frac{1}{2} \left( Z_{ij} - Z_{ji}^\ast \right).
\end{align}
$Z^H$ and $Z^A$ individually contain correlations: when one off-diagonal element is known, $Z^{H/A}_{ij}$, its off-diagonal partner, $Z^{H/A}_{ji}$, is also known. From the above, the low-order moments of $Z^H$ and $Z^A$ are as follows:
\begin{align}
\label{eqn_appb_8}
\langle Z^H_{ij} \rangle = \frac{1}{2} \left( \langle Z_{ij} \rangle  + \langle Z_{ji}^\ast \rangle \right) = 0 \\ \nonumber
\langle Z^A_{ij} \rangle  = \frac{1}{2} \left( \langle Z_{ij} \rangle  - \langle Z_{ji}^\ast \rangle \right) = 0.
\end{align}
and
\begin{align}
\label{eqn_appb_9}
\langle Z^H_{ij} Z^H_{kl} \rangle = \frac{1}{4} \langle \left( Z_{ij}  + Z_{ji}^\ast  \right) \left( Z_{kl}  + Z_{lk}^\ast  \right) \rangle = \sigma^2 \delta_{il} \delta_{jk} \\ \nonumber
\langle Z^A_{ij} Z^A_{kl} \rangle = \frac{1}{4}\langle \left( Z_{ij}  - Z_{ji}^\ast  \right) \left( Z_{kl}  - Z_{lk}^\ast  \right) \rangle = - \sigma^2 \delta_{il} \delta_{jk} \\ \nonumber
\langle Z^H_{ij} Z^A_{kl} \rangle = \frac{1}{4} \langle \left( Z_{ij}  + Z_{ji}^\ast  \right) \left( Z_{kl}  - Z_{lk}^\ast  \right) \rangle = 0.
\end{align}
The off-diagonal elements of $Z^H$ are perfectly correlated; the off-diagonal elements of $Z^A$ are perfectly anti-correlated; and the elements of $Z^H$ and $Z^A$ are uncorrelated.

The first-order moments of ${\bf N}^H$ take the form
\begin{align}
\label{eqn_appb_10}
N^H_{ij} & = \frac{1}{2} \left[ N_{ij} + N_{ji}^\ast \right] \\ \nonumber 
& = \frac{1}{2} \left[ (u_{ij} + v_{ji} + i( u_{ij} - v_{ji}) \right] \\ \nonumber
\langle N^H_{ij} \rangle & = 0,
\end{align}
and all of the elements remain zero-mean Gaussian processes. The second order moments are
\begin{align}
\label{eqn_appb_11}
\langle N^H_{ij} N^H_{kl} \rangle & =  \frac{1}{4} \langle \left[ N_{ij} + N_{ji}^\ast \right]\left[ N_{kl} + N_{lk}^\ast \right] \rangle \\ \nonumber
& =  \frac{1}{4} \left[   \langle N_{ij}N_{kl} \rangle+ \langle  N_{ij} N_{lk}^\ast \rangle + \langle  N_{ji}^\ast  N_{kl} \rangle +  \langle N_{ji}^\ast  N_{lk}^\ast \rangle \right] \\ \nonumber
& =  \frac{1}{4} \left[ \langle  N_{ij} N_{lk}^\ast \rangle + \langle  N_{ji}^\ast  N_{kl} \rangle \right] \\ \nonumber
& = \sigma^2 \delta_{il} \delta_{jk}
\end{align}
The third-order moments are zero 
\begin{align}
\label{eqn_appb_12}
\langle N^H_{ij} N^H_{kl} N^H_{mn} \rangle & =  \frac{1}{8} \langle \left[ z_{ij} + z_{ji}^\ast \right]\left[ z_{kl} + z_{lk}^\ast \right] \left[ z_{mn} + z_{nm}^\ast \right]\rangle \\ \nonumber
& = 0,
\end{align}
as are all odd-ordered moments.

Finally, the fourth-order moments are given by
\begin{align}
\label{eqn_appb_13}
\langle N^H_{ij} N^H_{kl} N^H_{mn} N^H_{rs} \rangle & =  \frac{1}{16} \langle \left[N_{ij} + N_{ji}^\ast \right]\left[ N_{kl} + N_{lk}^\ast \right] \left[ N_{mn} + N_{nm}^\ast \right]\left[ N_{rs} + N_{sr}^\ast \right] \rangle \\ \nonumber
& = \frac{1}{16} \langle \left[  N_{ij}N_{kl} + N_{ij} N_{lk}^\ast + N_{ji}^\ast  N_{kl}  +  N_{ji}^\ast  N_{lk}^\ast  \right] \\ \nonumber
& \times \left[  N_{mn}N_{rs} + N_{mn} N_{sr}^\ast + N_{nm}^\ast  N_{rs}  +  N_{nm}^\ast  N_{sr}^\ast  \right]  \rangle \\ \nonumber
& = \frac{1}{16} \langle \left[ \langle N_{ij}N_{kl} N_{nm}^\ast  N_{sr}^\ast \rangle + \langle N_{ij} N_{lk}^\ast   N_{mn} N_{sr}^\ast \rangle + \langle N_{ij} N_{lk}^\ast  N_{nm}^\ast  N_{rs} \rangle \right. \\ \nonumber
& \left.+ \langle N_{ji}^\ast  N_{kl} N_{mn} N_{sr}^\ast  \rangle + \langle N_{ji}^\ast  N_{kl}  N_{nm}^\ast  N_{rs} \rangle  +   \langle N_{ji}^\ast  N_{lk}^\ast  N_{mn}N_{rs}   \rangle  \right] \\ \nonumber
& =  \frac{1}{16} \langle \left[    \langle N_{ij}  N_{nm}^\ast \rangle   \langle  N_{kl} N_{sr}^\ast \rangle +  \langle N_{ij}   N_{sr}^\ast \rangle \langle  N_{kl}  N_{nm}^\ast  \rangle   \right. \\ \nonumber
& + \left.    \langle N_{ij}    N_{lk}^\ast \rangle \langle  N_{mn}  N_{sr}^\ast \rangle   +   \langle N_{ij} N_{sr}^\ast \rangle  \langle N_{mn} N_{lk}^\ast  \rangle \right. \\ \nonumber
& + \left.   \langle N_{ij} N_{lk}^\ast \rangle  N_{nm}^\ast  N_{rs}  \rangle +  \langle N_{ij}  N_{nm}^\ast \rangle \langle N_{lk}^\ast   N_{rs} \rangle \right. \\ \nonumber
& + \left. 
 \langle N_{ji}^\ast  N_{kl} \rangle \langle N_{mn} N_{sr}^\ast  \rangle  +  \langle N_{ji}^\ast  N_{mn} \rangle \langle  N_{kl} N_{sr}^\ast  \rangle \right. \\ \nonumber
 & + \left. \langle N_{ji}^\ast  N_{kl} \rangle \langle N_{nm}^\ast  N_{rs} \rangle +  \langle N_{ji}^\ast  N_{rs}  \rangle \langle N_{kl}  N_{nm}^\ast  \rangle  \right. \\ \nonumber
 & + \left.
   \langle N_{ji}^\ast  N_{mn} \rangle \langle N_{lk}^\ast  N_{rs} \rangle +  \langle N_{ji}^\ast N_{rs} \rangle \langle  N_{lk}^\ast  N_{mn}   \rangle  
 \right] \\ \nonumber
 \langle N^H_{ij} N^H_{kl} N^H_{mn} N^H_{rs} \rangle & = \frac{\sigma^4}{4} 
 \left[ \delta_{in} \delta_{jm}  \delta_{ks} \delta_{lr} + \delta_{is} \delta_{jr}  \delta_{kn} \delta_{lm}   \right. \\ \nonumber
& \left. + \delta_{il} \delta_{jk}  \delta_{ms} \delta_{nr} + \delta_{is} \delta_{jr}  \delta_{ml} \delta_{nk}   \right. \\ \nonumber
& \left. + \delta_{il} \delta_{jk}  \delta_{ms} \delta_{nr} + \delta_{in} \delta_{jm}  \delta_{lr} \delta_{ks}   \right. \\ \nonumber
& \left. + \delta_{jk} \delta_{il}  \delta_{ms} \delta_{nr} +  \delta_{jm} \delta_{in}  \delta_{ks} \delta_{lr}  \right. \\ \nonumber
& \left. + \delta_{jk} \delta_{il}   \delta_{nr}\delta_{ms} +  \delta_{jr} \delta_{is}  \delta_{kn} \delta_{lm}  \right. \\ \nonumber
& \left. + \delta_{jm} \delta_{in}   \delta_{lr}\delta_{ks} +  \delta_{jr} \delta_{is}   \delta_{lm}\delta_{kn}  \right] \\ \nonumber
& = \sigma^4 \left[ \delta_{il} \delta_{jk}  \delta_{ms} \delta_{nr} + \delta_{in} \delta_{jm}  \delta_{ks} \delta_{lr}
+  \delta_{is} \delta_{jr}  \delta_{kn} \delta_{lm}   \right] 
\end{align}

In conclusion, the moments are given by
\begin{align}
\label{eqn_appb_14}
\langle N^H_{ij} \rangle & = 0 \\ \nonumber
\langle N^H_{ij} N^H_{kl} \rangle & =  \sigma^2 \delta_{il} \delta_{jk} \\ \nonumber
\langle N^H_{ij} N^H_{kl} N^H_{mn} \rangle & =  0 \\ \nonumber
 \langle N^H_{ij} N^H_{kl} N^H_{mn} N^H_{rs} \rangle & =\sigma^4  \left[ \delta_{il} \delta_{jk}  \delta_{ms} \delta_{nr} + \delta_{in} \delta_{jm}  \delta_{ks} \delta_{lr}
+  \delta_{is} \delta_{jr}  \delta_{kn} \delta_{lm}   \right], 
\end{align}
where $\sigma^2$ is the variance of each of the real and imaginary parts of the original non-hermitian noise matrix.


\section{Perturbation}
\label{appendixC}

Suppose that a small change is made to a linear hermitian operator such that
\begin{equation}
\label{eqn_appc_1}
\hat{H} = \hat{H}^{0} + \hat{H}^{1},
\end{equation}
where $\hat{H}^{0}$ is the Hermitian operator associated with the unperturbed system, and $\hat{H}^{1}$ is the perturbation.  $\hat{H}^{0}$ has eigenvectors $| \phi^{0}_{n} \rangle$, where
\begin{equation}
\label{eqn_appc_2}
\hat{H}^{0} | \phi^{0}_{n} \rangle = E^{0}_{n} | \phi^{0}_{n} \rangle \hspace{10mm} {\rm where} \hspace{10mm}  \langle \phi^{0}_{m} | \phi^{0}_{n} \rangle = \delta_{mn};
\end{equation}
it will be assumed that these can be found or are already known. The aim is to find the eigenvectors and eigenvalues of the full operator,
\begin{equation}
\label{eqn_appc_3}
\hat{H} | \phi_{n} \rangle = E_{n} | \phi_{n} \rangle \hspace{10mm} {\rm where} \hspace{10mm}  \langle \phi_{m} | \phi_{n} \rangle = \delta_{mn},
\end{equation}
which implies that  the perturbation $\hat{H}^{1}$ is also Hermitian. 

Suppose that the perturbation can be gradually turned  on through some parameter $\lambda$, where $ 0 < \lambda < 1$:
\begin{equation}
\label{eqn_appc_4}
\hat{H} = \hat{H}^{0} + \lambda \hat{H}^{1}.
\end{equation}
The perturbation is fully operational when $\lambda = 1$. This process can be viewed in a mathematical sense, where $\lambda$ is a free parameter,  or in a temporal sense, where the perturbation is turned on adiabatically so that the system has time to respond giving a smooth transition to the new eigenvectors and eigenvalues:
\begin{equation}
\label{eqn_appc_5}
\left.
\begin{array}{c}
   | \phi^{0}_{n} \rangle \longrightarrow | \phi_{n} \rangle \\
   E^{0}_{n} \longrightarrow  E_{n}
\end{array}
\right\}  \hspace{5mm}  \forall \, n  \hspace{10mm} {\rm as} \hspace{5mm} \lambda = 0 \rightarrow 1 .
\end{equation}
$\lambda$ is a valuable mathematical tool because it can be used to test, to different orders, the sensitivity of the eigenvectors and eigenvalues of $\hat{H}$ to changes in the `magnitude' of $\hat{H}^{1}$. 

The eigenvalues must be real scalar functions of $\lambda$, and so the the perturbed values can be written as power series,
\begin{equation}
\label{eqn_appc_6}
E_{n} (\lambda) =  E^{(0)}_{n} + \lambda E^{(1)}_{n} + \lambda^{2} E^{(2)}_{n} \cdots = \sum_{m} \lambda^{m}_{n}  E^{(m)}_{n},
\end{equation}
where the $E^{(m)}_{n}$ are the correction factors that must be found for each eigenvalue $n$. Likewise for the eigenvectors, which are vector-valued functions of $\lambda$,
\begin{equation}
\label{eqn_appc_7}
| \phi_{n} \rangle (\lambda) = | \phi^{(0)}_{n} \rangle + \lambda | \phi^{(1)}_{n} \rangle + \lambda^{2} | \phi^{(2)}_{n} \rangle \cdots = \sum_{m} \lambda^{m}_{n}| \phi^{(m)}_{n} \rangle,
\end{equation}
where $|\phi^{(m)}_{n} \rangle$ are correction vectors that must be found for each eigenvalue $n$.

\subsection{First-order corrections}
\label{sec_foc}

Using (\ref{eqn_appc_4}), (\ref{eqn_appc_6}) and (\ref{eqn_appc_7}),
\begin{equation}
\label{eqn_appc_8}
\left( \hat{H}^{0} + \lambda \hat{H}^{1} \right) \left[ \sum_{m=0}^{\infty} \lambda^{m}| \phi^{(m)}_{n} \rangle \right] = \left[ \sum_{m=0}^{\infty} \lambda^{m}_{n}  E^{(m)}_{n} \right] \left[ \sum_{m=0}^{\infty} \lambda^{m}_{n}| \phi^{(m)}_{n} \rangle \right].
\end{equation}

In the spirit of {\em first-order perturbation theory}, extract only those terms that are linear in $\lambda$, remembering that  $\lambda^{0} = 1$:
\begin{equation}
\label{eqn_appc_9}
\hat{H}^{0} | \phi^{(1)}_{n} \rangle + \hat{H}^{1} | \phi^{(0)}_{n} \rangle = E^{(0)}_{n} | \phi^{(1)}_{n} \rangle + E^{(1)}_{n} | \phi^{(0)}_{n} \rangle.
\end{equation}

Taking the inner product with $| \phi^{(0)}_{n} \rangle$
\begin{align}
\label{eqn_appc_10}
\langle \phi^{(0)}_{n} | \hat{H}^{0} | \phi^{(1)}_{n} \rangle + \langle \phi^{(0)}_{n} | \hat{H}^{1} | \phi^{(0)}_{n} \rangle & = \langle \phi^{(0)}_{n} | E^{(0)}_{n} | \phi^{(1)}_{n} \rangle + \langle \phi^{(0)}_{n} | E^{(1)}_{n} | \phi^{(0)}_{n} \rangle,
\end{align}
and noting that $\langle \phi^{(0)}_{n} | \hat{H}^{0} = E^{(0)}_{n} \langle \phi^{(0)}_{n} |$, gives
\begin{equation}
\label{eqn_appc_11}
 E^{(1)}_{n} = \langle \phi^{(0)}_{n} | \hat{H}^{1} | \phi^{(0)}_{n} \rangle.
\end{equation}
Equation (\ref{eqn_appc_11}) is an expression for $E^{(1)}_{n}$ in terms of easily known quantities - the matrix element of $\hat{H}^{1}$ in terms of the eigenvector $| \phi^{(0)}_{n} \rangle$. This correction scales linearly with $\hat{H}^{1}$. When $\lambda = 1$, the perturbation is at full strength, and the change in eigenvalue $n$ is 
\begin{align}
\label{eqn_appc_12}
E^{(1)}_{n} & = \langle \phi^{(0)}_{n} | \hat{H}^{1} | \phi^{(0)}_{n} \rangle \\ \nonumber
& = {\rm Tr} \left[\hat{H}^{1} \hat{P}_n \right],
\end{align}
where 
\begin{align}
\label{eqn_appc_13}
\hat{P}_n  & = | \phi^{(0)}_{n} \rangle \langle \phi^{(0)}_{n} |, 
\end{align}
is the $n$'th unpertubed projector. 

The change in the eigenvectors can also be calculated. Taking the inner product of (\ref{eqn_appc_9}) with $| \phi^{(0)}_{m} \rangle$, where $m \neq n$:
\begin{align}
\label{eqn_appc_14}
\langle \phi^{(0)}_{m} | \hat{H}^{0} | \phi^{(1)}_{n} \rangle + \langle \phi^{(0)}_{m} | \hat{H}^{1} | \phi^{(0)}_{n} \rangle & = \langle \phi^{(0)}_{m} | E^{(0)}_{n} | \phi^{(1)}_{n} \rangle + \langle \phi^{(0)}_{m} | E^{(1)}_{n} | \phi^{(0)}_{n} \rangle,
\end{align}
giving
\begin{equation}
\label{eqn_appc_15}
 \langle \phi^{(0)}_{m} | \phi^{(1)}_{n} \rangle = \frac{ \langle \phi^{(0)}_{m} | \hat{H}^{1} | \phi^{(0)}_{n} \rangle }{ E_{n}^{(0)} - E_{m}^{(0)}} \hspace{10mm} {\rm for} \, \, m \neq n,
\end{equation}
but these are the expansion coefficients of $| \phi^{(1)}_{n} \rangle$ in the basis set  $| \phi^{(0)}_{m} \rangle$. The coefficient $\langle \phi^{(0)}_{n} | \phi^{(1)}_{n} \rangle$, however, is not known. It is not given by (\ref{eqn_appc_15}) which is singular, and therefore the expansion is not yet complete. Explicitly separating out this term:
\begin{align}
\label{eqn_appc_16}
| \phi^{(1)}_{n} \rangle & =  \langle \phi^{(0)}_{n} | \phi^{(1)}_{n} \rangle  | \phi^{(0)}_{n} \rangle  + \sum_{m \neq n}  \left\{ \langle \phi^{(0)}_{m} | \phi^{(1)}_{n} \rangle \right\}  | \phi^{(0)}_{m} \rangle \\ \nonumber
& =  \langle \phi^{(0)}_{n} | \phi^{(1)}_{n} \rangle  | \phi^{(0)}_{n} \rangle  + \sum_{m \neq n}  \frac{ \langle \phi^{(0)}_{m} | \hat{H}^{1} | \phi^{(0)}_{n} \rangle }{ E_{n}^{(0)} - E_{m}^{(0)}} | \phi^{(0)}_{m} \rangle.
\end{align}

The difficulty with  $\langle \phi^{(0)}_{n} | \phi^{(1)}_{n} \rangle$ is rooted in the normalisation of the perturbed eigenvectors. From (\ref{eqn_appc_7}), normalisation requires
\begin{align}
\label{eqn_appc_17}
\langle \phi_{n}   | \phi_{n} \rangle  =   \langle   \phi^{(0)}_{n}   | \phi^{(0)}_{n} \rangle   +    \sum_{ij} \lambda^{i}_{n}   \lambda^{j}_{n}   \langle   \phi^{(i)}_{n}   | \phi^{(j)}_{n} \rangle  = 1,
\end{align}
where the lowest-order term, corresponding to the length of the  unperturbed eigenvector, has been separated out. But because $ \langle   \phi^{(0)}_{n}   | \phi^{(0)}_{n} \rangle =1$, 
\begin{align}
\label{eqn_appc_18}
 \sum_{ij} \lambda^{i}_{n}   \lambda^{j}_{n}   \langle   \phi^{(i)}_{n}   | \phi^{(j)}_{n} \rangle  = 0 \qquad \forall \lambda^i_n, \lambda^j_n,
\end{align}
and so  $\langle \phi^{(i)}_{n}   | \phi^{(j)}_{n} \rangle  = 0 \, \, \, \forall i,j$ apart from $i=j=0$. Each of the correction vectors must be orthogonal to the unperturbed eigenvector: only rotations are allowed. To first order in $\lambda$,  $\langle   \phi^{(0)}_{n}   | \phi^{(1)}_{n} \rangle  = 0$. 
In conclusion,
\begin{equation}
\label{eqn_appc_19}
| \phi_{n}^{(1)} \rangle = \sum_{m \neq n}  \frac{ \langle \phi^{(0)}_{m} | \hat{H}^{1} | \phi^{(0)}_{n} \rangle }{ E_{n}^{(0)} - E_{m}^{(0)}} | \phi^{(0)}_{m} \rangle.
\end{equation}

It is convenient to use the notation
\begin{equation}
\label{eqn_appc_19b}
| \phi_{n}^{(1)} \rangle = \hat{W}_n^{(1)} | \phi^{(0)}_{n} \rangle,
\end{equation}
where
\begin{align}
\label{eqn_appc_19c}
\hat{W}_n^{(1)} & =  - \sum_{m \neq n} \frac{ |\phi^{(0)}_{m} \rangle \langle \phi^{(0)}_{m} | }{ E_{m}^{(0)} - E_{n}^{(0)}} \hat{H}^{1} | \phi^{(0)}_{n} \rangle \langle \phi^{(0)}_{n} | \\ \nonumber
& = - \hat{S}_n  \hat{H}^{1} \hat{P}_n.
\end{align}
Equations (\ref{eqn_appc_12}) and (\ref{eqn_appc_19b}) are the first-order corrections to the eigenvectors and eigenvalues respectively. 

To be valid, the shift in each eigenvalue must be smaller than the unperturbed eigenvalue. Using (\ref{eqn_appc_6}),
\begin{align}
\label{eqn_appc_20}
\sum_{m \neq 0} \lambda^{m}_{n}  E^{(m)}_{n}  & \ll   E_{n}^{(0)},
\end{align}
which to first order requires
\begin{align}
\label{eqn_appc_21}
E^{(1)}_{n} & \ll    E_{n}^{(0)}  \\ \nonumber
\langle \phi^{(0)}_{  n} | \hat{H}^{1} | \phi^{(0)}_{n} \rangle  & \ll   E_{n}^{(0)}.
\end{align}
Likewise, the change in each eigenvector must be smaller than the unperturbed eigenvector. Considering the coefficients in (\ref{eqn_appc_19}), to first order
\begin{align}
\label{eqn_appc_22}
\langle \phi^{(0)}_{m} | \hat{H}^{1} | \phi^{(0)}_{n} \rangle  & \ll  | E_{n}^{(0)} - E_{m}^{(0)} | \qquad \forall m \neq n.
\end{align}
The term on the right will be at it smallest for neighboring eigenvalues. 

Equations (\ref{eqn_appc_21}) and (\ref{eqn_appc_22}) impose calculable constraints on the matrix elements of the perturbation, and therefore on the allowable magnitude of the perturbation. The diagonal matrix elements of the perturbation in terms of the unperturbed eigenvectors, must be smaller than the unperturbed eigenvalues; and the off diagonal matrix elements in terms of the unperturbed eigenvectors, must be smaller than the differences between the unperturbed eigenvalues. The most closely spaced eigenvalues will impose the most demanding constraint, and lead to a failure of the first-order approximation. 


\subsection{Second-order corrections}
\label{sec_soc}

The second-order terms can be found in a similar way. Revisiting (\ref{eqn_appc_8}),
\begin{equation}
\label{eqn_appc_23}
\left( \hat{H}^{0} + \lambda \hat{H}^{1} \right) \left[ \sum_{m} \lambda^{m}| \phi^{(m)}_{n} \rangle \right] = \left[ \sum_{m} \lambda^{m}  E^{(m)}_{n} \right] \left[ \sum_{m} \lambda^{m}_{n}| \phi^{(m)}_{n} \rangle \right]
\end{equation}
and isolating those terms in $\lambda^{2}$, gives
\begin{equation}
\label{eqn_appc_24}
\hat{H}^{0} | \phi^{(2)}_{n} \rangle + \hat{H}^{1} | \phi^{(1)}_{n} \rangle = E^{(0)}_{n} | \phi^{(2)}_{n} \rangle +  E^{(1)}_{n} | \phi^{(1)}_{n} \rangle + E^{(2)}_{n} | \phi^{(0)}_{n} \rangle.
\end{equation}
As before, take the inner product with $| \phi^{(0)}_{n} \rangle$
\begin{equation}
\label{eqn_appc_25}
\langle \phi^{(0)}_{n} | \hat{H}^{0} | \phi^{(2)}_{n} \rangle + \langle \phi^{(0)}_{n} |\hat{H}^{1} | \phi^{(1)}_{n} \rangle = \langle \phi^{(0)}_{n} |E^{(0)}_{n} | \phi^{(2)}_{n} \rangle +  \langle \phi^{(0)}_{n} |E^{(1)}_{n} | \phi^{(1)}_{n} \rangle + \langle \phi^{(0)}_{n} |E^{(2)}_{n} | \phi^{(0)}_{n} \rangle,
\end{equation}
which leads to
\begin{equation}
\label{eqn_appc_26}
 E^{(2)}_{n} =  \langle \phi^{(0)}_{n} |\hat{H}^{1} | \phi^{(1)}_{n} \rangle - E^{(1)}_{n} \langle \phi^{(0)}_{n}  | \phi^{(1)}_{n} \rangle.
\end{equation}
Because only orthogonal corrections are allowed, $\langle \phi^{(0)}_{n} | \phi^{(1)}_{n} \rangle =0$,
\begin{equation}
\label{eqn_appc_27}
 E^{(2)}_{n} =  \langle \phi^{(0)}_{n} |\hat{H}^{1} | \phi^{(1)}_{n} \rangle,
\end{equation}
but $| \phi^{(1)}_{n} \rangle$ is already known, and therefore
\begin{align}
\label{eqn_appc_28}
 E^{(2)}_{n} & =  \langle \phi^{(0)}_{n} |\hat{H}^{1} \sum_{m \neq n}  \frac{ \langle \phi^{(0)}_{m} | \hat{H}^{1} | \phi^{(0)}_{n} \rangle }{ E_{n}^{(0)} - E_{m}^{(0)}} | \phi^{(0)}_{m} \rangle \\ \nonumber
 & = \sum_{m \neq n}  \frac{| \langle \phi^{(0)}_{m} | \hat{H}^{1} | \phi^{(0)}_{n} \rangle |^{2}}{ E_{n}^{(0)} - E_{m}^{(0)}}.
\end{align}
Introducing the reduced resolvent,
\begin{align}
\label{eqn_appc_29}
\hat{S}_n & =  \sum_{m \neq n} \frac{ | \phi^{(0)}_{m} \rangle\langle \phi^{(0)}_{m} |}{ E_{m}^{(0)} - E_{n}^{(0)}},
\end{align}
gives
\begin{equation}
\label{eqn_appc_30}
E^{(2)}_{n}   = - {\rm Tr} \left[  \hat{S}_n  \hat{H}^{1} P_{n} \hat{H}^{1} \right].
\end{equation}

Equation (\ref{eqn_appc_28}) shows that the second-order shift in the lowest-order eigenvalue, $ E^{(2)}_{1}$, is always negative. If the matrix elements of $\hat{H}^{1}$  are of comparable magnitude, neighbouring eigenvalues make a larger contribution than distant eigenvalues. If two of the original eigenvalues are close together, the values will `repel' as $\hat{H}^{1}$ is turned on, due to the sign changing in the denominator of (\ref{eqn_appc_26}). Eigenvalues tend to repel, forming anticrossings, retaining their ordering, and so can be traced individually as perturbations are applied.

To calculate the second-order corrections to the eigenvectors, start again with
\begin{equation}
\label{eqn_appc_31}
\hat{H}^{0} | \phi^{(2)}_{n} \rangle + \hat{H}^{1} | \phi^{(1)}_{n} \rangle = E^{(0)}_{n} | \phi^{(2)}_{n} \rangle +  E^{(1)}_{n} | \phi^{(1)}_{n} \rangle + E^{(2)}_{n} | \phi^{(0)}_{n} \rangle.
\end{equation}
Taking the inner product with  $| \phi^{(0)}_{m} \rangle$ where $m \neq n$ gives
\begin{align}
\label{eqn_appc_32}
\langle \phi^{(0)}_{m} | \phi^{(2)}_{n} \rangle & = - \frac{\langle \phi^{(0)}_{m} | \hat{H}^{1} | \phi^{(1)}_{n} \rangle}{ E_m^{(0)} - E_n^{(0)}} + E_n^1 \frac{\langle \phi^{(0)}_{m} | \phi^{(1)}_{n} \rangle}{ E_m^{(0)} - E_n^{(0)}} \\ \nonumber
 & = - \frac{\langle \phi^{(0)}_{m} | \hat{H}^{1} | \phi^{(1)}_{n} \rangle}{ E_m^{(0)} - E_n^{(0)}} + \langle \phi^{(0)}_{n} | \hat{H}^{1} | \phi^{(0)}_{n} \rangle \frac{\langle \phi^{(0)}_{m} | \phi^{(1)}_{n} \rangle}{ E_m^{(0)} - E_n^{(0)}} \\ \nonumber
  & = \left[ - \frac{\langle \phi^{(0)}_{m} | \hat{H}^{1} }{ E_m^{(0)} - E_n^{(0)}} + \langle \phi^{(0)}_{n} | \hat{H}^{1} | \phi^{(0)}_{n} \rangle \frac{\langle \phi^{(0)}_{m} |}{ E_m^{(0)} - E_n^{(0)}} \right] \left[  \sum_{m' \neq n}  \frac{ \langle \phi^{(0)}_{m'} | \hat{H}^{1} | \phi^{(0)}_{n} \rangle }{ E_{n}^{(0)} - E_{m'}^{(0)}} | \phi^{(0)}_{m'} \rangle \right],
\end{align}
but these are the coefficients of $| \phi^{(2)}_{n} \rangle$ in the basis set of  $|\phi^{(0)}_{m} \rangle$, and so
 \begin{equation}
\label{eqn_appc_33}
| \phi^{(2)}_{n} \rangle  = \sum_{m \neq n}  \sum_{m' \neq n}   \left[ - \frac{\langle \phi^{(0)}_{m} | \hat{H}^{1} | \phi^{(0)}_{m'} \rangle }{ E_m^{(0)} - E_n^{(0)}} + \langle \phi^{(0)}_{n} | \hat{H}^{1} | \phi^{(0)}_{n} \rangle \frac{\langle \phi^{(0)}_{m} | \phi^{(0)}_{m'} \rangle }{ E_m^{(0)} - E_n^{(0)}} \right] \left[   \frac{ \langle \phi^{(0)}_{m'} | \hat{H}^{1} | \phi^{(0)}_{n} \rangle }{ E_{n}^{(0)} - E_{m'}^{(0)}} \right]  |\phi^{(0)}_{m} \rangle.
\end{equation}
It can also be shown that if the intermediate normalisation $\langle \phi^{(1)}_{n}   | \phi^{(0)}_{n} \rangle$ is not imposed, then an additional term arises, giving
 \begin{align}
\label{eqn_appc_34}
| \phi^{(2)}_{n} \rangle  & = \sum_{m \neq n}  \sum_{m' \neq n}   \left[ - \frac{\langle \phi^{(0)}_{m} | \hat{H}^{1} | \phi^{(0)}_{m'} \rangle }{ E_m^{(0)} - E_n^{(0)}} + \langle \phi^{(0)}_{n} | \hat{H}^{1} | \phi^{(0)}_{n} \rangle \frac{\langle \phi^{(0)}_{m} | \phi^{(0)}_{m'} \rangle }{ E_m^{(0)} - E_n^{(0)}} \right] \left[   \frac{ \langle \phi^{(0)}_{m'} | \hat{H}^{1} | \phi^{(0)}_{n} \rangle }{ E_{n}^{(0)} - E_{m'}^{(0)}} \right]  |\phi^{(0)}_{m} \rangle \\ \nonumber
& - \frac{1}{2} \left[  \sum_{m \neq n}   \frac{ | \langle \phi^{(0)}_{m} |\hat{H}^{1} | \phi^{(0)}_{n} \rangle |^2}{( E_m^{(0)} - E_n^{(0)} )^2} \right] | \phi^{(0)}_{n} \rangle.
\end{align}
which is an expression for $| \phi^{(2)}_{n} \rangle$ in terms of known quantities.

Each of the terms in (\label{eqn_appc_35} can be expressed using the same notations as (\label{eqn_appc_19}. For the first quadratic term,
\begin{align}
\label{eqn_appc_35}
& \sum_{m \neq n}  \sum_{m' \neq n}  \frac{\langle \phi^{(0)}_{m} | \hat{H}^{1} | \phi^{(0)}_{m'} \rangle }{ E_n^{(0)} - E_m^{(0)}}   \frac{ \langle \phi^{(0)}_{m'} | \hat{H}^{1} | \phi^{(0)}_{n} \rangle }{ E_{n}^{(0)} - E_{m'}^{(0)}}  |\phi^{(0)}_{m} \rangle \\ \nonumber
& = \sum_{m \neq n}  \sum_{m' \neq n}  \frac{|\phi^{(0)}_{m} \rangle  \langle \phi^{(0)}_{m} | }{ E_m^{(0)} - E_n^{(0)}}  \hat{H}^{1}  \frac{| \phi^{(0)}_{m'} \rangle  \langle \phi^{(0)}_{m'} | }{ E_{m'}^{(0)} - E_{n}^{(0)}} \hat{H}^{1} | \phi^{(0)}_{n} \rangle  \\ \nonumber
& = \hat{S}_n  \hat{H}^{1}  \hat{S}_n  \hat{H}^{1} \hat{P}_n  | \phi^{(0)}_{n} \rangle \\ \nonumber
( \hat{W}_n^{(2)} )_a & = \hat{S}_n  \hat{H}^{1}  \hat{S}_n  \hat{H}^{1} \hat{P}_n 
\end{align}
For the second quadratic term
\begin{align}
\label{eqn_appc_36}
& - \sum_{m \neq n}  \frac{ \langle \phi^{(0)}_{m} | \hat{H}^{1} | \phi^{(0)}_{n} \rangle }{ E_{n}^{(0)} - E_{m}^{(0)}}   \frac{  \langle \phi^{(0)}_{n} | \hat{H}^{1} | \phi^{(0)}_{n} \rangle }{ ( E_n^{(0)} - E_m^{(0)} )}    |\phi^{(0)}_{m} \rangle \\ \nonumber
& =  - \sum_{m \neq n}  \frac{   |\phi^{(0)}_{m} \rangle  \langle \phi^{(0)}_{m} |  }{( E_{n}^{(0)} - E_{m}^{(0)})^2} \hat{H}^{1} | \phi^{(0)}_{n} \rangle   \langle \phi^{(0)}_{n} | \hat{H}^{1} | \phi^{(0)}_{n} \rangle\\ \nonumber
( \hat{W}_n^{(2)} )_b & = - \left( \hat{S}_n \right)^2 \left( \hat{H}^{1} \hat{P}_n \right)^2.
\end{align}

For the third term, 
\begin{align}
\label{eqn_appc_37}
& - \frac{1}{2} \sum_{m \neq n}   \frac{ | \langle \phi^{(0)}_{m} | \hat{H}^{1} | \phi^{(0)}_{n} \rangle |^2  }{( E_{n}^{(0)} - E_{m}^{(0)})^2}   | \phi^{(0)}_{n} \rangle \\ \nonumber
 & =  - \frac{1}{2} \sum_{m \neq n}   \frac{ \langle \phi^{(0)}_{n} | \hat{H}^{1} | \phi^{(0)}_{m} \rangle  \langle \phi^{(0)}_{m} | \hat{H}^{1} | \phi^{(0)}_{n} \rangle}{( E_{n}^{(0)} - E_{m}^{(0)})^2}   | \phi^{(0)}_{n} \rangle \\ \nonumber
  & =  - \frac{1}{2} \sum_{m \neq n}   | \phi^{(0)}_{n} \rangle  \langle \phi^{(0)}_{n} | \hat{H}^{1}   \frac{| \phi^{(0)}_{m} \rangle  \langle \phi^{(0)}_{m} | }{( E_{n}^{(0)} - E_{m}^{(0)})^2} \hat{H}^{1} | \phi^{(0)}_{n} \rangle   \langle \phi^{(0)}_{n}  | \phi^{(0)}_{n} \rangle\\ \nonumber
( \hat{W}_n^{(2)} )_c & = - \frac{1}{2} \hat{P}_n \hat{H}^{1} \left( \hat{S}_n \right)^2 \hat{H}^{1} \hat{P}_n.
\end{align}

Overall
\begin{align}
\label{eqn_appc_38}
 \hat{W}_n^{(2)} & = \left(  \hat{S}_n  \hat{H}^{1} \right)^2 \hat{P}_n  - \left( \hat{S}_n \right)^2 \left( \hat{H}^{1} \hat{P}_n \right)^2 - \frac{1}{2} \hat{P}_n \hat{H}^{1} \left( \hat{S}_n \right)^2 \hat{H}^{1} \hat{P}_n.
\end{align}


\subsection{Third-order corrections}
\label{sec_toc}

Starting again with
\begin{equation}
\label{eqn_appc_39}
\left( \hat{H}^{0} + \lambda \hat{H}^{1} \right) \left[ \sum_{m} \lambda^{m}| \phi^{(m)}_{n} \rangle \right] = \left[ \sum_{m} \lambda^{m}  E^{(m)}_{n} \right] \left[ \sum_{m} \lambda^{m}_{n}| \phi^{(m)}_{n} \rangle \right],
\end{equation}
and isolating those terms involving $\lambda^{3}$ gives
\begin{equation}
\label{eqn_appc_40}
\hat{H}^{0} | \phi^{(3)}_{n} \rangle + \hat{H}^{1} | \phi^{(2)}_{n} \rangle = E^{(0)}_{n} | \phi^{(3)}_{n} \rangle +  E^{(1)}_{n} | \phi^{(2)}_{n} \rangle + E^{(2)}_{n} | \phi^{(1)}_{n} \rangle+ E^{(3)}_{n} | \phi^{(0)}_{n} \rangle.
\end{equation}
Taking the inner product with  $| \phi^{(0)}_{n} \rangle$, and remembering that only orthogonal corrections are allowed to ensure normalisation, leads to
\begin{align}
\label{eqn_appc_41}
E^{(3)}_{n} & =  \langle  \phi^{(0)}_{n} | \hat{H}^{1} | \phi^{(2)}_{n} \rangle,
\end{align}
giving an expression for $E^{(3)}_{n}$ in terms of the $ | \phi^{(2)}_{n} \rangle $ given by (\ref{eqn_appc_33}).

Considering each of the terms in (\ref{eqn_appc_36}) separately. For the first term
\begin{align}
\label{eqn_appc_42}
& \sum_{m \neq n}  \sum_{m' \neq n}   \left[ - \frac{\langle \phi^{(0)}_{m} | \hat{H}^{1} | \phi^{(0)}_{m'} \rangle }{ E_m^{(0)} - E_n^{(0)}} \right] 
\left[   \frac{ \langle \phi^{(0)}_{m'} | \hat{H}^{1} | \phi^{(0)}_{n} \rangle }{ E_{n}^{(0)} - E_{m'}^{(0)}} \right] \langle  \phi^{(0)}_{n} | \hat{H}^{1} |\phi^{(0)}_{m} \rangle \\ \nonumber
= &
  | \phi^{(0)}_{n} \rangle \langle  \phi^{(0)}_{n} | \hat{H}^{1}  \left[ \sum_{m \neq n} \frac{ |\phi^{(0)}_{m} \rangle\langle \phi^{(0)}_{m} | }{ E_m^{(0)} - E_n^{(0)}} \right]  \hat{H}^{1}  \left[  \sum_{m' \neq n}    \frac{ | \phi^{(0)}_{m'} \rangle \langle \phi^{(0)}_{m'} |  }{ E_{m'}^{(0)} - E_{n}^{(0)}} \right] \hat{H}^{1} \\ \nonumber
& = {\rm Tr} \left[  \hat{H}^{1}  \hat{S}_n \hat{H}^{1} \hat{S}_n \hat{H}^{1} \hat{P}_n    \right],
\end{align}
and for the second term
\begin{align}
\label{eqn_appc_43}
& \sum_{m \neq n}  \sum_{m' \neq n}    \langle \phi^{(0)}_{n} | \hat{H}^{1} | \phi^{(0)}_{n} \rangle \langle  \phi^{(0)}_{n} | \hat{H}^{1} |\phi^{(0)}_{m} \rangle  \left[ \frac{\langle \phi^{(0)}_{m} | \phi^{(0)}_{m'} \rangle }{ E_m^{(0)} - E_n^{(0)}} \right] \left[   \frac{ \langle \phi^{(0)}_{m'} | \hat{H}^{1} | \phi^{(0)}_{n} \rangle }{ E_{n}^{(0)} - E_{m'}^{(0)}} \right]   \\ \nonumber
&  = {\rm Tr} \left[-  | \phi^{(0)}_{n} \rangle  \langle \phi^{(0)}_{n} | \hat{H}^{1} | \phi^{(0)}_{n} \rangle \langle  \phi^{(0)}_{n} | \hat{H}^{1}  \left[  \sum_{m \neq n} \frac{ |\phi^{(0)}_{m} \rangle \langle \phi^{(0)}_{m} }{ E_m^{(0)} - E_n^{(0)}} \right] \left[ \sum_{m' \neq n}    \frac{  | \phi^{(0)}_{m'} \rangle\langle \phi^{(0)}_{m'} |}{ E_{m'}^{(0)} - E_{n}^{(0)}} \right]  \hat{H}^{1} \right] \\ \nonumber
& = - {\rm Tr} \left[     \hat{H}^{1} \hat{S}_n^2 \hat{H}^{1}   \hat{P}_n  \hat{H}^{1} \hat{P}_n \right].
\end{align}
Overall, the change in the eigenvalues at third order is
\begin{align}
\label{eqn_appc_44}
E^{(3)}_{n} & =  {\rm Tr} \left[  \hat{H}^{1}  \hat{S}_n \hat{H}^{1} \hat{S}_n \hat{H}^{1} \hat{P}_n    \right]  - {\rm Tr} \left[     \hat{H}^{1} \hat{S}_n^2 \hat{H}^{1}   \hat{P}_n  \hat{H}^{1} \hat{P}_n \right].
\end{align}

\subsection{Summary}
\label{sec_sum}

Using (\ref{eqn_appc_12}), (\ref{eqn_appc_30}), and  (\ref{eqn_appc_44}), the total deviation in eigenvalue $n$ is
\begin{align}
\label{eqn_appc_45}
\delta E_{n} (\lambda) & \approx \lambda E^{(1)}_{n} + \lambda^{2} E^{(2)}_{n} + \lambda^{3} E^{(3)}_{n} + \mathcal{O} (\lambda^4)\\ \nonumber
\delta E_{n} (\lambda) & \approx \lambda {\rm Tr} \left[\hat{H}^{1} \hat{P}_n \right] \\ \nonumber
& - \lambda^2 {\rm Tr} \left[   \hat{H}^{1} \hat{S}_n  \hat{H}^{1} P_{n} \right] \\ \nonumber
& + \lambda^3 {\rm Tr} \left[  \hat{H}^{1}  \hat{S}_n \hat{H}^{1} \hat{S}_n \hat{H}^{1} \hat{P}_n    \right]  - \lambda^3 {\rm Tr} \left[     \hat{H}^{1} \hat{S}_n^2 \hat{H}^{1}   \hat{P}_n  \hat{H}^{1} \hat{P}_n \right] \\ \nonumber
& + \mathcal{O} (\lambda^4).
\end{align}
Likewise, 
\begin{equation}
\label{eqn_appc_46}
| \phi_{n} \rangle (\lambda) = \left[ P_n + \lambda W_n^{(1)} + \lambda^2 W_n^{(2)} + .... \right] | \phi^{(0)}_{n} \rangle,
\end{equation}
where, according to (\ref{eqn_appc_19c}),
\begin{align}
\label{eqn_appc_47}
\hat{W}_n^{(1)} & =  - \hat{S}_n  \hat{H}^{1} \hat{P}_n \\ \nonumber
 \hat{W}_n^{(2)} & = \left(  \hat{S}_n  \hat{H}^{1} \right)^2 \hat{P}_n  - \left( \hat{S}_n \right)^2 \left( \hat{H}^{1} \hat{P}_n \right)^2 - \frac{1}{2} \hat{P}_n \hat{H}^{1} \left( \hat{S}_n \right)^2 \hat{H}^{1} \hat{P}_n.
\end{align}
which reproduces eqn 4.23, p 103, of Kato\cite{Kat1}. 

\end{document}